\documentstyle[11pt]{article}
\newcommand{\setthmnum}[2]{%
\renewcommand{\thesection}{\arabic{section}}%
\setcounter{section}{#1}%
\setcounter{THEOREM}{#2}%
}

\newcounter{avgSecNum}
\newcounter{avgThmNum}
\newcounter{compSecNum}
\newcounter{compThmNum}
\newcounter{secStack}
\newcounter{thmStack}

\newcommand{\pushsec}{\setcounter{secStack}{\value{section}}}
\newcommand{\popsec}{\setcounter{section}{\value{secStack}}}
\newcommand{\pushthm}{\setcounter{thmStack}{\value{THEOREM}}}
\newcommand{\popthm}{\setcounter{THEOREM}{\value{thmStack}}}
\newcommand{\othm}[2]{\pushsec\pushthm\setthmnum{\value{#1}}{\value{#2}}}
\newcommand{\eothm}{\popthm\popsec}
\newcommand{\opro}[2]{\pushsec\pushthm\setthmnum{\value{#1}}{\value{#2}}}
\newcommand{\eopro}{\popthm\popsec}

\advance \topmargin by -\headheight
\advance \topmargin by -\headsep

\evensidemargin \oddsidemargin

\newcommand{\tempvar}{Z}

\newcommand{\gv}{\ | \ }
\newcommand{\lgv}{\ \ \vline \ \ }
\newcommand{\abs}[1]{\left|{#1}\right|}

\newcommand{\real}{\mb{R}}

\newcommand{\mprop}[1]{Proposition~\ref{#1}}
\newcommand{\mlem}[1]{Lemma~\ref{#1}}
\newcommand{\mthm}[1]{Theorem~\ref{#1}}

\newcommand{\sct}[1]{Section~\ref{#1}}

\newcommand{\eps}{\varepsilon}

\newcommand{\pcr}{C_p}
\newcommand{\qcr}{C_q}
\newcommand{\qrr}{R_q}

\newcommand{\cb}[2]{\scriptsize \left(\!\!\begin{array}{c}#1\\#2\end{array}\!\!\right)}

\newcommand{\sbset}{\subseteq}

\newcommand{\mc}[1]{{\cal #1}}
\newcommand{\mr}[1]{\mbox{#1}}

\newcommand{\mi}[1]{{\it #1}}
\newcommand{\mb}[1]{{\bf #1}}

\newcommand{\ms}[1]{{\sf #1}}

\newcommand{\emph}[1]{{\em #1\/}}

\newcommand{\tb}[1]{{\bf #1}}

\newcommand{\tsc}[1]{{\sc #1}}

\newcommand{\textit}[1]{{\it #1}}

\newcommand{\rar}{\rightarrow}

\newcommand{\bfloor}[1]{\left\lfloor {#1} \right\rfloor}

\newcommand{\bceil}[1]{\left\lceil {#1} \right\rceil}

\newcommand{\ktime}{\bceil{\frac{k}{2}}}

\newcommand{\commentout}[1]{}
\newcommand{\etc}{{etc.}}
\newcommand{\Ocal}{\mc{O}}
\newcommand{\set}[1]{\{#1\}}
\newcommand{\cjk}{C_3(j,k)}
\newcommand{\cijk}{F_3(i,j,k)}
\newcommand{\start}{\mi{start}}
\newcommand{\finish}{\mi{finish}}
\newcommand{\cjkp}{F_3(j,k)}
\newcommand{\fract}[2]{\frac{\stackrel{}{{#1}}}{#2}}

\newcommand{\sd}{\mi{sender}}

\newcommand{\Hm}{\mi{m}}
\newcommand{\Lm}{\mbox{\mi{m}}}

\renewcommand{\mr}[1]{{\rm #1}}
\renewcommand{\cb}[2]{\left(\!\begin{array}{c} #1 \\ #2 \end{array}\!\right)}

\newcommand{\aeq}{\approx}

\newcommand{\tcmp}{\#\mbox{-}{\sdrc}}
\newcommand{\tcst}{\mb{c}\mbox{-}{\sdrc}}
\newcommand{\tchb}{\mb{c}\mbox{-}{\hbmsg}}
\newcommand{\nhb}{\#\mbox{-}{\hbmsg}}
\newcommand{\nhbav}{\nhb^{\mr{avg}}}

\newcommand{\totc}{\mb{c}^\mr{total}}
\newcommand{\cterm}{\bceil{\frac{2\ttime}{\hbt}}}

\newcommand{\rb}{\ms{SR}_{hb}}

\newcommand{\spc}{\mb{S}}

\newcommand{\blst}{\begin{list}{}{}}
\newcommand{\elst}{\end{list}}
\newcommand{\bit}{\begin{itemize}}
\newcommand{\eit}{\end{itemize}}
\newcommand{\ben}{\begin{enumerate}}
\newcommand{\een}{\end{enumerate}}
\newcommand{\comm}[1]{}
\newcommand{\ds}{\displaystyle}
\newcommand{\ts}{\textstyle}

\newcommand{\hbmsg}{\mbox{\tsc{hbmsg}}}
\newcommand{\hbt}{\delta}

\newcommand{\sfa}{\ms{a}}

\newcommand{\percent}{\%}
\newcommand{\sdrc}{\ms{SR}}

\newcommand{\lf}{\gamma}
\newcommand{\cm}{\mb{c}\mbox{-}\psend}
\newcommand{\csend}{\mb{c}_0}
\newcommand{\csendav}{\mb{c}^{\mr{avg}}}
\newcommand{\csendst}{\mb{c}_1}
\newcommand{\cw}{\mb{c}\mbox{-}\mi{wait}}

\newcommand{\cpsend}{\mb{c}}

\newcommand{\msgt}{{\delta}}

\newcommand{\ptc}{\prt}

\newcommand{\numprt}{{\#}\mbox{-}{\psend}}

\newcommand{\prt}{\sdrc}

\newcommand{\tow}{t\mbox{-}{\mi{wait}}}

\newcommand{\imsg}{\frac{1}{(1-\lf)}}

\newcommand{\jmsg}{\frac{1}{(1-\lf)^2}}

\newcommand{\cpl}[1]{\overline{#1}}
\newcommand{\spb}{\sigma}

\newcommand{\psend}{\ms{send}}
\newcommand{\prcv}{\ms{receive}}
\newcommand{\rsend}{\ms{SEND}}
\newcommand{\rrcv}{\ms{RECEIVE}}

\newcommand{\rrcving}{\ms{RECEIV}${ing}$ }

\newcommand{\prcving}{\ms{receiv}${ing}$ }
\newcommand{\prcvd}{\ms{receive}${d}$ }
\newcommand{\prcvs}{\ms{receive}${s}$ }

\newcommand{\rsending}{\ms{SEND}${ing}$ }

\newcommand{\psending}{\ms{send}${ing}$ }
\newcommand{\psent}{\ms{sent}}
\newcommand{\psends}{\ms{send}${s}$ }

\newcommand{\req}{\mbox{\tsc{req}}}

\newcommand{\pf}{\beta}
\newcommand{\pct}{\alpha}
\newcommand{\ttime}{\tau}

\newcommand{\E}{\mb{E}}

\newcommand{\ub}{\mbox{\tb{meet}-\tb{Bob}}}
\newcommand{\cow}{\mbox{\tb{c}-\tb{wait}}}
\newcommand{\ack}{\mbox{\tsc{ack}}}

\newcommand{\sdp}{{\ptc}_s^{\hbt}}
\newcommand{\rdp}{{\ptc}_r^{\hbt}}
\newcommand{\triv}{{\ptc}_{{\it tr}}}
\newcommand{\esd}{\E^{\sdp}}

\newtheorem{THEOREM}{Theorem}[section]
\newenvironment{theorem}{\begin{THEOREM} \hspace{-.85em} {\bf :} }%
                        {\end{THEOREM}}
\newtheorem{LEMMA}[THEOREM]{Lemma}
\newenvironment{lemma}{\begin{LEMMA} \hspace{-.85em} {\bf :} }%
                      {\end{LEMMA}}
\newtheorem{COROLLARY}[THEOREM]{Corollary}
\newenvironment{corollary}{\begin{COROLLARY} \hspace{-.85em} {\bf :} }%
                          {\end{COROLLARY}}
\newtheorem{PROPOSITION}[THEOREM]{Proposition}
\newenvironment{proposition}{\begin{PROPOSITION} \hspace{-.85em} {\bf :} }%
                            {\end{PROPOSITION}}
\newtheorem{DEFINITION}[THEOREM]{Definition}
\newenvironment{definition}{\begin{DEFINITION} \hspace{-.85em} {\bf :} \rm}%
                            {\end{DEFINITION}}
\newtheorem{CLAIM}[THEOREM]{Claim}
\newenvironment{claim}{\begin{CLAIM} \hspace{-.85em} {\bf :} \rm}%
                            {\end{CLAIM}}
\newtheorem{EXAMPLE}[THEOREM]{Example}
\newenvironment{example}{\begin{EXAMPLE} \hspace{-.85em} {\bf :} \rm}%
                            {\end{EXAMPLE}}
\newtheorem{REMARK}[THEOREM]{Remark}
\newenvironment{remark}{\begin{REMARK} \hspace{-.85em} {\bf :} \rm}%
                            {\end{REMARK}}
\newcommand{\thm}{\begin{theorem}}
\newcommand{\lem}{\begin{lemma}}
\newcommand{\pro}{\begin{proposition}}
\newcommand{\dfn}{\begin{definition}}
\newcommand{\rem}{\begin{remark}}
\newcommand{\xam}{\begin{example}}
\newcommand{\cor}{\begin{corollary}}
\newcommand{\prf}{\noindent{\bf Proof:} }
\newcommand{\ethm}{\end{theorem}}
\newcommand{\elem}{\end{lemma}}
\newcommand{\epro}{\end{proposition}}
\newcommand{\edfn}{\bbox\end{definition}}
\newcommand{\erem}{\bbox\end{remark}}
\newcommand{\exam}{\bbox\end{example}}
\newcommand{\ecor}{\end{corollary}}
\newcommand{\eprf}{\bbox\vspace{\topsep}}
\newcommand{\beqn}{\begin{equation}}
\newcommand{\eeqn}{\end{equation}}
\newcommand{\bbox}{\vrule height7pt width4pt depth1pt}

\newcommand{\clm}{\begin{claim}}
\newcommand{\eclm}{\end{claim}}
\newcommand{\inter}{\cap}
\renewcommand{\phi}{\varphi}
\newcommand{\eg}{{e.g.}}
\newcommand{\ie}{{i.e.}}

\begin{document}
\thispagestyle{empty}
\title{A Decision-Theoretic Approach to Reliable Message Delivery\thanks{This
work was supported in part by NSF
grant IRI-96-25901 and by the Air Force Office of
Scientific Research grant F49620-96-1-0323.}}
\author{Francis C. Chu \and
%joe12 moved this down; I think it's more standard
% and Joseph Halpern \\
%{\tt \{fcc,halpern\}@cs.cornell.edu}
%}
Joseph Y. Halpern}

\date{
Department of Computer Science \\
Upson Hall, Cornell University \\
Ithaca, NY  14853-7501, USA
\\[.4cm]
{\tt \{fcc,halpern\}@cs.cornell.edu}
}

\maketitle

%fcc15  Do we want to put this back in?
\hfill
\begin{minipage}{7.5cm}
\small
To be, or not to be: that is the question:\\
Whether 'tis nobler in the mind to suffer\\
The slings and arrows of outrageous fortune,\\
Or to take arms against a sea of troubles,\\
And by opposing end them?
%\hfill \emph{Hamlet} Act III, Scene i

\hfill \emph{Hamlet} (III, i)
\end{minipage}

\begin{abstract}
We argue that the tools of decision theory
%fcc15 need to
should
be taken more seriously in the specification and analysis of systems.  We
illustrate this by considering a simple problem involving reliable
communication, showing how considerations of utility and probability can be
used to decide when it is worth sending heartbeat messages and, if they are
sent, how often they should be sent.
\end{abstract}


\section{Introduction}
\label{s:intro}

%fcc23  rephrased
%In designing and running systems, decisions must always be made:
%joe23; I don't like ``replete'' in this context:
%Systems design and implementation are replete with choices:
In designing and implementing systems, choices must always be made:
When should we
garbage collect?  Which transactions should be
%fcc15 aborted?
aborted (to remove a deadlock)?
How big should the page table be?  How often should we resend a message that is
not acknowledged?  Currently, these decisions seem to be made based on
intuition and experience.  However, studies suggest that decisions made in this
way are prone to inconsistencies and other
%fcc11
%pitfalls.
%(See \cite{rs89} for some common mistakes
%people make when making decisions.)
pitfalls~\cite{rs89}.
Just as we would like to formally verify critical programs in order to avoid
bugs, we would like to apply formal methods when making important decisions in
order to avoid making suboptimal decisions.  Mathematical logic has given us
the tools to verify programs, among other things.  There are also standard
mathematical tools for making decisions, which come from \emph{decision
theory}~\cite{res}.  We believe that these tools need to be taken more
seriously in systems design.
%fcc23 rephrased
%This paper can be viewed
We view this paper
as a first step towards showing how this can be done and the benefits of so
doing.

\label{ss:ex1}

Before we delve into the technical details, let us consider a motivating
example.  Suppose Alice made an appointment with Bob and the two are supposed
to meet at five. Alice shows up at five on the dot but Bob is nowhere in sight.
At 5:20, Alice is getting restless.
%fcc23 a little tie-in
%Should she stay or leave?
The question is ``To stay or not to stay?''
%fcc23
%That depends.
The answer, of course, is 
%fcc25 that it depends.
``It depends.''
Clearly, if Bob is an important business client and they are about to close a
deal, she might be willing to wait longer.  On the other hand, if Bob is an
in-law she never liked, she might be happy to have an excuse to leave.  At a
more abstract level, the \emph{utility} of actually having the meeting is (or,
at least, should be) an important ingredient in Alice's calculations.  But
there is another important ingredient: likelihood.  If 
%fcc25 Alice meets Bob
%joe25: typo
%Alice and bob meet
Alice and Bob meet
frequently, she may know something about how prompt he is.  Does he typically
arrive more or less on time (in which case the fact that he is twenty minutes
late might indicate that he is unlikely to come at all) or is he someone who
quite often shows up half an hour late?  Not surprisingly, utilities and
probabilities (as measures of likelihood) are the two key ingredients in
decision theory.

While this example may seem far removed from computer systems, it can actually
be viewed as capturing part of \emph{atomic commitment}~\cite{sks97}.  To see
this, suppose there is a coordinator $p_c$ and two other processes $p_a$ and
$p_b$ working on a transaction.  To commit the transaction, the coordinator
must get a {\sc yes} vote from both $p_a$ and $p_b$.  Suppose the coordinator
gets a {\sc yes} from $p_a$, but hears nothing from $p_b$.  Should it continue
to wait or should it abort the transaction?
%joe10
%To make this decision, we would
%claim that the type of information we need
The
%fcc23 type (two types of information needed)
types
of information we need to make this decision
%fcc23 is precisely that
are precisely those
considered in the Alice-Bob example above: probabilities and utilities.  While
it is obvious
%fcc23 added
that
the amount of time Alice should wait depends on the situation,
%fcc17
atomic commit protocols typically have a context-independent timeout period.
If $p_c$ has not heard from all the processes by the end of the timeout period,
then the transaction is aborted.  Since the importance of the transaction and
the cost of waiting are context-\emph{dependent}, the timeout period would not
be appropriate in every case.

Although it is not done in atomic commit protocols, there certainly is an
awareness that we need to take utilities or costs into account elsewhere in the
database literature.\footnote{Awareness of cost is by no means limited to the
database community.
%joe14:
For example, a sampling of the papers at a recent DISC (Distributed Computing)
Conference, showed that cost was mentioned in at least seven of
them~\cite{bmpp98,cm98,ehw98,fms98,mib98,tar98,yaw98}.  Cost and utility are
also discussed, for example, in~\cite{kes97} and~\cite{kl95,ls98}.}
%joe14: S. Keshav's book on networks certainly mentions utility (I have
%a copy I can show you).  You should also check the Lazar stuff that I
%pointed out too.
For example, when a deadlock is detected in a database system, some
transaction(s) must be rolled back to break the deadlock.  How do we decide
which ones?  The textbook
%joe18: back to original; see below
%fcc17 response~\cite[p.~497]{sks97} is that ``[we] should
%response is that ``[we] should
response~\cite[p.~497]{sks97} is that ``[we] should
roll back those transactions that will incur the minimum cost.
Unfortunately,
%joe18: I actually preferred the original.  In any case, you don't want
%two consecutive periods.  Went back to the original to avoid this problem.
%the term minimum cost is not a precise one.''~\cite[p.~497]{sks97}.
the term minimum cost is not a precise one.''
Typically, costs have been quantified in this context by considering things
like how long the transaction has been running and how much longer it is likely
to run, how many data items it has used, and how many transactions will be
involved in a rollback.  This is precisely the type of analysis to which the
tools of decision theory can be applied.
Ultimately we are interested in when each transaction of interest will complete
its task.  However, some transactions may be more important than others.  Thus,
ideally, we would like to attach a utility to each vector of completion times.
Of course, we may be uncertain about the exact outcome (\eg, the exact running
time of a transaction).  This is one place where likelihood enters the picture.
Thus, in general, we will need both probabilities and utilities to decide which
are the most appropriate transactions to abort.
Of course, obtaining the probabilities and utilities may in practice be
difficult.  Nevertheless, we may often be able to get reasonable estimates of
them (see \sct{s:conrem} for further discussion of this issue), and use them to
guide our actions.

In this paper, we illustrate how decision theory
%joe10
%might be used---and some of
%the subtleties that arise in using it---in making systems decisions, by
can be used and some of the subtleties that arise in using it.  We focus on one
simple problem involving reliable communication.  For ease of exposition, we
make numerous simplifying assumption in our analysis.  Despite these
simplifying assumptions, we believe our results show that decision theory can
be used in the specification and design of systems.

We are not the first to attempt to apply decision theory in computer science.
Shenker and his colleagues~\cite{BBS98,BS98}, for example, have used ideas from
decision theory to analyze various network protocols; Microsoft has a Decision
Theory and Adaptive Systems group that has successfully used decision theory in
a number of applications, including troubleshooting problems with printers and
intelligent user interfaces in
%fcc19 Office
%fcc11
%'97 (for further details, see %http://www.microsoft.com/research/dtg).
Office~'97.
(See {\tt http:/$\!$/research.microsoft.com/dtas/} for further details.)
%fcc21
Mikler et al.~\cite{mhw96} have looked at network routing from a
utility-theoretic 
%fcc25 made the following sentence into a footnote: it doesn't go well with
%the next sentence which begins with ``however''.
%perspective.
%fcc23
%joe25: unfootnoted and removed the however
%perspective.\footnote{One important difference between our paper and
%theirs is
perspective.  One important difference between our paper and theirs is
that they do not treat the utility function as a given: Their aim is to
find a good utility function so that the routing algorithm would exhibit
the desired 
behavior (of avoiding the hot spot).
%joe25
%However, 
More generally,
our focus on writing specifications in terms of utility, and the subtleties
involved with the particular application we consider here---reliable
communication---make the thrust of this paper quite different from others in
the literature.

The rest of this paper is organized as follows.  We
%joe10: line shaving
%first give a brief review of
briefly review
some decision-theoretic concepts in
\sct{s:primer}.  In \sct{s:relcom} we describe the basic model and introduce
the communication problem that serves as
%joe10
%a running example throughout the paper.
our running example.  We show that the expected cost of even a single attempt
at reliable communication is infinite if there is uncertainty about process
failures.  We then show in \sct{s:hb1} how we can achieve reliable
communication with finite expected cost by augmenting our system with
%joe12
%a \emph{heartbeat failure detector}, in the spirit of \cite{act97}.
%fcc17
\emph{heartbeat} messages, in the spirit of Aguilera, Chen, and
Toueg~\cite{act97}. However, the heartbeat messages themselves come at a cost;
%joe10
%In \sct{s:hb2} we investigate the cost of sending heartbeats.
this cost is investigated in \sct{s:hb2}.
%joe10: there are other assumptions too; why focus on these?
%under the assumption that a link
%may drop a particular message with a fixed probability $\lf$,
%independent of the state of the link.
%Finally, in \sct{s:conrem} we give some concluding remarks.
We offer some conclusions in \sct{s:conrem}.
%joe22
Some proofs are relegated to the appendix.

\section{A Brief Decision Theory Primer}
\label{s:primer}

The aim of decision theory is to help agents make rational decisions.  There
are a number of equivalent ways of formalizing the decision process.
%fcc19 Here,
In this paper,
we assume that (a) we have a set $\Ocal$ of possible states of the world or
%fcc17
\emph{outcomes}, (b) the agent can assign a \emph{utility}
%fcc21 (from $\real \cup \{\infty, -\infty\}$),
 from $\real \cup \{\infty, -\infty\}$ (denoted $\real^*$)
to each outcome in $\Ocal$, and (c)
each action or choice $\sfa$ of the agent can be associated with a subset
$\Ocal_{\sfa}$ of $\Ocal$ and
%fcc17
a probability measure $\Pr_{\sfa}$ on $\Ocal_{\sfa}$.
%joe22
(This is essentially equivalent to viewing $\Pr_{\sfa}$ as a probability
measure on $\Ocal$ which assigns probability 0 to the outcomes in
$\Ocal - \Ocal_{\sfa}$.)
%this says if action $\sfa$ is performed, then outcomes in $\Ocal -
%\Ocal_{\sfa}$ are impossible.  It is stronger than saying that outcomes in
%$\Ocal - \Ocal_{\sfa}$ occur with probability 0.}

Roughly speaking, the utility associated with an outcome measures how happy the
agent would be if that outcome occurred.  Thus, utilities quantify the
preferences of the agent.  The agent prefers outcome $o_1$ to outcome $o_2$ iff
the utility of $o_1$ is higher than that of $o_2$.  The set $\Ocal_{\sfa}$ of
outcomes associated with an action or choice $\sfa$ are the outcomes that might
arise if $\sfa$ is performed or chosen; the probability
%joe22: called it a probability measure consistently
%distribution on
measure on
$\Ocal_{\sfa}$ represents how likely each outcome is if $\sfa$ is performed.
These are highly nontrivial assumptions, particularly the last two.  We discuss
them (and to what extent they are attainable in practice) in \sct{s:conrem}.
For now, though, we just focus on their consequences.

Recall that a
%fcc19 random variable
\emph{random variable}
on the set $\Ocal$ of outcomes is a function from
$\Ocal$ to
%fcc19 the reals
%joe20
%$\real$.
%fcc21
$\real^*$.
Given a random variable $X$ and a probability measure $\Pr$ on the
outcomes, the
%fcc19 expected value
\emph{expected value}
of $X$ with respect to $\Pr$, denoted 
%fcc25 (we use superscript everywhere else)
%$\E_{\Pr}(X)$, is
$\E^{\Pr}(X)$, is
%fcc24  (We never actually use this definition in the paper.)
%$\sum_{o \in \Ocal} \Pr(o) X(o)$.  (We drop the subscript $\Pr$ if it is clear
%from the context.)
$\sum_{v \in X(\Ocal)} v\Pr(X=v)$, where $X(\Ocal)$ is the range of $X$ and
$X=v$ denotes the set $\{ o \in \Ocal : X(o) = v \}$.  We drop the 
%fcc subscript
superscript 
$\Pr$ if it is clear from the context.  Note that utility is just a random
variable on outcomes.  Thus, with each action or choice, we have an associated
%fcc19 expected utility,
\emph{expected utility},
where the expectation is taken with respect to
%fcc19
%the probability distribution associated with the
%choice.
$\Ocal_{\sfa}$ and $\Pr_{\sfa}$.
%fcc21
Since utilities can be infinite, we need some conventions to handle infinities
in arithmetic expressions.
%joe22: reordered a little
 If $x > 0$, we let $x \cdot \pm \infty = \pm \infty$;
if $x < 0$, we let $x \cdot \pm \infty = \mp \infty$.
For all $x \in \real$, we let $x + \pm \infty = \pm \infty$.
%If $x$ is a positive real number, we let  .
Finally, we let $0 \cdot \infty = 0$.
%joe22
We assume that $+$ and
%fcc23 used $\cdot$ for times
%$\times$
$\cdot$
remain commutative on $\real^*$, so this covers all the cases but $\infty + (-
\infty)$, which we take to be undefined.

The ``rational choice'' is
%fcc21 then
typically taken to be the one that maximizes
expected utility.  While other notions of rationality are clearly possible, for
the purposes of this paper, we focus on expected utility maximization.  Again,
see \sct{s:conrem} for further discussion of this issue.

We can now apply these notions to the Alice-Bob example from the introduction.
One way of characterizing the possible outcomes is as pairs $(m_a, m_b)$, where
$m_a$ is the number of minutes that Alice is prepared to
%fcc15 wait
wait,
and $m_b$ is the time that Bob actually arrives.  (If Bob does not arrive at
all, we take $m_b = \infty$.)  Thus, if $m_a \ge m_b$, then Alice and Bob meet
at time $m_b$ in the outcome $(m_a, m_b)$.  If $m_a < m_b$, then Alice leaves
before Bob arrives.  What is the utility of the outcome $(m_a, m_b)$?  Alice
and Bob may well assign different utilities to these outcomes.  Since we are
interested in Alice's decision, we consider Alice's utilities.  A very simple
assumption is that there is a fixed positive benefit $\ub$ to Alice if she
actually meets Bob and a cost of $\cow$ for each minute she waits, and that
these utilities are additive.
%fcc17
We assume here that $\cow \leq 0$.
%joe22
%\footnote{
(In general, \emph{costs}  are
%joe22
described by
non-positive utilities.)
Under this assumption, the utility of the outcome $(m_a, m_b)$ is
%fcc10 $\ub + m_b(\cow)$ if $m_a \ge m_b$ and $m_a(\cow)$ if $m_a < m_b$.
$\ub + m_b \cow$ if $m_a \ge m_b$ and $m_a \cow$ if $m_a < m_b$.

Of course, in practice, the utilities might be much more complicated and need
not be additive.
%joe10
%It could be that Alice has a magazine to read, so waiting for
%fcc19 If
For example, if
Alice has a magazine to read, waiting for the first fifteen minutes might be
relatively painless, but after that, she might get increasingly frustrated and
the cost of waiting might increase exponentially, not linearly.  The benefit to
meeting Bob may also depend on the time they meet, independent of Alice's
frustration.  For example, if they have a dinner reservation for
%fcc15 6 PM
6~p.m.~at a restaurant half an hour away, the utility of meeting Bob may drop
drastically after 5:30.  Finally,
%joe10
%we can well imagine that
the utility of $(m_a, m_b)$ might depend on $m_b$ even if $m_a < m_b$.  For
example, Alice might feel happier leaving
%joe10
%after 15 minutes if she knew that Bob would
%arrive an hour late than if she know that Bob arrived at 5:16.
at 5:15 if she knew that Bob would arrive at 6:30 than if she knew he
would arrive at 5:16.

Once Alice has decided on a utility function, she has to decide what action to
take.  The only choice that Alice has is how long to wait.  With each choice
$m_a$, the set of possible outcomes consists of those of the form $(m_a, m_b)$,
for all possible choices of $m_b$.  Thus, to compute the expected utility of
the choice $m_a$, she needs a probability measure over this set of
outcomes, which effectively means a probability measure over Bob's
possible arrival times.

This approach of deciding at the beginning how long to wait may seem far
removed from actual practice, but suppose instead Alice sent her assistant
Cindy to meet Bob.  Knowing something about Bob's timeliness
%fcc15 (and
(or
lack thereof), she may well want to give Cindy instructions for how long to wait.
Taking the cost of waiting to be linear in the amount of time that Cindy waits
is now not so unreasonable, since while Cindy is tied up waiting for Bob, she
is not able to help Alice in other ways.  If Cindy goes to meet Bob frequently
for Alice, it may make more sense for Alice just to tell Cindy her utility
function, and let Cindy decide how long to wait based on the information she
acquires regarding
%fcc23 the likelihood of Bob arriving.
Bob's punctuality.
Of course, once we think in
terms of Alice sending an assistant, it is but a small step to think of Alice
running an application, and giving the application instructions to help it
decide how to act.

\section{Reliable Communication}
\label{s:relcom}

We now consider a problem that will serve as a running example throughout the
rest of the paper.  Consider a system
%fcc23  never used again (and it looks the same as the strong failure detector)
%$\sysi$
consisting of a sender $p$ and a receiver $q$ connected by an unreliable
%fcc17
%bidirectional link,
bidirectional link.
%which drops messages with probability $0 < \lf < 1$.  If a
%message is not dropped, the transmission delay is $\ttime$.
%fcc11 	added for $\spc_0$
%Furthermore, the link does not create messages out of thin air.
%fcc17
We assume that the link satisfies the following properties:
\bit
\item The transmission delay of the link is $\ttime$.
\item The link can only fail by losing (whole) messages and the probability of a
message loss is $\lf$.
\eit
%joe22
%Note
We assume
that the transmission delay and the probability of message loss are
independent of the state of the system.%
%\footnote{Our approach would work if
\footnote{The results of this paper hold even if
these quantities do depend on the state of the link.  For example, $\lf$ may be
a function of the number of messages in transit.  We stick to the
simpler model for ease of exposition.}
A process is \emph{correct} if it never crashes.
%fcc23 Let $x \in \set{p, q}$ and
For $x \in \set{p, q}$,
let $\pct_x$ be the probability that $x$ is correct
%fcc19  I'm not so sure about this.  Are we fixing the set of runs and changing
%the distribution or are we varying the set of runs?
%That is, if we select a
%run (uniformly) at random, the probability that $x$ is correct in that run is
%$\pct_x$.
%joe20
%More precisely, the probability of the runs in which $x$ is correct is
%$\pct_x$.
(more precisely, the probability of the set of runs in which $x$ is
correct).
In runs in which $x$ is not correct, $x$ crashes in each time unit with
%fcc23  just say that it is independent of everything else
%probability $\pf_x > 0$.  We assume that events of $p$ being correct
%or of crashing if it is not correct are both independent of the
%corresponding events for
%fcc15 $q$.
%$q$ (and the behavior of the link).
%joe22: We need this in the proof of Proposition 3.5
%We further assume that the probability of $p$ crashing
%at time $t$ given that it
%has not crashed earlier is independent of any time $t-1$ events.
probability $\pf_x > 0$, independent of all other events in the system (such as
the events that occurred during the previous time unit).

%joe23
%We remark that
%fcc23
%joe23
%we believe the assumption
The assumptions that seems most reasonable to us is
%fcc24
%that $\pct_p = \pct_q = 0$---in practice,
that $\pct_p = \pct_q = 0$: in practice,
%is the one that
%%fcc23 seems most reasonable to us
%seems most reasonable
%in practice
there is always a positive probability that a process will crash
in any given round.\footnote{We assume that \emph{round $k$} takes place
between time $k-1$ and $k$.}  We allow the possibility that $\pct_x \ne 0$ to
facilitate comparison to most of the literature, which does not make
probabilistic assumptions about failure.  It also may be a useful way of
modeling
%fcc23 that
the scenario in which
processes stay up forever ``for all practical purposes'' (for
%joe10
%example, that the system has been replaced before the process crashes).
example, if the system 
%fcc25 is replaced before the
is scheduled to be taken off-line before the
%fcc23 process crashes).
processes crash).

We want to implement a reliable link on top of the unreliable
%fcc23 one.
%joe23
link
provided by the system.
%fcc17
That is, we want to implement
%joe22: rewrote and simplified
a reliable send-receive protocol $\sdrc$
using the (unreliable) sends and receives provided by the
link, denoted $\psend$ and $\prcv$.
$\sdrc$ is a \emph{joint} protocol, consisting of a {\rsend} protocol
for the sender and a {\rrcv} protocol for the receiver.
%reliable send and receive
%(denoted as $\rsend$
%and $\rrcv$) as a sequence of the unreliable
%%joe18
%send and receive provided by the
%link (denoted as $\psend$ and $\prcv$).
%joe12: te referee is right here
%More precisely, suppose we use $\psend$/$\prcv$ to denote the (unreliable) send
%and receive provided by the link.
%joe12: added
%We want to implement reliable high-level $\rsend$/$\rrcv$ actions as a
%sequence of $\psend$/$\prcv$ events.
%%fcc17
%Suppose $p$ wants to $\rsend$ a message to $q$
%fcc19
%(or $q$ wants to $\rrcv$ a message from $p$)
%\footnote{We allow receiver-driven
%protocols in which $q$ queries $p$ for the message.}
%joe22: cut the two different sets; we never use it.  (We seem quite
%inconsistent with our notation anyway.) It also causes
%problems  the one place we might consider using it (see below).
% from the set
%$\Mh$.
%%\footnote{We can think of $\Mh$ as application messages or messages
%from
%%higher layers in the protocol stack.}
%To this end, $p$ and $q$ would run some
%protocol $\prt$ that would have them exchange messages from the set
%$\Mp{\prt}$.  (Depending on $\prt$, $\Mh$ may or may not be a subset of
%$\Mp{\prt}$.)  We use $\Hm$ to denote members of $\Mh$ and $\Lm$ to
%denote members of $\Mp{\prt}$ (which may include $\Mh$ as a subset).
%Note that $\prt$ is a joint protocol
%which has two parts: one part for $p$ and one part for $q$.  We refer
%to $p$'s part as $\rsend$ and $q$'s part as $\rrcv$
%joe22
%This distinction is important since in a receiver-driven protocol, $q$
%actually
%initiates the $\sdrc$ protocol.
$\sdrc$ can be initiated by either $p$ or $q$.  A send-receive protocol is said
to be \emph{sender-driven} if it is initiated by $p$ and \emph{receiver-driven}
if it is initiated by $q$.  (Web browsing can be viewed as an instance of a
receiver-driven
%fcc23
%protocol.
activity.
The web browser queries the web server for the content of the page.)  We assume
that $\psend$s and $\prcv$s take place at a time $t$, while $\rsend$s and
$\rrcv$s take place over an interval of time (since, in general, they may
involve a sequence of $\psend$s and $\prcv$s).

%Given an $\Hm$, we use $\prt(\Hm)$ to denote a
%member of $\Mp{\prt}$ that $p$ could $\psend$ if $p$ tries to
%$\rsend(\Hm)$.\footnote{Note that $\Hm$ may or may not be the same as $\prt(\Hm)$.
%$\prt(\Hm)$ may be different from $\Hm$, for example, if $\prt$ adds a header
%to $\Hm$.} For ease of exposition, we will assume that $\prt$ is in fact a 1-1
%function (so for each $\Hm$ there is exactly one possible $\prt(\Hm)$ and if
%$\Hm \neq \Hm'$ then $\prt(\Hm) \neq \prt(\Hm')$).  For brevity, we will
%abbreviate $\prt(\Hm)$ with $\Hmp$.
%and $\rsend$/$\rrcv$ to denote the (reliable) send and receive
%implemented by some protocol.

%joe22
%Note that
We assume that
$\psend$ and $\prcv$ satisfy the following two
%fcc24   If we want to start with ``We assume that'', we don't want to say
%``given our assumptions''.
%properties, given our assumptions about the link:
properties:
\bit
\item If $q$ $\prcvs$ $\Lm$ at time $t$, then $p$ $\psent$ $\Lm$ at time
$t-\ttime$ and $\Lm$ was not lost (since the link cannot create messages or
duplicate messages and the transmission delay is known to be $\ttime$).
\item If $p$ $\psends$ $\Lm$ at time $t$, then with probability $1-\lf$,
$q$ will $\prcv$ $\Lm$ at time $t+\ttime$;
%fcc24  (so that $q$ does not receive m at some other time)
if $q$ does not $\prcv$ $\Lm$ at time $t+\ttime$, $q$ will never $\prcv$ it.
\eit
What specification should $\sdrc$ satisfy?
%joe12
Clearly we do not want
%fcc23 them
the processes
to create messages out of whole cloth.  Thus, we certainly want the following
requirement:
\blst
%fcc11	rephrased (dropped the first requirement)
%	Under the new spc_0, a protocol that does not \psend(m) will never
%	\prcv(m) and so never \rrcv(m).
%fcc17
%\item[$\spc_0$]  If $p$ $\rsend$s $m$  then $p$ must also $\psend$ $m$
%(possibly at some
%fcc17 earlier
%later time); if $q$ $\rrcv$s $m$ from $p$, then $q$
%must also $\prcv$ $m$ from $p$ and at some earlier time $p$ must
%$\rsend$ $m$ to $q$.%
\item[$\spc_0$.]
%joe22: there is a nontrivial problem here.  None of the properties
%connect SEND/RECEIVE to send/receive.  It is perfectly consistent with
%our properties that q RECEIVES at time t + \tau if p SENDS at time t,
%regardless of whether q actually received a message.  (So, in fact, S_0
%does not stop q from creating messages out of whole cloth.) Rewrote.
%Note that it would be difficult to state this with two different
%message sets.
%If $q$ finishes $\rrcving$ $\Hm$ at time $t$, then $p$ started
%$\rsending$ $\Hm$ at time $t'
If $q$ finishes $\rrcving$ $\Hm$ at time $t$, then
%fcc23
%fcc26  I think we need p \rsending...
%$\sdrc$ must have started with $\Hm$ as the message at some time 
$p$ must have started $\rsending$ $\Hm$ at some time
$t' \le t$ and
$q$ must have $\prcvd$ $\Hm$ at some time $t'' \le t$.
\elst
We shall implicitly assume $\spc_0$ without further comment throughout the paper.
%fcc13
%fcc17 $m$
%We also assume that $p$ wants to $\rsend(\Hm)$ for some $\Hm \in \Mh$ at time 0 in
%every run.

The more interesting question is what
%fcc11	In a way the safety requirements are also reliability requirements.
%reliability
liveness
%joe14: wording
%requirements should $\rsend$/$\rrcv$ satisfy.  Perhaps the most obvious
requirements $\sdrc$ should  satisfy.  Perhaps the most obvious
requirement
%fcc11	is
is:
\blst
\item[$\spc_1$.]
If $p$ and $q$ are correct and
%fcc21  $p$
%fcc17
%starts $\rsending$ $\Hm$ then $q$ eventually
$\sdrc$ is started with $\Hm$ as the message, then $q$ eventually
%fcc17
finishes  $\rrcving$ $\Hm$.
\elst
Although $\spc_1$ is very much in the spirit of typical specifications, which
%joe22: only should preceded the word it modifies
%only focuses
focus only
on what happens if processes are correct, we would argue that it
is rather uninteresting, for two reasons (which apply equally well to many
other similar specifications).  The first shows that it is too weak: If $\pct_p
= \pct_q = 0$, then $p$ and $q$ are correct (\ie, never crash) with probability
0.  Thus, specification $\spc_1$ is rather uninteresting in this case:  It is
saying something about a set of runs with vanishingly small likelihood.  The
second problem shows that $\spc_1$ is too strong: In runs where $p$ and $q$ are
correct, there is a chance (albeit a small one) that
%fcc17 $\psend$s the message $m$
%joe18: typo
%the link may loss all messages. In this case, $q$ cannot finish $\rrcv(m)$,
the link may lose all messages. In this case, $q$ cannot finish $\rrcving$ $\Hm$,
%fcc23
%since it cannot distinguish such runs from the ones in which $p$ crashes before
%starting $\rsending$ $\Hm$.
since it cannot $\prcv$ $\Hm$ (as all the messages are lost).
Thus $\spc_1$ is not satisfied.

Of course, both of these problems are well known.  The standard way to
strengthen $\spc_1$ to deal with the first problem is to require only that $p$
and $q$ be correct for ``sufficiently long'', but then we need to quantify
%joe10
%this.  It is far from clear how to specify ``sufficiently long'' in an
%application such as this one.
this; it is far from clear how to do so.  The standard way to deal with the
second problem is to restrict attention to \emph{fair} runs, according to some
notion of fairness~\cite{Francez86}, and require only that $q$ finishes
$\rrcving$ $\Hm$ in fair runs.
%joe10
%While fairness may make it in some cases easier to verify certain
%properties of programs and is a useful abstraction for helping us characterize
%conditions necessary to prove certain properties, one of the things that makes
Fairness is a useful abstraction for helping us characterize conditions
necessary to prove certain properties.  However, what makes fairness of
practical interest is that, under reasonable probabilistic assumptions, it
holds with probability 1.

Our interest here, as should be evident from the introduction,
is to make more explicit use of probability in writing a specification.  For
example, we can write a probabilistic specification like the following:
%fcc12
%\blst
%\item[$\spc_2$.]
%$\ds \lim_{t \rightarrow \infty} \Pr(q$ $\rrcv$s $m$ by time $t$
%$\gv$ $p$  $\rsend$s $m$ at time 0 and $p$ and $q$
%%fcc11  do not  crash up to
%are up at
%time $t) = 1$.
%\elst
\blst
%fcc17
%\item[$\spc_2$.]
%$\ds \lim_{t \rightarrow \infty} \Pr(q$ $\rrcv$s $m$ no later than $t$ time
%units after $p$ $\rsend$s $m$ $\gv$ $p$ and $q$ are up $t$ time units after $p$
%$\rsend$s $m) = 1$.
\item[$\spc_2$.]
$\lim_{t \rightarrow \infty} \Pr(q$ finishes $\rrcving$ $\Hm$ no later than $t$ time
units after the start of $\sdrc \gv p$ and $q$ are up $t$ time units after the
start of $\sdrc) = 1$.
\elst
Requirement $\spc_2$ avoids the two problems we saw with $\spc_1$.  It says, in
a precise sense, that if $p$ and $q$ are up for sufficiently long, then $q$
will
%fcc19 $\rrcv(\Hm)$
$\rrcv$ $\Hm$ with high probability
(where 
%fcc25 now 
``sufficiently long'' is quantified probabilistically).  Moreover,
by making only a probabilistic statement, we do not have to worry about unfair
runs: They occur with probability 0.

%fcc17 took out and rewritten
%joe15: added the rest of this paragraph
%joe22: reinstated some of the following; I think the current writeup
%misses an important point.
The traditional approach has been to separate specifying the properties that a
protocol must satisfy from the problem of finding the best algorithm that meets
the specification.  But that approach typically assumes that properties are
all-or-nothing propositions.  That is, it implicitly assumes that a
desirable
%fcc17
%property must be true in every single run (or perhaps every single fair run) of
property must be true in every run (or perhaps every fair run) of
a protocol.  It does not allow a designer to specify that it may be
acceptable for a desirable property to sometimes fail to hold, if that results
in much better properties holding in general.
We believe that, in
general,
issues of cost should not be separated from the problem of specifying the
behavior of an algorithm.
%One of the shortcomings of the traditional approach is that it limits
%us to
%specifying properties of individual runs only.  A statement like ``if a
%majority of processes is correct, then the correct processes eventually reach
%consensus'' talks about a property of individual runs and may appear as part of
%a traditional specification.  On the other hand, a statement like ``if
%no more than half of the processes may fail, then the correct
%processes will eventually
%reach an agreement $95\percent$ of the time'' talks about a property of the set
%of all runs in which at least half the processes are correct.
%joe18: I wouldn't talk about ``global'' properties in this way; it's
%nonstandard
%It is conceivable that every algorithm that satisfies a particular
%traditional specification also has a somewhat undesirable ``global''
%property (\ie, a property on the set of runs).
A protocol that satisfies a particular traditional specification may
do so at the price of having rather undesirable behavior on a
significant fraction of runs.
%joe22: combined next few sentences
For example,
%a protocol that is absolutely
%safe may have to sacrifice liveness $20\percent$ of the time.
to ensure safety, a protocol may block $20\percent$ of the time.  There may be
an alternate protocol that is unsafe only $2\percent$ of the time but also
blocks only $2\percent$ of the time.
%joe18: this is a distraction
%\footnote{Of
%course, in a traditional specification, liveness is only required if certain
%conditions are satisfied, such as ``the majority of processes is correct.''  So
%the algorithm could still be correct and still fail to satisfy liveness in some
%runs, as long as those runs also fail to satisfy the conditions.}
%If we remove
%the requirement of absolute safety, we may get an algorithm that always
%satisfies liveness but may fail to satisfy safety (say) $2\percent$ of
%the time. Note that a safe algorithm will prevent processes from taking
%a certain
%step $s$ (say making a decision) in a run
%%joe18: typo
%%$r$ in which the processes cannot distinguish it
%$r$ which the processes cannot distinguish
%from a run $r'$ in which a safety violation will result if step
%$s$ were taken, even though in $r$ no such violation will result.  Thus the
%probability
%%joe18: typo
%of
%a safe algorithm blocking could be (much) higher than the
%probability of a safety violation if the processes were allowed to
%proceed despite the possibility of a safety violation.
Whether it is better to violate safety $2\percent$ of the time
%joe22: added
and liveness $2\percent$ of the time or to never violate safety but violate
liveness
$20\percent$ of the time obviously depends on the context.  The problem with
the traditional approach is that this comparison is never even considered (any
algorithm that does not satisfy safety is automatically
%joe22: if you're going to mention FLS, you need to say a little (one
%sentence should suffice) about why they do not always require safety.
dismissed).
%fcc23 for now
%\footnote{Fekete, Lynch, and Shvartsman~\cite{fls97} also considers
%specifications which do not always require safety.}

%A quantative approach would allow us to write more precise specifications.

%Thus it is possible that we have an algorithm $A$ which satisfies a certain
%traditional specification but

%The traditional approach specifies properties of individual runs (e.g., the
%correct processes never disagree)
% run (or fair run).  It

%It does not allow us to specify properties of sets of runs (e.g., the

%fcc17
%While we believe $\spc_2$ is a better specification of what is desired than
While we believe $\spc_2$ is a better specification of what is desired than
$\spc_1$, it is still not good enough for our purposes, since it does not
take costs into account.  Without costs, we still cannot decide if it is better
to violate liveness $20\percent$ of the time or to violate safety $2\percent$ of
%fcc23 for consistency
the time and liveness $2\percent$ of
the time.
%fcc17 cut
%For example, it is possible to satisfy $\spc_2$ using
%an algorithm that has $p$ $\psend$ing its $n\th$ message only at time $2^n$.
%This does not seem so
%fcc17 reasonable.
%reasonable.\footnote{This algorithm is not an exponential back off algorithm,
%since it says $p$ should $\psend(m)$ only at times $2^n$ no matter what
%happens.  Exponential back off algorithms only perform exponential back off
%when collision (or some other problem) is detected.  Furthermore, exponential
%back off algorithms do not increase the exponent indefinitely.}
As a first step to thinking in terms of costs, consider the following
specification:
\blst
\item[$\spc_3$.]
%fcc17 The expected cost of a $\rsend$ is finite.
%joe18: added
For each message $\Hm$,
the expected cost of $\sdrc(\Hm)$ is finite.
\elst
As stated, $\spc_3$ is not 
%joe25: actually, I think it should be two words;  it's a well-defined
%concept, but a concept that is well defined.
%fcc25 well defined, 
%well-defined,
well defined,
since we have not specified the
%joe22
%cost model.
cost function.
We now consider a particularly simple
cost function, much in the spirit of
the Alice-Bob example discussed in \sct{s:primer}.
%fcc23  $\sdrc$ is a protocol
%Suppose we have a protocol
%that implements $\sdrc$.
Let $\sdrc$ be a send-receive protocol.
Its outcomes are just the possible \emph{runs} or \emph{executions}.  We want
to associate with each run its utility.  There are two types of costs we will
take into account: sending messages and waiting.  The intuition is that each
attempt to send a message consumes some system resources and each time unit
spent waiting costs the user.  The total cost is a weighted sum of the two.
%joe22: I think you negate a proposition, not a number.  Cut this anyway.
%the utility of a run is
%fcc17 its negative cost.
%the negation of its cost.

More precisely, let $\cm$ and $\cw$ be constants representing the cost of
$\psending$ a message and of waiting one time unit,
%fcc19
respectively.
%\footnote{Note that we are assuming that, for simplicity, the cost
%of $\psend(\Lm)$ is the same for
%%joe20
%%any
%all $\Lm \in \Mp{\prt}$.}
Given a run $r$, let $\numprt(r)$ be the number
%joe22
(possibly $\infty$)
of
%joe18: ``$\psend{\Lm}$ made by the processes'' sounds funny
%$\psend(\Lm)$ made by the
%%fcc17 protocol ****
%processes
%fcc17 during run $r$.
$\psend$s
%fcc23 that occur in
done by the protocol in
run $r$.
%\footnote{Note that $\numprt(r)$ could be infinite: It is
%possible that for all $t$ there is a $t' > t$ such that
%%joe18
%%a process $\psend(\Lm)$ at time $t$.}
%$\psend(\Lm)$ occurs at time $t'$ for some message $\Lm$.}
%fcc17
%If one of the processes crashes in run $r$ before $q$ $\rrcv$s
%$m$, let $\tow(r)$ be the time of the crash.  Otherwise, if $q$ eventually
%$\rrcv$s $m$ in $r$, let $\tow(r)$ be the time $q$ $\rrcv$s $m$. Let $\tow(r) =
%\infty$ if both processes are correct and $q$ never $\rrcv$s $m$ in $r$.  The
%cost of waiting in run $r$ is then $\tow(r) \cw$.
We now want to define $\tow(r)$, which intuitively is the amount of time $q$
spends waiting to $\rrcv$ $\Hm$.  When should we start counting?  In the
Alice-Bob example, it was clear, since Alice starts waiting for Bob at 5:00.
We do not want to start counting at a fixed time, since we do not assume that
the processes will start their protocol at a particular time.  What we want is
to start at the time when $\sdrc$ is invoked.  When do we stop counting,
assuming we started?  If there are no process crashes, then we stop counting
when $q$ finishes $\rrcving$ $\Hm$.  What if there are process crashes?  In
traditional specifications (such as $\spc_1$), the protocol has no obligations
once a process fails.  To facilitate comparison between our approach and the
traditional approach, we stop counting at the time of a process crash if it
happens before $q$ finishes $\rrcving$ $\Hm$.  (Note that $q$ may
never finish
%joe22
%in case
$\rrcving$ if
a process crashes.)

Let $t_s$ be the time $\sdrc$ is invoked.  (If no such time exists, we let
$\tow(r) = 0$.)  Let $t_p$ be the time $p$ crashes ($t_p = \infty$ if $p$ does
not crash); let $t_q$ be the time $q$ crashes ($t_q = \infty$ if $q$ does not
crash); let $t_f$ be the time $q$ finishes $\rrcving$ $\Hm$ ($t_f = \infty$ if
$q$ does not finish).  Finally let 
%joe24: this could be negative if, say, t_p < t_s
%$\tow(r) = %\min\{t_p, t_q, t_f\} - t_s$.
$\tow(r) = \max\{\min\{t_p, t_q, t_f\}, t_s\} - t_s$.
We take the
%fcc15
(total) cost of run $r$ to be 
%fcc25 $\csend(r) = \numprt(r) \cm + \tow(r) \cw$.
$$\csend(r) = \numprt(r) \cm + \tow(r) \cw.$$
Note that
%joe22: no need for the argument
%$\csend(r)$
$\csend$
is a random variable on runs.
%joe22
%fcc25 For this particular cost function,
If $\csend(r)$ captures the cost of run $r$ 
%joe15
%(which we assume it does),
(as we are assuming here it does), then
$\spc_3$ says that we want
%joe22: here you definitely don't want the r.  You take expectation of a
%random variable (a function).  Changed this everywhere (there were lots
%of occurrences; I may have missed some).
%$\E(\csend(r))$ to be finite.
$\E(\csend) = \E(\numprt) \cm + \E(\tow) \cw$ to be finite.

%fcc21
Note that, if $\sdrc$ is \emph{not} invoked in a run $r$, then $\csend(r) = 0$.
Since we are interested in the expected cost of $\sdrc$, we
%joe22: grammar
%will only consider
consider only
runs in which $\sdrc$ is actually invoked.  Also, since we are interested in
the expected cost of a \emph{single} invocation in this (and the next) section,
we assume
%joe22
for ease of exposition
that the protocol is invoked at time 0 (so $\tow(r) = \min \{t_p,
t_q, t_f\}$) throughout these two sections without further comment.
%If this holds for all runs in the system, then
%$\E(\csend) = 0$ no matter what protocol we are using.  (The other
%specifications are also trivially satisfied by any protocol in a system in
%which $\sdrc$ is never invoked in any run.  Such a system is possible since we
%do not require the process to have something to $\sdrc$: Just because Alice has
%a telephone does not mean she will call Bob.)  We would like to focus on
%systems in which $\sdrc$ actually gets invoked.  Let $\sysi_t$ be a system like
%$\sysi$ with the additional property that $\sdrc$ is invoked no later than time
%$t$ in every run $r$ of $\sysi_t$ such that the invoking process is up at time
%$t$ in $r$.  We say that a protocol \emph{satisfies $\spc_i$ in $\sysi_t$} iff
%all the runs of the protocol in $\sysi_t$ satisfies $\spc_i$.  We say that a
%protocol \emph{satisfies} $\spc_i$ iff it satisfies $\spc_i$ in $\sysi_t$ for
%all $t$.
%
%Note that $\sysi_0$ is a subset of all $\sysi_t$, so if $\spc_i$ is not
%satisfiable in $\sysi_0$, then it is no
%in $\sysi_0$ the only runs with $\csend(r) = 0$ are those with a
%process crash at time 0.

%fcc21
\pro
%joe22
%If $\csend(r)$ characterizes the cost of a run,
%%and $\sdrc$ is invoked at time 0,
$\spc_2$ and $\spc_3$ are incomparable
%joe22
under cost function $\csend$.
\epro
\prf
%fcc15
%Note that
%in this cost function, $\spc_2$ and $\spc_3$ are incomparable.
%fcc17 some rewrites
%fcc21 rewrites
Suppose $\pct_p = \pct_q = 1$.  Consider a
%joe22: added
send-receive
protocol $\prt_0$ in which $p$
$\psends$ $\Hm$ in every round until it $\prcvs$ $\ack(\Hm)$, and $q$
$\psends$ its $k$th $\ack(\Hm)$ $N^k$ rounds after $\prcving$ $m$ for the $k$th
time, where $N \lf > 1$.  (Recall that $\lf$ is the probability of message loss.)
It is easy to see that $\prt_0$ satisfies $\spc_2$.  We show that it does not
satisfy $\spc_3$ by showing that $\E(\numprt) = \infty$.

The basic idea is that $q$ is not acknowledging the receipt of $\Hm$ in a
timely fashion, so $p$ will $\psend$ too many copies of $\Hm$.  Let $A_k = \{r
: $ $q$'s first $k$ $\ack$s are lost and the $(k+1)$st $\ack$ makes it in
$r\}$;
%joe22: we need this too
let $A_\infty = \{r:$ all of $q$'s $\ack$s are lost$\}$.
Note that $\Pr(A_k) = \lf^k (1-\lf)$
%joe22
%fcc24
%and $\Pr(A_\infty) = 0$.
and $\Pr(A_\infty) = 0$ (so we can ignore runs in $A_\infty$ for the purpose of
computing expected cost, since we adopted the convention that $0 \cdot \infty
= 0$).
Note also that $\E(\numprt \gv A_k) \geq N^k$, since $p$ cannot possibly get
its first $\ack(\Hm)$ before time $N^k$
%joe22
in runs in $A_k$.
Thus
\[
\E(\numprt) = \sum_{k=0}^\infty \E(\numprt \gv A_k) \Pr(A_k)
\geq  \ds \sum_{k=0}^\infty N^k \lf^k (1-\lf).
\]
It is clear that the last sum is not finite, since $N \lf > 1$; thus
the algorithm fails to satisfy $\spc_3$.
%\eprf\\
%fcc17 changed to reflect the change to \tow

%fcc21
Suppose $\pct_p = \pct_q = 0$.
%Then it is easy to see that the trivial
%protocol (\ie, the ``do nothing'' protocol) satisfies $\spc_3$
%If in run $r$ $p$ crashes at time 0, $\csend(r) = 0$, since $p$ is not correct
%and $p$ does not complete $\rsend(\Hm)$ (because $p$ does not start
%$\rsend(\Hm)$).  If $p$ does not crash at time 0, then $p$ starts and completes
%$\rsend(\Hm)$ at time 0, since the trivial protocol finishes as soon as it
%starts.  In that case, $\csend(r) = \rrfin$, which is essentially the time $q$
%crashes.  Since $q$ is expected to crash by
%fcc19 $1/\pf_q$,
%$\frac{1}{\pf_q}$,
%we have that
%fcc19 $\E(\csend) = (1-\pf_p)/\pf_q$.
%$\E(\csend) = \frac{1-\pf_p}{\pf_q}$.
%Since $q$ never starts $\rrcving$ $\Hm$ (since nothing ever happens),
%$\csend(r) = 0$ for all $r$.  However, the trivial protocol clearly fails to
%satisfy $\spc_2$.
Consider the trivial protocol (\ie, the ``do nothing'' protocol).  In a round
in which both $p$ and $q$ are up, one of $p$ or $q$ will crash
%fcc23 added
in the next round
%fcc25
%with probability $\pf = \pf_p + \pf_q - \pf_p\pf_q$, so one of them is, so one
%of them is expected to crash by time
with probability $\pf = \pf_p + \pf_q - \pf_p\pf_q$.
So the probability that the first crash happens at time $k$ is $(1-\pf)^k \pf$.
Thus one of them is expected to crash at time
%fcc25: no, it's k (1-\pf)^k \pf
%*****
%joe24: I think this was right before:  The expected time to crash is
%\sum k (1-\pf)^{k-1} \pf = \pf/(1-\pf} \sum k (1-\pf)^k
%                         = \pf/(1-\pf) (1-\pf)/\pf^2 = 1/\pf
%fcc24 $\frac{1}{\pf}$.
%fcc25 added
\[
\begin{array}{lll}
\ds \sum_{k=0}^\infty k (1-\pf)^k \pf 
& = & \ds \frac{\pf(1-\pf)}{(1-(1-\pf))^2} \\
& = & \ds \frac{1-\pf}{\pf}.\\
\end{array}
\]
%fcc25 moved here
(Here and elsewhere in this paper we use the well-known fact that $\sum_{k=0}^\infty
k x^k = \frac{x}{(1-x)^2}$.)
Thus, $\E(\csend) = \frac{1-\pf}{\pf} \cw$ for the
%$\frac{1}{\pf}$.
%Thus, $\E(\csend) = \frac{\cw}{\pf}$ for the
trivial protocol, so the trivial protocol satisfies $\spc_3$, although it
clearly does not satisfy $\spc_2$.~\eprf

The following theorem characterizes when $\spc_3$ is implementable
%fcc23 using
with respect to the
cost function
%joe22
%above
$\csend$.
Moreover, it shows that with this cost function,
when
$\spc_3$
is satisfiable, there are in fact protocols that satisfy $\spc_3$ and $\spc_2$
simultaneously.
%fcc19
%Before we begin, note that if for all runs $r$, $p$ and $q$ never starts
%$\rsend$ and $\rrcv$ in $r$, then any protocol will satisfy both $\spc_2$ and
%$\spc_3$ vacuously.  This is, however, an uninteresting scenario.  So in the
%following, we assume that $p$ will start the protocol at time 0 (in the
%sender-driven case) or $q$ will start the protocol at time 0 (in the
%receiver-driven case).
\thm
\label{thm:s}
%joe22
%If $\csend(r)$ characterizes the cost of a run, then
%$\spc_3$ is implementable in $\sysi$ iff $\pct_p = 0$ or $\pct_q = 0$
Under cost function $\csend$, there is a send-receive protocol
satisfying
$\spc_3$ iff $\pct_p = 0$ or $\pct_q = 0$
or $\pct_q = 1$ or $\pct_p = 1$.  Moreover, if $\pct_p = 0$ or $\pct_q = 0$ or
$\pct_q = 1$ or $\pct_p = 1$, then there is a
%joe22
send-receive
protocol that satisfies both
$\spc_2$ and $\spc_3$.
\ethm
\prf
Suppose $\pct_q = 1$ or $\pct_p = 0$.  Consider the
%fcc23
(sender-driven)
protocol $\prt_1$ in which $p$
%joe12: expanded
%$\psend$s $m$ to $q$ until $p$
%fcc23
%$\rsends$ $\Hm$ by $\psending$ $\Hm$ to $q$ at every round
$\psends$ $m$ to $q$
%fcc11 	added
until
$p$
%fcc21 gets
$\prcvs$ an
$\ack(\Hm)$ from $q$, and $q$ $\psends$ $\ack(\Hm)$
whenever it $\prcvs$ $\Hm$.
%fcc23
$\prt_1$ starts when $p$ first $\psends$ $m$ and $q$ finishes $\rrcving$
$m$ when it first $\prcvs$ $m$.
%joe12: added
%fcc19
%Thus, the $\rsend$ starts when $p$ first  $\psend(m)$ and ends when
%fcc21 Thus,
%$\rsend(\Hm)$ starts when $p$ first $\psends$ $\Hm$ and ends when $p$ $\prcvs$
%$\ack(\Hm)$ for the first time from $q$; $\rrcv(\Hm)$ starts and ends the first
%time $q$ $\prcvs$ $\Hm$.  To see that
%joe22
%$\sdrc$ starts when $p$ first $\psends$ $\Hm$.  $q$ finishes $\rrcving$ $\Hm$
%when it first $\prcvs$ $\Hm$.
To see that
%fcc17 this works,
%joe22
%the algorithm
$\prt_1$
is correct, first consider the case that $\pct_q = 1$.
%fcc17 deleted mention of crash
Let $C_p = \{r : p$ $\prcvs$ $\ack(\Hm)$ at least once from $q$ in $r\}$.
%fcc23
%Clearly
%fcc25  the statements only hold before $p$ $\prcvs$ $\ack$.
%For $r \in C_p$, clearly
%joe25: I don't see why; of course, $p$ won't send after he receives an
%ack, but that's irrelevant
%For $r \in C_p$, 
%fcc26  I was confusing p *receiving* the ack with q *sending* the ack
%After q sends an ack not dropped by the link, it may crash before
%receiving additional messages.  Thus the probability is not just 
%(1-\lf).  It's (1-\lf)*(probability that q stays alive), which is \leq
%(1-\lf).  It doesn't affect too much of the proof, since we just want to 
%show things are finite, but it is technically incorrect to say that
%given any message sent by p q will get it with probability (1-\lf) in
%runs of C_p.  (It is true before q sends an ack that gets through, since
%q *is* guaranteed to stay alive that long in C_p.)
%
%before $p$ $\prcvs$ the first $\ack(\Hm)$ in $r$,
%before $q$ $\psends$ the first $\ack(\Hm)$ $\prcvd$ by $p$ in $r$,
%the probability that $p$ $\psends$ a message $\Hm$ $\prcvd$ by $q$ is 
%the probability that a message $\Hm$ $\psent$ by $p$ is$\prcvd$ by $q$ is 
%fcc25 $1-\lf$;
%$1-\lf$ and
%the probability that both
%$\psend(\Hm)$ and the corresponding
%$\psend(\ack(\Hm))$ are both successful is $(1 - \lf)^2$.
%$\Hm$ is $\prcvd$ by $q$ and the corresponding $\ack(\Hm)$ is $\prcvd$ by $p$
%is $(1-\lf)^2$.
%joe22: N_1 and N_2 are random variables; you should make that clear
%Suppose the $N_1$th copy of $\Hm$
Let $N_1(r)=k_1$ if the $k_1$th copy of $\Hm$
is the first $\prcvd$ by $q$
%and the $N_2$th copy whose
and let $N_2(r) = k_2$ if the $k_2$th copy of $\Hm$ is the one whose
corresponding $\ack(\Hm)$ is the first
$\prcvd$ by $p$.

%joe22
%We expect $N_1$ to be $\imsg$ in runs of $C_p$ and $N_2$ to be $\jmsg$.
%fcc26
%It is straightforward to show that $\E(N_1 \gv C_p) = \imsg$ and $\E(N_2 \gv
%C_p) = \jmsg$.
Since the probability that the link may drop a particular message is $\lf$,
\[
\E(N_1 \gv C_p) 
 =  \sum_{k=1}^\infty k \lf^{k-1} (1-\lf) 
 =  \frac{1-\lf}{\lf} \sum_{k=1}^\infty k \lf^k 
 =  \frac{1-\lf}{\lf} \frac{\lf}{(1-\lf)^2} 
 =  \frac{1}{1-\lf}.
\]
An analogous argument shows that $\E(N_2 \gv C_p) = \jmsg$.
%fcc25 cut  (the original was correct)
\commentout{
Since
a message takes $\ttime$
%fcc23
time units
to arrive and messages are
%fcc24 sent
$\psent$
in every round, it follows that
%fcc25 \tow is when when q gets the message, not when p gets the ack
%fcc24
%joe24: This is not quite right.
%Here's the calculation I did:
%E = \sum_k (\ttime + k) (1-\gamma)^2 (2\gamma - \gamma^2)^k
%   = \ttime (1-\gamma)^2 \sum_k  (2\gamma - \gamma^2)^k +
%     (1-\gamma)^2 \sum_k k (2\gamma - \gamma^2)^k +
%   = \ttime + (2\gamma - \gamma^2)/(1-\gamma)^2
%   = \ttime - 1 +  1/(1-\gamma)^2
%$\E(\tow \gv C_p) = \imsg + \ttime$.
}%\end{commentout}
%fcc25 added explanation
Note that
$\tow(r) = N_1(r) + \ttime - 1$ for $r \in C_p$, so
$\E(\tow \gv C_p) = \E(N_1 \gv C_p) + (\ttime - 1) = \imsg + \ttime - 1$.
%$\E(\tow \gv C_p) = \ttime - 1 + \imsgs$.
Moreover, since $p$ stops
$\psending$ $\Hm$ when it $\prcvs$ $\ack(\Hm)$ from $q$, it will stop
%fcc23 moved to the end:  r 2 \ttime looks bad
%joe22
%in $r$
$2 \ttime$
rounds after the
%joe22
%$N_2$th $\psend$ $\Hm$.
$N_2(r)$th $\psend$ of
%fcc23 $\Hm$.
$\Hm$ in run $r$.
Thus $\jmsg + 2 \ttime - 1$ is the number of
times $p$ is expected to $\psend$ $\Hm$ in runs of $C_p$.  We expect $1-\lf$ of
these to be successful, 
%fcc26
%so $\imsg + (2 \ttime - 1)(1 - \lf)$ is the number of
%times $q$ is expected to $\psend$ $\ack(\Hm)$ in runs of $C_p$.  
so the number of times $q$ is expected to $\psend$ $\ack(\Hm)$ is at most
$\imsg + (2\ttime-1)(1-\lf)$. (The actual expected value is slightly less since
$q$ may crash shortly after $\psending$ the first $\ack(\Hm)$ $\prcvd$ by $p$
in runs of $C_p$).  
We conclude
%fcc26 that $\E(\numprt \gv C_p) = \imsg + \jmsg + (2 \ttime - 1) (2 -\lf)$.
that $\E(\numprt \gv C_p) \leq \imsg + \jmsg + (2 \ttime - 1) (2 -\lf)$.
%fcc17 changed
%It is easy to see that $\E(\tow \gv \cpl{C_p}) \le \E(\tow \gv
%C_p)$, since in runs of $\cpl{C_p}$, $p$ crashes before it completes
%$\rsend(\Hm)$ (\ie, before $p$ $\prcv(\ack(\Hm))$) with probability 1.
%\footnote{We adopt the convention
%that $0 \cdot \infty = 0$}
%Suppose $r \in \cpl{C_p}$.
%Then we have three possibilities:
%fcc23
Thus $\E(\csend \gv C_p)$ is finite, since both $\E(\numprt \gv C_p)$ and
$\E(\tow \gv C_p)$ are finite.

%joe22
%We can partition $\cpl{C_p}$ into two sets:
%fcc23
%To compute $\E(\numprt \gv \cpl{C_p})$, we
We now turn to $\E(\csend \gv \cpl{C_p})$.
We first partition $\cpl{C_p}$ into two sets:
\bit
%fcc24
%joe24: I preferred the original; reinstated
\item $F_1 = \{r : p$ crashes 
%fcc25 and does not $\prcv$ $\ack(\Hm)$ from
before $\prcving$ an $\ack(\Hm)$ from
%fcc25 only two items (delete comma)
% $q\}$, and
$q\}$ and
%\item $F_1 = \{r : p$ crashes and does not $\prcv$ $\ack(\Hm)$ from $q\}$.
\item $F_2 = \{r : p$ does not crash and does not $\prcv$ $\ack(\Hm)$ from $q\}$.
\eit
Note that $\Pr(F_2) = 0$ and $\Pr(F_1) = 1 - \Pr(C_p)$.
%joe22
%Thus we may ignore runs of $F_2$,
We may ignore runs of $F_2$ for the purposes of computing the expected
cost
since we adopted the convention that $0 \cdot \infty = 0$.  In runs
%joe22
$r$
of $F_1$, $\tow(r)$ is
%fcc23 no longer than
at most
the time it takes for $p$ to crash,
which is expected to occur at time
%fcc19 $1/\pf_p$.
%fcc24 $\frac{1}{\pf_p}$.  Thus $\E(\tow \gv F_1) \leq \frac{1}{\pf_p}$.
$\frac{1-\pf_p}{\pf_p}$.   Thus $\E(\tow \gv F_1) < \frac{1}{\pf_p}$.
Furthermore, if $p$ crashes at time $t_c$
%joe22
in $r \in F_1$,
it $\psends$ $\Hm$
%fcc23
exactly
$t_c$ times
%joe22
in $r$
(since $p$ does not $\prcv$ $\ack(\Hm)$ in runs of $F_1$).  In that case,
$q$ $\psends$ $\ack(\Hm)$ at most $t_c$ times.  So $\numprt(r) \leq 2 t_c$ if $p$
crashes at time $t_c$ in $r \in F_1$.  Thus
%fcc24 $\E(\numprt \gv F_1) \leq \frac{2}{\pf_p}$.
$\E(\numprt \gv F_1) < \frac{2}{\pf_p}$.
%fcc23  This is not correct (and we are done anyway)
%So
%\[
%\begin{array}{lll}
%\E(\csend) & = & \ds \Pr(C_p) \E(\csend \gv C_p) + \Pr(F_1) \E(\csend \gv F_1) \\
%5& \leq &\ds \imsg + \jmsg + (2 \ttime - 1) (2 - \lf) + \frac{3}{\pf_p}.\\
%\end{array}
%\]
%Then, with probability 1, $p$ crashes before
%it $\prcv(\ack(\Hm))$.  Thus $rsfin$ is the time $p$ crashes; $p$ is expected
%to crash by time $1/\pf_p$.  Thus $\E(\tow \gv \cpl{C_p}) \leq 1/\pf_p$.
%It follows then that $\E(\numprt \gv \cpl{C_p}) \leq 2/\pf_p$, since
%at each time unit there are at most two invocations of $\psend()$.
%fcc23
%It follows that $\E(\csend)$ is finite,
It follows that $\E(\csend \gv \cpl{C_p})$ is finite.  Since
both $\E(\csend \gv C_P)$ and $\E(\csend \gv \cpl{C_p})$ are finite,
$\E(\csend)$ is finite;
% \le (\imsg + \jmsg + (2
%\ttime - 1)(2 - \lf)) \cm + (\imsg +
%\ttime) \cw$.
so
%fcc23 this protocol
$\prt_1$
satisfies $\spc_3$.
%fcc17
To see that the protocol satisfies $\spc_2$, note that for $t \ge \ttime$, the
probability that $q$ does \emph{not} finish $\rrcving$ $\Hm$ by time $t$ given
that both $p$ and $q$ are still
%fcc16 up,
up is $\lf^{t-\ttime}$.  Thus $\spc_2$ is also satisfied.

%fcc17
Now consider the case that $\pct_p = 0$.  Note that in this case, $p$ is
expected to crash at time
%fcc24
%$\frac{1}{\pf_p}$.
$\frac{1-\pf_p}{\pf_p}$.
%fcc24
%Thus, $\E(\tow) \leq \frac{1}{\pf_p}$ and $\E(\numprt) \leq \frac{2}{\pf_p}$
Thus, $\E(\tow) < \frac{1}{\pf_p}$ and $\E(\numprt) < \frac{2}{\pf_p}$
(for the same reason as above), regardless of whether $q$ is correct.  Thus
$\E(\csend)$ is again finite.  The argument that $\spc_2$ is satisfied is the
same as before.

Now suppose $\pct_p = 1$ or $\pct_q = 0$.  These cases are somewhat analogous
to the ones above, except we need a
%fcc17 ``receiver-driven''
%joe22: took out italics; already defined
%\emph{receiver-driven}
receiver-driven
protocol.
Consider a protocol $\prt_2$ in which $q$
%fcc23 continually queries $p$ until it
queries $p$ in every round until it
gets a message from $p$.
%fcc11	Aren't we only looking at single-shot case though?  See discussion
%	after the proof of Thm 2 (about the size of \hbt).
%joe14: you're right.
%joe12
%More precisely,
%if we want to $\rsend$/$\rrcv$ a number of messages, we assume that each
%query is numbered.  The $k$th $\rsend$ starts
%when $p$ gets the first query numbered $k$ from $q$ and ends with the
%$\psend$ event just prior to when $p$ receives the first query numbered
%$k+1$ from $q$.  (Note that this means that $p$ does not know that the
%$k$th $\rsend$ has ended until the $(k+1)$th begins.)  The $k$th $\rrcv$
%starts when $q$ sends the first query numbered $k$ to $p$ and ends when
%$q$ $\prcv$s the $k$th message from $p$.
%
%fcc15 again, we assume single instance and now we assume $p$ wants to send in
%every run (so $q$ knows it).
%
%we assume that $q$ knows that $p$ has a message to send.  (This makes sense,
%for example, if $p$ is sending a stream of messages, $q$ has received the $k^\th$
%message in the stream has arrived, but still has not received the $(k+1)$st
%message.)  In that case,
%$q$ continually $\psend$s a query to $p$ until it $\prcv$s the message.
%joe20: this doesn't right.  It seems to aassume that $q$ knows what
%high-level message it is supposed to get.  If that were true, there
%would be no need for p to send at all.  If there's only one message,
%then q just needs to send  query.  If there's more than one, it needs
%to send an indication of what it got last (which means we need to
%assume that p does not want to send the same high-level message twice
%in a row).  Anyway, this needs to be rewritten
%
%fcc21
More precisely, let $\req$ denote a \emph{request message}. 
$q$ $\psends$ $\req$ to $p$ every time unit until it $\prcvs$ $\Hm$ and $p$
$\psends$ $\Hm$ every time it $\prcvs$ $\req$.
%fcc23 $\sdrc$
$\prt_2$
starts when $q$ $\psends$ the first $\req$ and $q$ finishes $\rrcving$ $\Hm$
%fcc19 This is problematic:  if p crashes then p ``finishes'' SEND(m), since
%      the last send(m) would be the ``end'' of SEND(m).  This wasn't a problem
%      before, since if p crashes, we stop.
%
% $p$'s last $\psend(\Hm)$;
%joe20: what's wrong with what you commented out?  Why should the SEND
%end with the last psend (or when p crashes, if it that happens before
%it psends?  Your current formulatoin clearly doesn't handle the case
%where q crashes.
%fcc21 $q$'s first $\prcv(\Hm)$.
when $q$ $\prcvs$ $\Hm$ for the first time.
%fcc17 the $\rrcv(\Hm)$ starts and ends with the $\prcv$.  By
%$\rrcv(\Hm)$ starts at the time $q$ first $\psend(\req(\Hm))$ and ends with the
%$q$'s first $\prcv(\Hm)$.
By reasoning similar to the previous cases, we can
show that $\E(\numprt)$ and $\E(\tow)$ are both finite
%fcc16
(so $\spc_3$ is satisfied)
and that $\spc_2$ is satisfied.

%fcc23
\newcommand{\infimp}{\stackrel{\infty \ }{\Longrightarrow}}
%fcc17
We now turn to the negative result.  It turns out that the negative result is
much more general than the positive result.  In particular, it holds for any
cost function with a certain property.  In the following, we 
%fcc25 say that
use
%joe22: redefined to match usage; note we don't need limits
%``$f$ is an unbounded function of $x$'' to mean that
% $\lim_{x\rar\infty} f(x) = \infty$.
%
%fcc23  It's not clear in what sense is f ``a function of g'' given this definition.
%       I was trying to come up with an intuitively appealing sufficient
%       condition.  The condition would have been sufficient if we assume the
%       function is ``continuous at infinity'' with the appropriate
%       definition (i.e., the value of the function is the (left) limit).
%       I used limits because in general a function could be unbounded but not
%       infinite at infinity.
%
%       Now that I think about it, it's best to use symbols instead of words.
%``$f$ is an unbounded function of $g$'' to mean that
\emph{$g \infimp f$} to 
%fcc25 mean 
denote
that if $g(x) = \infty$ then $f(x) = \infty$.

\lem
\label{pro:neg}
Let $\cpsend(r)$ be a cost function
%joe22
%such that the cost of waiting is an unbounded
%function of $\tow(r)$ and that the cost of sending is an unbounded
%function of $\numprt(r)$.
%fcc23
%that is an unbounded function of both $\tow$ and $\numprt$.
such that $\tow(r) \infimp \cpsend(r)$ and $\numprt(r) \infimp \cpsend(r)$.
If $0 < \pct_p < 1$ and $0 < \pct_q < 1$, then for any
%joe22
%$\cpsend(r) = \infty$ with some probability $\eps > 0$.
send-receive protocol $\prt$, $\Pr(\{r: \cpsend(r) = \infty\}) > 0$.
\elem
\prf
Suppose $\prt$ is a send-receive protocol for $p$ and $q$.  Let $R_1 =
\{r : $ $q$ crashes at time 0 and $p$ is correct in $r\}$.  Note that $p$ will
do the same thing in all runs in $R_1$: Either $p$ stops $\psending$ after some
time $t$ or $p$ never stops $\psending$.  If $p$ never stops, then $\numprt(r)
= \infty$ for all $r \in R_1$.
%In that case,
%joe22
Since, by assumption,
%fcc23 $\cpsend$ is an unbounded function of $\numprt$, we have
$\numprt(r) \infimp \cpsend(r)$, we have
that $\cpsend(r) = \infty$ for each $r \in R_1$.
%since we assumed that the cost of sending is an unbounded function of
%$\numprt(r)$.
%joe22
%We may take $\eps$ to be $\pct_p (1-\pct_q) \pf_q$, the probability of
%$R_1$.
Since $\Pr(R_1) = \pct_p (1-\pct_q) \pf_q > 0$, we are done.
Now suppose $p$ stops $\psending$ after time $t$.  Let $R_2 = \{r : $
$p$ crashes at time 0 and $q$ is correct in $r\}$.  Note that $q$ will do the
same thing in all runs of $R_2$:  Either $q$ stops $\psending$ after
some time $t'$ or $q$ never stops $\psending$.  If $q$ never stops,
%joe22
%the same reasoning as above works.
then $\cpsend(r) = \infty$ for all $r \in R_2$ and $\Pr(R_2) = \pct_q(1
- \pct_p)\pf_p > 0$, so again we are done.
%joe22
%So suppose $q$ stops after time $t'$. Let
Finally, suppose that $q$ stops $\psend$ing at time $t'$ in runs of
$R_2$. Let
%fcc19 $t'' \geq t, t'$.
$t'' = 1 + \max \{t, t'\}$.
Consider $R_3 = \{r : $ both processes are correct and all
messages up to time $t''$ are lost in $r\}$.  Then $\tow(r) = \infty$ for all
$r \in R_3$.
%Let $n_p$ be the number of messages sent by $p$ and $n_q$ be the
%number of messages sent by $q$.
%joe22
By assumption,
%fcc23 $\cpsend$ is an unbounded function of $\tow$,
$\tow(r) \infimp \cpsend(r)$,
so $\cpsend(r) = \infty$ for all $r \in R_3$.
%Since the cost of waiting is an unbounded function of $\tow(r)$, the
%cost of
%waiting is infinite for all runs of $R_3$ and we can take $\eps$ to be
Let $n_p$ and $n_q$ be the number of invocations of $\psend$ by $p$ and $q$,
respectively, in runs of $R_3$ (note that $p$ and $q$ do the same thing in all
runs of $R_3$).  Then $\Pr(R_3) = \pct_p \pct_q \lf^{n_p + n_q} > 0$, completing
%fcc24 the proof.~\eprf \ \ [\mlem{pro:neg}]
the proof.~\eprf~~(\mlem{pro:neg})
%$\eprf_{\mbox{\scriptsize  \mlem{pro:neg}}}$
%where $n_p$ is the number of invocations of $\psend$
%made by $p$ and $n_q$ is the corresponding quantity for $q$.

%joe22
%It is clear that in the cost function $\csend(r)$, the cost of sending
%is linear in
%$\numprt(r)$ and the cost of waiting is linear in $\tow(r)$, and so the Lemma applies.
Clearly
%fcc23 $\csend(r)$ is an unbunded function of both $\numprt(r)$ and $\tow(r)$,
$\numprt(r) \infimp \csend(r)$ and $\tow(r) \infimp \csend(r)$,
so Lemma~\ref{pro:neg} applies immediately and we are
%fcc24 done.~\eprf \ \ [\mthm{thm:s}]
done.~\eprf~~(\mthm{thm:s})
%fcc9
%joe10: Sorry; I know I told you to add this.  I forgot we say it in the
%next section.
%The negative part of the above theorem is very much in the spirit of
%impossibility results in \cite{act97}.

Of course, once we think in terms of utility-based specifications like
$\spc_3$, we do not want to know just whether a protocol implements 
%fcc25 $\spc_3$.
%joe25: I think semicolon is better than colon here
%$\spc_3$:
$\spc_3$;
we are in a position to compare the performance of different protocols that
implement $\spc_3$ (or of variants of one protocol that all implement $\spc_3$)
by considering their expected utility.  Let $\sdp$ and $\rdp$ be
generalizations (in the sense that they send messages every $\msgt$ rounds,
where $\msgt$ need not be 1) of the sender-driven and receiver-driven protocols
from \mthm{thm:s}, respectively.  Let $\triv$ denote the trivial (\ie, ``do
nothing'') protocol.  We use $\E^\ptc$ to denote the expectation operator
determined by the probability measure on runs induced by using protocol $\ptc$.
Thus, for example, $\E^{\sdp}(\numprt)$ is the expected number of messages sent
%joe12: corrected typo spotted by referee
%by $\sdp$.  If $\pct_p = \pct_1 = 0$, then $\sdp$, $\rdp$, and $\triv$
by $\sdp$.  If $\pct_p = \pct_q = 0$, then $\sdp$, $\rdp$, and $\triv$ all
satisfy $\spc_3$ (although $\triv$ does not satisfy $\spc_2$).  Which is
better?

In practice, process failures and link failures are very unlikely events.
We assume in the rest of the paper that $\pf_p$, $\pf_q$, and $\lf$ are all
very small, so that we can ignore sums of products of these terms (with
coefficients like $2\ttime^2$, $\msgt$, \etc).
%To make our presentation more legible, we would like to ignore ``small'' terms.
%More precisely, we give a definition of the \emph{degree} of a term.
%for a product involving $\pf_p$, $\pf_q$, or $\lf$, let the
%\emph{degree} of that product be the sum of the exponents of $\pf_p$, $\pf_q$,
%and $\lf$.
%If the degree is postive, then there are more terms in the
%numerator than in the denominator and we treat the term as small.  The degree
%of a sum is the minimum of all its summands.
%joe22: cut
%\footnote{Of course, with a
%little more effort, we can do more exact calculations, but doing this
%yields no further insight and results in unwieldy expressions.}
%fcc24
%(Informally,
%joe10
%we can think of
%fcc24
One way to formalize this is to say that
products involving $\pf_p$, $\pf_q$, and $\lf$ are $O(\eps)$ terms and
$2\ttime^2$, $\msgt$, etc., are $O(1)$ terms.  We write $t_1 \aeq t_2$ if
%fcc24
%$t_1$ and $t_2$ differ by an expression that is the sum of products involving
%(at least) one of $\pf_p$, $\pf_q$, or $\lf$.
$\abs{t_1 - t_2}$ is $O(\eps)$.
Note that we do not assume expressions like $\frac{\pf_p}{\pf_q}$ and
$\frac{\pf_q}{\pf_p}$ are small.

%fcc24
%joe24: slightly rewrite
For the following result only, we assume that
%$\frac{1}{\pf_p}$ and $\frac{1}{\pf_q}$ are $O(\eps^{-1})$, so
not only are $\pf_p$ and $\pf_q$ $O(\eps)$, they are also
%fcc25 $\Theta(\eps)$,
$\Theta(\eps)$,\footnote{Recall that $x$ is $\Theta(\eps)$ iff $x$ is $O(\eps)$
and $x^{-1}$ is $O(\eps^{-1})$.}
 so that  if
%one of them is multiplied
$\frac{1}{\pf_p}$ or $\frac{1}{\pf_q}$ is multiplied
by an expression that is $O(\eps^2)$, then the result is $O(\eps)$, which
can then be ignored.
%fcc23
%Suppose we have a product involving $\pf_p$, $\pf_q$, or $\lf$.  We say that
%the \emph{degree} of the product is $k$ if there are $k$ factors which are one
%of $\pf_p$, $\pf_q$, or $\lf$ (and that there is no factor with $\pf_p$,
%$\pf_q$, or $\lf$ in the denominator).  If we have a fraction whose numerator and
%denominator are both products involving $\pf_p$, $\pf_q$, or $\lf$, we assume
%that the fraction is small if the degree of the numerator is larger than the
%degree of the denominator.
%fcc19
%joe20: Where did you assume it before?  I though you cut that stuff.
%If you want to assume it here, explain why it doesn't hurt
%For the following result, we again assume that $p$ starts $\rsend(\Hm)$ at time
%0 or $q$ starts $\rrcv(\Hm)$ at time 0 (if the appropriate process is up at
%time 0).
%
%fcc10
%joe12: bad idea to put this in until we hit the final version
%\newpage
%
%fcc23
\setcounter{compSecNum}{\value{section}}
\setcounter{compThmNum}{\value{THEOREM}}
\pro
\label{p:comparison}
If $\pct_p = \pct_q = 0$,
then
\begin{center}
\begin{tabular}{lll}
%fcc24
%$\E^{\triv}(\tow) \aeq \frac{1}{\pf_p + \pf_q}$, &  &
$\E^{\triv}(\tow) = \frac{1-(\pf_p+\pf_q-\pf_p\pf_q)}{\pf_p + \pf_q-\pf_p\pf_q}$, &  &
$\E^{\triv}(\numprt) = 0$,\\
$\E^{\sdp}(\tow) \aeq \ttime$, &  &
$\E^{\sdp}(\numprt) \aeq \frac{(\ttime+1) \pf_q}{\msgt \pf_p} + 2 \cterm$,\\
$\E^{\rdp}(\tow) \aeq 2\ttime$, & &
$\E^{\rdp}(\numprt) \aeq \frac{(\ttime+1) \pf_p}{\msgt\pf_q} + 2 \cterm$.\\
\end{tabular}
\end{center}
\commentout{
If $\pct_p = \pct_q = 0$,
% and $\sdrc$ is invoked at time 0,
then\\[.25cm]
%fcc17
\begin{tabular}{lllll}
%fcc17
%fcc21 added commas
$\E^{\triv}(\tow) \aeq \frac{1}{\pf_p + \pf_q}$, &  &
%
%Note that the trivial protocol now has a higher cost, since p could ``finish''
%without q receiving the message.
%$\E^{\triv}(\tow(r)) = \frac{1 - \pf_p}{\pf_q}$ &  &
$\E^{\sdp}(\tow) \aeq \ttime$, &  &
$\E^{\rdp}(\tow) \aeq 2\ttime$,\\

$\E^{\triv}(\numprt) = 0$, &  &
$\E^{\sdp}(\numprt) \aeq \frac{\ttime \pf_q}{\msgt \pf_p} + 2 \cterm$,  &  &
$\E^{\rdp}(\numprt) \aeq \frac{\ttime\pf_p}{\msgt\pf_q} + 2 \cterm$.\\
\end{tabular}
%\begin{tabular}{lll}
%$\E^{\triv}(\tow(r)) \aeq (1-\pf_p-\pf_q)(1+\frac{1}{\pf_p + \pf_q})$ &  &
%$\E^{\triv}(\numprt(r)) = 0$ \\[.1in]
%\vspace{.125in}
%$\E^{\sdp}(\tow(r)) \aeq \ttime$ &  &
%$\E^{\sdp}(\numprt(r)) \aeq \frac{\ttime \pf_q}{\msgt \pf_p} + 2 \cterm$ \\[.1in]
%\vspace{.125in}
%$\E^{\rdp}(\tow(r)) \aeq 2\ttime$ &  &
%$\E^{\rdp}(\numprt(r)) \aeq \frac{\ttime\pf_p}{\msgt\pf_q} + 2 \cterm$.\\
%\end{tabular}
}%\end{commentout}
\epro

%joe22
\prf The relatively straightforward (but tedious!) calculations are
relegated to the appendix.~\eprf

%joe10
%Note that the expected cost of messages for $\sdp$ and $\rdp$ are the same,
Note that the expected cost of messages for $\sdp$ is the same as that for
$\rdp$, except that the roles of $\pf_p$ and $\pf_q$ are reversed.
The expected time cost of $\rdp$ is roughly $\ttime$ higher than that of $\sdp$,
because $q$ cannot
%fcc23 $\rrcv$ $m$
finish $\rrcving$ $m$
before time $2\ttime$ with a receiver-driven
protocol, whereas $q$ may
%fcc23 $\rrcv$ $m$
finish $\rrcving$ $m$
as early as $\ttime$ with a sender-driven
protocol.  This says that the choice between the sender-driven and
receiver-driven protocol should be based largely on the relative probability of
failure of $p$ and $q$.  It also suggests that we should take $\msgt$ very large
to minimize
%joe12: corrected typo spotted by referee
%costs.  (Intuitively, the large $\msgt$, the lower the message costs in
costs.  (Intuitively, the larger $\msgt$
%fcc11 	added
is,
the lower the message costs in
the case that $q$ crashes before acknowledging $p$'s message.)
%joe10:
%This may not seem so
%reasonable; essentially it is
This conclusion (which may not seem so reasonable) is essentially due to the
fact that we are examining a single
%fcc23
invocation of
$\sdrc$ in isolation.
%joe10: The $O(\eps)$ terms are irrelevant here
%(and also, to some extent, the fact that we are
%ignoring sums of products involving $O(\eps)$ terms).
%joe10
%We consider repeated
%invocations of $\rsend$s/$\rrcv$s in Section~\ref{s:hb2}.
As we shall see in Section~\ref{s:hb2}, this conclusion is no longer justified
once we consider repeated invocations of
%fcc10 	$\rsend$/s/$\rrcv$s.
$\sdrc$.  Finally, note that if the cost of messages is
%fcc24 high,
high and
waiting is cheap,
%fcc24 and $q$ is quite likely to fail (relative to $p$), then
%fcc24 It's a joint protocol now.
%$p$ is
the processes are
better off (according to this cost function) using $\triv$.
%fcc24
%not sending any messages
%at all
%(\ie, using $\triv$) than sending them in the hope of getting a response
%from $q$.

Thus, as far as $\spc_3$ is concerned, there are times when $\triv$ is better
than $\sdp$ or $\rdp$.  How much of a problem is it that $\triv$ does not
satisfy $\spc_2$?  Our claim is that if this desideratum (\ie, $\spc_2$) is
important, then it should be reflected in the cost function.  While the
cost
function
%fcc23 we have chosen
in our example
does take into account waiting time, it does not
penalize it sufficiently to give us $\spc_2$.  It is not too hard to find a
cost function that captures $\spc_2$.  For example, suppose we take
$\csendst(r) = N^{\tow(r)}$, where
%fcc17 $N(1- \pf_p - \pf_q + \pf_p\pf_q) > 1$.\footnote{Note
$N(1- \pf_p - \pf_q + \pf_p\pf_q) > 1$.
%joe22
%\footnote{Note
%that $\spc_3$ may be unsatisfiable for certain values of $N$.}

\pro
\label{thm:spc2}
%joe22
%If $\csendst(r)$ characterizes the cost of a run, then
Under cost function $\csendst$,
%and $\sdrc$ is invoked at time 0,
$\spc_3$ implies $\spc_2$.
\epro
%joe12: for WDAG
%\begin{prf}
%See the full paper.
%\end{prf}
%fcc11
%joe14:
\prf %(of \mthm{thm:spc2})
%fcc17 ***
%This proof may still need to be fixed.
%
%Note that $\spc_3$ is not satisfiable if $0 < \pct_p, \pct_q < 1$ (the proof
%for the negative part of \mthm{thm:s} applies).  Thus we assume $\pct_p = 0$ or
%$\pct_q = 0$ or $\pct_p = 1$ or $\pct_q = 1$.
%fcc17 rewritten (added ``starts'' and ``finishes'', among other things)
%joe22: rewrote and simplified proof
%We first introduce a few propositions to make the proof more succinct.
%Let $X(t)$
%%be the proposition ``$q$ finishes $\rrcving$ $\Hm$ no later than $t$
%consist of those runs where $q$ finishes $\rrcving$
%$\Hm$ no later than $t$ time
%units after the start of $\sdrc$ and let $Y(t)$ consist of those
%runs where $p$ and
%$q$ are up $t$ time units after the start of $\sdrc$.
%Let $\pcr(t)$
%consist of those runs where
%$p$ crashes at time $t$ and let $\qcr(t)$ consist of those runs where
%$q$ crashes at time $t$.
Suppose $\prt$ is a protocol that does not satisfy $\spc_2$; we show it
does not satisfy $\spc_3$ (under  cost function $\csendst$).
Let $C_p(t)$ and $C_q(t)$ consist of those runs of $\prt$ where
$p$ and
$q$,
respectively, are up for $t$ time units after the start of $\sdrc$ (and
perhaps longer). Let $\qrr(t)$ consist of the runs of $\prt$ where $q$
finishes $\rrcving$ $\Hm$ no later than time $t$ units after the start
of $\sdrc$.
%joe22
%\footnote{We could also require that $q$ $\rrcvs$ $\Hm$ at most once.}
%and this is the the first time $q$ receives $\Hm$.
%$\spc_2$ says that $\lim_{t \rar \infty} \Pr(\qrr(t) \gv
%C_p(t) \inter C_q(t)) = 1$.
Since $\prt$ does not satisfy $\spc_2$, there exists
$\eps > 0$
%joe22
%such that there is
and
an increasing infinite sequence of times
%fcc23 $t_0, t_1, \ldots$
$t_0, t_1, \ldots$,
such
%joe22: X(t_i) is a set, not a formula
%fcc24: It was originally a proposition.  (It doesn't matter now anyway.)
%that $\Pr(\neg X(t_i) \gv Y(t_i)) > \eps$ for all $i$.
that $\Pr(\cpl{\qrr(t_i)} \gv C_p(t_i) \inter C_q(t_i)) > \eps$ for
all $i$.
%where $\cpl{\qrr(t_i)}$ is the complement of $\qrr(t_i)$.
%We have three cases to
%consider: $\pct_p < 1$, $\pct_q < 1$, and $\pct_p = \pct_q = 1$.
We consider the case $\pct_p=\pct_q=1$ and $\pct_p \pct_q < 1$ separately.

%fcc24
Suppose $\pct_p=\pct_q=1$. Then $\Pr(\pcr(t) \inter \qcr(t)) = 1$ for all $t$.
So $$\Pr(\tow > t_i) = \Pr(\cpl{\qrr(t_i)}) = \Pr(\cpl{\qrr(t_i)} \gv
\pcr(t_i) \inter \qcr(t_i)) > \eps$$
for all $i$.  Let $V_i = \{ r : \tow(r) > t_i\}$ and $V_\infty = \{ r :
\tow(r) = \infty \}$.  Note that $V_\infty = \bigcap_{i=0}^\infty V_i$ and
that $V_i \supseteq V_{i'}$ for $i' > i$.  Thus $\Pr(V_\infty) =
\Pr(\bigcap_{i=0}^\infty V_i) > \eps$.  So $\E(\csendst) \geq
\Pr(V_\infty) N^\infty = \infty$.
%Since $\{ r : \tow(r) > t \} \supseteq \{ r : \tow(r) > t'\}$
%for $t' > t$,
%the above is equivalent to
%$\Pr(\cpl{\qrr(t_i)}) > \eps$ for all $i$, since $\Pr(\pcr(t_i) \inter
%\qcr(t_i)) = 1$ for all $i$.
%We also have $\Pr(\pcr(t) \inter \qcr(t)) = 1$ for all $t$, so
%$\Pr(\tow \geq t) = \Pr(\cpl{\qrr(t)})$.   Thus $\Pr(\tow \geq t_i) > \eps$ for
%all $i$.

%Suppose $\pct_p < 1$.
%Consider a run in which $p$ is not correct. Note that
%fcc19
%$\Pr(\neg X(t_i) \wedge \pcr(t_i+1) \gv Y(t_i)) \leq \Pr(\neg X(t_i) \wedge
%(\pcr(t_i+1) \vee \qcr(t_i+1) \vee \qrr(t_i+1)) \gv Y(t_i)) = \Pr(\tow(r) = t_i  \gv
%Y(t_i))$.  Recall that the event $\pcr(t)$ is independent of anything else and
%fcc24
Now we turn to the case that $\pct_p \pct_q < 1$.  Let $W(t) = \{ r : \tow(r) =
t \}$.  Note that $\tow(r) = t_i+1$ for all runs $r \in \cpl{\qrr(t_i)} \inter
\cpl{C_p(t_i+1)} \inter C_p(t_i) \inter C_q(t_i)$.  Thus,
\[
%\begin{array}{lll}
\Pr(W(t_i +1) \gv C_p(t_i) \inter C_q(t_i)) \ge
\Pr(\cpl{C_p(t_i+1)} \inter \cpl{\qrr(t_i)}
\gv C_p(t_i) \inter C_q(t_i)).\]
%joe22: unnecessary
%& \leq & \Pr(\neg X(t_i) \wedge (\pcr(t_i+1) \vee \qcr(t_i+1) \vee
%\qrr(t_i+1))
%\gv Y(t_i))\\
%joe22: I think this should be t_i+1
%& = & \Pr(\tow(r) = t_i \gv Y(t_i)).
%\end{array}
%\]
%joe22:  What's "anything else"?  You need to be much clearer.  In any
%case, it's conditional independence that you need
%Recall that the event $\pcr(t)$ is independent of anything else and
%happens with probability $\pf_p$.
Given our independence assumptions regarding process failures,
%fcc23 line too long
%$$\Pr(\cpl{C_p(t_i+1)} \inter \cpl{\qrr(t_i)}
%\gv C_p(t_i) \inter C_q(t_i)) =
%\Pr(\cpl{C_p(t_i+1)} \gv C_p(t_i))
%\Pr(\cpl{\qrr(t_i)} \gv C_p(t_i) \inter C_q(t_i)) >
%(1 - \pct_p) \pf_p \eps.$$
$$
%fcc23 changed to match format
%fcc24 changed back to 3 columns---re-spaced some equations to fit
\begin{array}{lll}
%\begin{array}{ll}
%&
 \Pr(\cpl{C_p(t_i+1)} \inter \cpl{\qrr(t_i)} \gv C_p(t_i) \inter C_q(t_i))
%\\
&
 = & \Pr(\cpl{C_p(t_i+1)} \gv C_p(t_i)) \Pr(\cpl{\qrr(t_i)} \gv
C_p(t_i) \inter C_q(t_i)) \\
&
 > & (1 - \pct_p) \pf_p \eps.\\
\end{array}
$$
A similar argument
%fcc23 (exchanging the roles of $p$ and $q$)
(exchanging the roles of $C_p$ and $C_q$)
shows that
%fcc23
%$$\Pr(\cpl{\qrr(t_i)} \gv C_p(t_i) \inter C_q(t_i)) >
%(1 - \pct_q) \pf_q \eps.$$
%fcc24  This is not what we ultimately want.
%$$\Pr(\cpl{C_q(t_i+1)} \inter \cpl{\qrr(t_i)} \gv C_p(t_i) \inter C_q(t_i)) >
$$\Pr(W(t_i+1) \gv \pcr(t_i) \inter \qcr(t_i)) >
(1 - \pct_q) \pf_q \eps.$$
%Thus $\Pr(\tow(r) = t_i \gv Y(t_i))
%\geq
%\Pr(\neg X(t_i) \gv Y(t_i)) \pf_p > \eps \pf_p$.
%Furthermore, since the probability of $p$ crashing at any time
%unit is $\pf_p$, independent of $Y(t_i)$ (or anything else), we have
%$\Pr(\tow(r) = t_i \gv Y(t_i)) \geq \Pr(\neg X(t_i) \gv Y(t_i)) \pf_p > \eps \pf_p$.
%
%Let $Z(t)$ be the proposition that says ``$p$ and $q$ are up $t$ time units
%after $p$ $\rsend$s $m$ and $p$ crashes at the next time unit''.
%Note that $\Pr(\neg X(t) \gv Y(t)) = \Pr(\neg X(t) \gv Z(t))$, since the protocols
%are deterministic and $p$ is equally likely to crash wether $q$ $\rrcv$d $m$ or
%not.  Note also that $\Pr(\neg X(t) \gv Z(t)) = \Pr(\tow(r) = t \gv Z(t))$.
%$\eps > 0$ such that $\Pr(\tow(r) > t_i \gv Y(t_i)) > \eps$ for all $i$.
%\Pr(\neg X(t) AND (crash at t_1+1 OR RECV at t_i+1) \gv Y(t_i)) = \Pr(\tow(r) = t_i \gv Y(t_i))
% \gv V
%\Pr(\neg X(t) \gv Y(t_i)) \pf_p
So
$$
\begin{array}{lll}
%\begin{array}{ll}
% &
\E(\csendst)
%\\
&
\geq &
%joe22
%\infty \Pr(\tow = \infty) +
%fcc24
%\ds \sum_{k=0}^\infty \Pr(\tow = k) N^k
\ds \sum_{k=0}^\infty \Pr(W(k)) N^k
\\
%joe22
% &\geq &\ds \sum_{i=0}^\infty \Pr(\tow(r) = t_i) N^{t_i} \\
&
 \geq &
%\infty \Pr(\tow = \infty) +
% \ds \sum_{i=0}^\infty \Pr(\tow = t_i+1) N^{t_i+1} \\
 \ds \sum_{i=0}^\infty \Pr(W(t_i+1)) N^{t_i+1} \\
&
 \geq &
%\infty \Pr(\tow = \infty) +
% \ds \sum_{i=0}^\infty \Pr(\tow = t_i+1 \inter C_p(t_i)
 \ds \sum_{i=0}^\infty \Pr(W(t_i+1) \inter C_p(t_i)
\inter C_q(t_i)) N^{t_i+1} \\
&
 = &
%\infty \Pr(\tow = \infty) +
%\ds \sum_{i=0}^\infty \Pr(\tow = t_i+1 \gv C_p(t_i)
\ds \sum_{i=0}^\infty \Pr(W(t_i+1) \gv C_p(t_i)
\inter C_q(t_i)) \Pr(C_p(t_i) \inter C_q(t_i))) N^{t_i+1} \\
&
 > & \ds
%\infty \Pr(\tow = \infty) +
\max\{(1-\pct_p) \pf_p,(1-\pct_q) \pf_q\} \eps \sum_{i=0}^\infty
(1-\pf_p-\pf_q+\pf_p\pf_q)^{t_i} N^{t_i+1}.
\end{array}
$$
Since $(1-\pf_p-\pf_q+\pf_p\pf_q) N > 1$ by assumption,
%joe22
%and $(1-\pct_p) \eps \pf_p > 0$,
%$\E(\csendst)$ is infinite and $\ptc$ fails to satisfy $\spc_3$.
%fcc24
we are done.~\eprf
\commentout{
if either $\pct_p < 1$ or $\pct_q < 1$, then
$\E(\csendst)$ is infinite.
If $\pct_p = \pct_q = 1$, then, as we now show, $\Pr(\tow = \infty) >
\eps$, so again $\E(\csendst)$ is infinite.  In any case, $\prt$ does
not satisfy $\spc_3$.

%joe22
%The case that $\pct_q < 1$ is analogous.

%Now suppose $\pct_p = \pct_q = 1$.
To see that $\Pr(\tow = \infty) > \eps$ if $\pct_p = \pct_q = 1$,
observe that in this case, $C_p(t) \inter C_q(t)$
consists of all runs.
Thus, $\spc_2$ becomes $\lim_{t\rar\infty} \Pr(\qrr(t)) = 1$.  Since
$\prt$ does not satisfy $\spc_2$, we have that
$\Pr(\cpl{\qrr(t_i}))
> \eps$ for all $i \geq 0$.
Moreover, since we have no process crashes,
%$\Pr(\qrr(t)) \geq \Pr(\qrr(t'))$ for $t' \leq t$, because if $q$
%finishes
%$\rrcving$ $\Hm$ no later than $t'$ in a run $r$ then $q$ finishes
%$\rrcving$ $\Hm$ no later than $t$ in $r$.
%%fcc19 The following depends on certain runs to not exist (not just
%having
%%probability 0).
%joe22: ??  Runs don't ``disappear''.
%(and $r$ will not ``disappear'' because of a process crash).
%Thus, if $\Pr(\cpl{\qrr_q(t_i)}) > \eps$ for all $i \geq 0$, then
%$\Pr(\cpl{\qrr_q(t)}) <
%\eps$ for all $t$.
%fcc23 repeats the above
%since no process crashes,
it follows that for all runs $r \in
\cpl{\qrr(t_i)}$, we have $\tow(r) > t_i$.  Thus, we have
$\Pr(\tow > t_i) > \eps$ for all $i$.  Since $\{r: \tow(r) > t_i \}
\supseteq \{r: \tow(r) > t_{i'}\}$ for $i' > i$, it follows that
$\Pr(\tow = \infty) = \Pr(\bigcap_i \tow > t_i) > \eps$.  This
completes the proof.~\eprf
}%\end{commentout}
%$\neg X(t)$ implies $\tow(r) \geq t$; that
%is, if $\neg X(t)$ is true in $r$, then $\tow(r) \geq t$.  So
%$\Pr(X(t)) < 1-
%\eps$ for all $t$ means $$\sum_{t=0}^\infty \Pr(\tow(r) < t) \leq
%1-\eps.$$
%Thus $\Pr(\tow(r) = \infty) \geq \eps$.  So $\spc_3$ is not satisfied,
%since the cost of waiting is infinite.
%\eprf

The moral here is that $\spc_3$ gives us the flexibility to specify what really
matters in a protocol, by appropriately describing the cost function.
%fcc23 Referee A suggested that we should discuss the relative merit of the
%      cost functions.  I think there is nothing to discuss, since we are
%      not the ones to choose the cost functions.  However, it might be
%      worthwhile to point this out (this is a key difference between our
%      paper and Mikler's).
We would like to remind the reader that the cost functions are not ours to
choose: They reflect the user's preferences.  (Thus we are not saying that
$\csendst$ is better than $\csend$ or vice versa, since each user is entitled
to her own preferences.)  What we are really saying here is that if $\spc_2$
matters to the user, then her cost function would force $\spc_3$ to imply
$\spc_2$---in particular, her cost function could not be $\csend$.

%joe12: we don't implemenet a heartbeat failure detector;
%\section{The Heartbeat Failure Detector}
\section{Using Heartbeats}
\label{s:hb1}

We saw in \sct{s:relcom} that $\spc_3$ is not implementable if we are not
certain about the correctness of the processes (\ie, if the probability that
they are correct is strictly between 0 and 1)
%fcc19 added
%fcc23 made more precise
%if the cost of sending and waiting are unbounded.
and the cost function $\cpsend(r)$ has the property that $\numprt(r) \infimp
\cpsend(r)$ and $\tow(r) \infimp \cpsend(r)$.
Aguilera, Chen, and Toueg~\cite{act97} (ACT from now on) suggest an approach
that
%fcc23 avoids
circumvents
this problem, using \emph{heartbeat} messages.
%joe12: added
%joe22: cut
%\footnote{The heartbeat messages are used to implement what they call a
%\emph{heartbeat failure detector}, which is a distributed oracle.}
Informally, a heartbeat from process $i$ is a message sent by $i$ to all other
processes to tell them that it is still alive.
%fcc19 Should ACT be considered singular or is it still plural?
%joe20: I think it's plural
%ACT show
ACT show
that there is a protocol using heartbeats that achieves \emph{quiescent}
reliable communication; \ie, in every run of the protocol, only finitely many
messages are required to achieve reliable communication (not counting the
heartbeats).  Moreover, they show that, in a precise sense, quiescent reliable
communication is not possible
%joe12: made more precise
%without the help of (something like) heartbeats, a
if
%fcc23 processes may
%fcc15 not crash
%be correct
we are not certain about the correctness of the processes
and communication is unreliable, a result much in the spirit of the negative
part of Theorem~\ref{thm:s}.%
%joe12: added details
\footnote{ACT actually show that their impossibility result holds even
if there is only one process failure, only finitely many messages can be lost,
and the processes have access to
%fcc17 a \emph{strong} failure detector, which means
$\mc{S}$ (a \emph{strong} failure detector), which means
that eventually every faulty process is permanently suspected and at least one
correct process is never suspected.  The model used by ACT is somewhat
different from the one we are considering, but we can easily modify their
results to fit our model.}  In this section, we show that (using
%fcc23 the linear cost function of the previous section)
the cost function $\csend$)
we can use heartbeats to implement $\spc_3$ for all values of $\pct_p$ and
$\pct_q$.

For the purposes of this paper, assume that processes send a message we call
$\hbmsg$ to each other every $\hbt$ time units.  Protocol $\rb$ in
Figure~\ref{algo1} is a protocol for reliable communication based on ACT's
protocol.  (It is not as general as theirs, but it retains all the
%joe10
%relevant features for our purpose.)
features relevant to us.)  Briefly, what happens according to this protocol is
that the failure detector layer of $q$
%fcc15 sends
$\psends$ $\hbmsg$ to the corresponding layer of $p$ periodically.  If $p$
wants to $\rsend$ $\Hm$, $p$ checks to see if any (new) $\hbmsg$ has arrived;
if so, $p$ $\psends$ $\Hm$ to $q$, provided it has not already $\prcvd$
$\ack(\Hm)$ from $q$; $q$ $\psends$ $\ack(\Hm)$ every time it $\prcvs$
%fcc23 $\Hm$.
$\Hm$ and $q$ finishes $\rrcving$ $\Hm$ the first time it $\prcvs$ $\Hm$.
%joe24: the wording was clunky
%Note that according to this protocol, $q$ does
Note that $q$ does
%fcc17 not
\emph{not}
%fcc15 send
$\psend$ any
%fcc24 $\hbmsg$s:
$\hbmsg$s as part
%joe24
%of the $\sdrc$ protocol:
of $\rb$.
That is the job of the failure-detection layer, not the job of the protocol.
(We assume that the protocol is built on top of a
%joe22
%failure detection
failure-detection
%fcc17 mechanism.
service.)
%fcc23  We are not ``using'' the cost function:  we just assume that \csend is
% the user's cost function.
%joe23: nevertheless, we need to say something here
%By using the cost function of the previous section,
%we acutally do not count the cost of $\hbmsg$s.
The cost function of the previous section does ot count the costs of
$\hbmsg$s.   That is,
since $\numprt(r)$ is the number of messages $\psent$ by the protocol,
$\csend(r)$ is not affected by the number of $\hbmsg$s $\psent$ in run $r$.
It is also worth noting that this is a sender-driven protocol, quite like that
given in the proof of
%fcc23  Theorem~\ref{thm:s}.
Theorem~\ref{thm:s}.\footnote{The reader might notice that the runs induced by
this protocol actually resemble those of the \emph{receiver-driven} protocol in
the proof of \mthm{thm:s} (if we identify $\hbmsg$ with $\req$).  The
difference is that in the receiver-driven protocol in the proof of
\mthm{thm:s}, the protocol for the receiver actually $\psends$ the $\req$s
whereas here the $\hbmsg$s are $\psent$ not by the protocol but by an
underlying heartbeat layer, independent of the protocol.}  It is
straightforward to also design a receiver-driven protocol using heartbeats.
%joe12: It was a bad idea to move this; it ended up in a funny location
%moved it to a better place
%fcc10	\begin{figure}[tb]
%fcc10	moved
\begin{figure}
\begin{center}
\begin{minipage}[t]{7.5cm}
The sender's protocol $(\rsend)$:
\begin{enumerate}
\setlength{\itemsep}{0cm}
\setlength{\parsep}{0cm}
\item \tb{while} $\neg \prcv(\ack(\Hm))$ \tb{do}
\item \verb+  +\tb{if} $\prcv(\hbmsg)$ \tb{then}
\item \verb+    +$\psend(\Hm)$
\item \verb+  +\tb{fi}
\item \tb{od}
\end{enumerate}

\end{minipage}
\verb+  +
\begin{minipage}[t]{7.5cm}
The receiver's protocol $(\rrcv)$:
\begin{enumerate}
\setlength{\itemsep}{0cm}
\setlength{\parsep}{0cm}
\item \tb{while} $\mb{true}$ \tb{do}
\item \verb+  +\tb{if} $\prcv(\Hm)$ \tb{then}
\item \verb+    +$\psend(\ack(\Hm))$
\item \verb+  +\tb{fi}
\item \tb{od}
\end{enumerate}

\end{minipage}
\end{center}
\hrule

\caption{Protocol $\rb$}
\label{algo1}
\end{figure}

We now want to show that $\rb$ implements $\spc_3$ and get a good estimate of
the actual expected cost.
\thm
%\begin{theorem}
\label{t:impl2}
%If $\csend(r)$ characterizes the cost of a run, then Protocol $\rb$
%implements
Under cost function $\csend$, Protocol $\rb$ satisfies
$\spc_3$.
%in $\sysi$.
Moreover, $\E(\tow) \aeq 2\ttime$ and $\E(\numprt)
\aeq 2 \cterm$, so that $\E(\csend) \aeq 2\ttime \cw + 2\cterm \cm$.
%\end{theorem}
\ethm

\prf
%fcc11
%See the full paper.
%joe22
%By a similar argument as in the proof of \mprop{p:comparison}, we can
Using arguments similar to those of the proof of \mprop{p:comparison}, we can
show that
$\E(\tow) \aeq 2\ttime$ and $\E(\numprt) \aeq 2 \cterm$.
%joe22
We leave details to the reader.~\eprf

The analysis of $\rb$ is much like that of $\sdp$ in \mprop{p:comparison}.
Indeed, in the case that $\pct_p = \pct_q = 0$, the two protocols are almost
identical.  The waiting time is roughly $\ttime$ more
%fcc17 in $\rb$,
for $\rb$,
%fcc17 This needs rewrite:  If we stay with our model that $\rsend(m)$ starts
%when $p$ $\psend(m)$ the first time, they are ``the same'', except processes
%may crash, so the cost isn't really quite the same...
since $p$ does not start $\psending$ until it $\prcvs$ the first $\hbmsg$ from $q$.
On the other hand, we are better off using $\rb$ if $q$ crashes before
acknowledging $p$'s message.  In this case, with $\sdp$, $p$ continues to $\psend$
until it crashes, while with $\rb$, it stops $\psending$ (since it does not get any
$\hbmsg$s from $q$).  This leads to an obvious question: Is it really worth
sending heartbeats?  Of course, if both $\pct_p$ and $\pct_q$ are between 0 and
1, we need heartbeats or something like them to get around the impossibility
result of \mthm{thm:s}. But if $\pct_p = \pct_q = 0$, then we need to look
carefully at the relative size of $\cm$ and $\cw$ to decide which protocol has
the lower expected cost.

This suggests that the decision of whether to implement a heartbeat layer must
take probabilities and utilities seriously, even if we do not count either the
overhead of building such a layer or the cost of heartbeats.  What happens if
we take the cost of heartbeats into account?  This is the subject of the next
section.

\section{The Cost of Heartbeats}
\label{s:hb2}

In the previous section we showed that $\spc_3$ is achievable with the help of
heartbeats.  When we computed the expected costs, however,
%fcc23 we were not counting the cost of $\psending$ $\hbmsg$ because the
%heartbeats were not $\psent$ by the protocol.
we did so with the cost function $\csend$, which does not count the cost of
heartbeats.
%fcc23
%I am not sure if we want to say that certain cost functions are
%``appropriate''.  This sounds like we get to choose the cost functions
%and we should choose the appropriate ones, which doesn't square with one
%of the philosophies of the paper (which is that the utilities come from
%users).
%
%This cost function is appropriate if we consider the heartbeat layer as
%given, and are just interesting in designing a protocol that takes advantage of
%it.  But it is not appropriate if we are trying to decide whether
While someone who takes the heartbeat layer for granted (such as an application
programmer or end-user) may have $\csend$ as their cost function,
%joe23: redundant
%the cost function of
someone who has to decide whether
%fcc19 to use a heartbeat protocol at all
to implement a heartbeat layer
or how frequently heartbeats should be
%fcc23 sent.
sent (such as a system designer) is likely to have a different cost
function---one which takes the cost of heartbeats into account.

As evidence of this, note that it is immediate from \mthm{t:impl2} that under
the cost function
%fcc23 we have been using,
$\csend$,
the choice of $\hbt$ that minimizes the expected cost is clearly at most
$2\ttime + 1$.  Intuitively, if we do not charge for heartbeats, there is no
incentive to space them out.  On the other hand, if we do charge for
heartbeats, then typically we will be charging for heartbeats that are sent
long after a given invocation of $\rb$ has completed.

The whole point of
%fcc23 heartbeat messages is that they are meant to be used, not
having a heartbeat layer is that heartbeats are meant to be used, not
just by one invocation of a
%fcc23 added
single
protocol, but by multiple invocations of (possibly) many protocols.  We would
expect that the optimal frequency of heartbeats should depend in part on how
often the protocols that use them are invoked.  The picture we have is that the
$\rb$ protocol is invoked from time to time, by different processes in the
system.  It may well be that various invocations of it are running
simultaneously.  All these invocations share the heartbeat messages, so their
cost can be spread over all of them.  If invocations occur often, then there
will be few ``wasted'' heartbeats between invocations, and the analysis of the
previous subsection gives a reasonably accurate reading of the costs involved.
On the other hand, if $\hbt$ is small and invocations are infrequent, then
there will be many ``wasted'' heartbeats.  We would expect that if there are
infrequent invocations, then heartbeats should be spaced further apart.

We now consider a setting that takes this into account.  For simplicity, we
continue to assume that there are only two processes, $p$ and $q$, but we now
allow both $p$ and $q$ to invoke $\rb$.
%joe12
%(It is possible to do this with $n$ processes, but the two-processor
(It is possible to do this with $n$
%fcc23 processes,
processes and more than one protocol,
but the two-process
%fcc23
and single protocol
case suffices to illustrate the main point, which is that the optimal $\hbt$
should depend on how often
%fcc23 $\rb$
the protocol
is invoked.)  We assume that each process,
while it is running, invokes $\rb$ with probability $\spb$ at each time unit.
Thus, informally, at every round, each running process tosses a coin with
probability of $\spb$ of landing heads.  If it lands heads, the process then
invokes $\rb$ with the other as the recipient.
%fcc21
(Note that we no longer assume that the protocol is invoked at time 0 in
this section.)

Roughly speaking, in computing the cost of a run, we consider the cost of each
invocation of $\rb$ together with the cost of all the heartbeat messages sent
in the run.  Our interest will then be in the cost \emph{per invocation of
$\rb$}.  Thus, we apportion the cost of the heartbeat messages among the
invocations of $\rb$.  If there are relatively few invocations of $\rb$, then
there will be many ``wasted'' heartbeat messages, whose cost will need to be
shared among them.

For simplicity, let us assume that each time $\rb$ is invoked, a different
message is sent.  (For example, messages could be numbered and include the name
of the sender and recipient.)  We say \emph{$\rb(m)$ is invoked at time $t_1$
in $r$} if at time $t_1$ some process $x$ first executes line 1 of the code of
the sender with message $m$.  This invocation of $\rb$ \emph{completes} at time
$t_2$ if the last message associated with the invocation (either a copy of $m$
or a copy of $\ack(m)$) is sent at time $t_2$.  If
%fcc24 minor correction
%the receiver crashed before
$x$ received the last heartbeat message from the receiver before
%fcc24
%$x$ invokes
invoking $\rb(m)$,
%joe10
%let $t_2 = t_1$.
we take $t_2 = t_1$ (that is, the invocation completes as soon as it starts in
this case).

%joe20: this doesn't read right.  I preferred the original.  Why did you
%change it?
The processes will (eventually) stop
$\psending$ $m$ or $\ack(m)$ if either
%$\psend(\Hm)$ or $\psend(\ack(\Hm))$ if either
process crashes or if
%fcc19 $x$ stops $\psend$ing $m$ because it $\prcv$d $\ack(m)$.
%fcc21 the sender $\prcv(\ack(\Hm))$.
the sender $\prcvs$ $\ack(\Hm)$.
Thus, with probability 1, all invocations
%joe10
of $\rb$
will eventually complete.
%fcc23   I fail to see the point of this remark.  (Maybe it made sense a while
%        ago.)
%(Note that we are still not counting heartbeat messages.)
%fcc19 T -> t
Let $\tcmp(r,t)$ be the number of invocations of $\rb$ that have
completed by time $t$ in $r$; let $\tcst(r,t)$ be the cost of these
invocations.  Let $\tchb(r,t)$ be the cost of
%fcc19 the $\hbmsg$s up to time $t$ in
%joe20: this doesn't make sense.  What is the cost of $psend(\hbmsg)$?
%$\psend(\hbmsg)$ up to time $t$ in
$\psending$ $\hbmsg$ up to time $t$ in
$r$.  This is simply the number of $\hbmsg$s sent up to time $t$ (which we
denote by $\nhb(r,t)$) multiplied by $\cm$.  Let $\totc(r,t) = \tcst(r,t) +
\tchb(r,t)$.  Finally, let
%fcc19 $\csendav(r) = \limsup_{t\rar\infty} (\totc(r,t) /(\tcmp(r,t)+1))$,
$$\csendav(r) = \limsup_{t\rar\infty} \frac{\totc(r,t)}{\tcmp(r,t)+1},$$
where
%fcc23 $\limsup$
``$\limsup$''
denotes the limit of the supremum, that is,
%fcc19
%$\lim_{t'\rar\infty} \sup_{0 \le t \le t'} (\totc(r,t) /
%(\tcmp(r,t)+1))$.
$$\csendav(r) = \lim_{t'\rar\infty} \sup_{0 \le t \le t'} \frac{\totc(r,t)}
{\tcmp(r,t)+1}.\footnote{By adding 1 to the denominator, we guarantee it is
never 0;
%joe12
%adding 1 also simplifies one of the technical calculations below.}
adding 1 also simplifies one of the technical calculations needed in the
proof of Theorem~\ref{thm:avg}.}$$
Thus $\csendav(r)$ is essentially the average
cost per invocation of $\rb$, taking heartbeats into account.  We write
%fcc21 added quotes
``$\limsup$'' instead of ``$\lim$'' since the limit may not exist in general.
(However, the proof of the next theorem shows that in fact, with probability 1,
the limit does exist.)
%joe10: moved here; we need to say it before the proof, not in the
%middle of it.
For the following result only, we assume that $\sqrt{\pf_p}$ and $\sqrt{\pf_q}$ are
also $O(\eps)$.

\newcommand{\limt}{\lim_{t\rar\infty}}

%fcc23
\setcounter{avgSecNum}{\value{section}}
\setcounter{avgThmNum}{\value{THEOREM}}
\thm
%joe22: I think there's a minor bug in the calculation
%fcc24: changed back.  I think it was fine.
%\frac{1}{\hbt\spb}\cm$, where $0 < \lambda < 1$ is a constant that
%joe24: ?? the previous comment seems out of context
%\begin{theorem}
%fcc25:  \frac{hbt}{2} -> \frac{\hbt-1}{2}
%        The time to the first \hbmsg is between 0 and \hbt - 1, not 0 and \hbt,
%        so the average was off.
\label{thm:avg}
Under the cost function $\csendav$, Protocol $\rb$ satisfies $\spc_3$.
Furthermore,
$\E(\csendav) \aeq ((1-\pct_p)(1-\pct_q)\lambda + \pct_p \pct_q)
\left(2\cterm\cm +
%fcc24
%\left(2\ttime + \frac{\hbt-1}{2}\right)\cw \right) +
\left(\ttime + \frac{\hbt-1}{2}\right)\cw \right) +
\frac{1}{\hbt\spb}\cm$, where $0 < \lambda < 1$.
%joe24: cut; it turns out it depends on \spb, \cm, and \tow too.  By the
%time it depends on all the constants, there's no point in saying
%anything
% is a constant that depends only
%fcc24
%on $\pf_p$ and $\pf_q$.\footnote{Intuitively, $\lambda$ depends on the
%probability that both $p$ and $q$ are up relative to the probability that one
%is up and the other down.}
%on $\pf_p$, $\pf_q$, $\ttime$, and $\hbt$.
%\end{theorem}
\ethm
%joe22
\prf  See the appendix.~\eprf

Note that with this cost function, we have a real decision
%joe10:
to make
in terms of how
frequently to send heartbeats.  As before, there is some benefit to making
%fcc25 
%$\hbt > 2 \ttime$, to minimize the number of useless messages sent when $\rb$
$\hbt > 2 \ttime$: it minimizes the number of redundant messages sent when $\rb$
is invoked (that is, messages sent by the sender before receiving the
receiver's acknowledgment).  Also, by making $\hbt$ larger we will send fewer
heartbeat messages between invocations of $\rb$.  On the other hand, if we make
$\hbt$ too large, then the sender may have to wait a long time after invoking
$\rb$ before it can send a message to the receiver (since messages are only
sent upon receipt of a heartbeat).  Intuitively, the greater $\cw$ is relative
to $\cm$, the smaller we should make $\hbt$.  Clearly we can find an optimal
choice for $\hbt$ by standard calculus.

In the model just presented, if $\cw$ is large enough relative to $\cm$, we
will take $\hbt$ to be 1.  Taking $\hbt$ this small is clearly inappropriate
once we consider a more refined model, where there are buffers that may
overflow.  In this case, both the probability of message loss and the time for
message delivery will depend on the number of messages in transit.  The basic
notions of utility still apply, of course, although the calculations become
more complicated.  This just emphasizes the obvious point is that in deciding
what value (or values) $\hbt$ should have, we need to carefully look at the
actual system and the cost function.

\section{Discussion}
\label{s:conrem}

We have tried to argue here for the use of decision theory both in the
specification and the design of systems.  Our (admittedly rather simple)
analysis already shows both how decision theory can help guide the decision
made and how much the decision depends on the cost function.  None of our results
are deep; the cost function just makes precise what could already have been seen
from an intuitive calculation.  But this is precisely the
%fcc19 point.
point:
By writing our specification in terms of costs, we can make the intuitive
calculations precise.  Moreover, the specification forces us to make clear
exactly what the cost function is and encourages the elicitation of utilities from
users.  We believe
%joe22:
that
these are both important features.  It is important for the
user (and system designer) to spend time thinking about what the important
attributes of the system are and to decide on preferences between various
tradeoffs.

%fcc9
%fcc10
A possible future direction is to study standard problems in the literature
(\eg, Consensus, Byzantine Agreement, Atomic Broadcast, \etc) and recast the
specifications in utility-theoretic terms.
%For example, we may
%charge a Consensus algorithm each time unit before the decision is made and
%charge the algorithm a large penalty if Agreement is violated.
One way to do this is to replace a liveness requirement by an unbounded
increasing cost function (which is essentially the ``cost of waiting'') and
replace a safety requirement by a large penalty.
%, which ensures an
%algorithm with small expected cost will rarely violate the safety condition.
Once we do this, we can analyze the algorithms
%joe11
%people use to solve these
that have been used to solve these problems,
%joe11
%We expect that the algorithms people use will reflect their implicit
%assumptions about probabilities and utilities.  If not, we may be able to give
%an algorithms which performs better. We may also be able to devise new
%algorithms from the utility-theoretic specifications.
and see to what extent they are optimal given reasonable assumptions
about probabilities and utilities.

%Once we fix a cost function for a problem, we may analyze standard algorithms
%which solve these problems in terms of their expected utility.  We expect that
%the standard algorithms should perform

%It may be interesting to rewrite the specifications of standard problems, such
%as Consensus and Byzantine Agreement,

%We believe that while decision theory is not explicitly used in systems design,
%some of the things people do in practice are implicitly motivated by
%utility-theoretic considerations.

%In general, we may be able to replace liveness requirements by an unbounded
%increasing cost function, so that if nothing ``good'' happens, the cost is
%infinite.  A safety requirement may be replaced with a large penalty for its
%violation.

%For example, one may optimize for the
%failure-free runs when doing Consensus or Byzantine Agreement.  This reflects
%the believe of the protocol designer that failures are unlikely in practice.

While we believe that there is a great deal of benefit
%joe11
%from analyzing systems in these terms,
to be gained from analyzing systems in terms of utility,
it is quite often a nontrivial matter.
%joe11
%We discuss some of the key difficulties here:
Among the most significant difficulties are the following:
\begin{enumerate}
\item Where are the utilities coming from?  It is far from clear that a
user can or is willing to assign a real-valued utility to all possible outcomes
in practice.  There may be computational issues (for example, the set of
outcomes can be enormous) as well as psychological issues.  While the agent may
be prepared to assign qualitative utilities like ``good'', ``fair'', or
``bad'', he may not be prepared to assign
%fcc19 .7.
$20.7$.
While to some extent the system can convert qualitative utilities to a
numerical representation, this conversion may not precisely captures the user's
intent.
%joe12
%More generally, it is clear that there are
There are also
nontrivial user-interface issues involved in eliciting
utilities from users.  In light of this, we need to be very careful if results
depend in sensitive ways on the details of the utilities.
\item Where are the probabilities coming from?  We do not expect users
to be experts at probability.  Rather, we expect the system to be gathering
statistics and using them to estimate the probabilities.  Of course, someone
still has to tell the system what statistics to gather.  Moreover, our
statistics may be so sparse that we cannot easily obtain a reliable estimate of
the probability.
\item Why is it even appropriate to maximize expected utility?  There
are times when it is far from clear that this is the best thing to do,
especially if our estimates of the probability and utility are suspect.  For
example, suppose one action has a guaranteed utility of 100 (on some
appropriate scale), while another has an expected utility of 101, but has a
nontrivial probability of having utility 0.  If the probabilities and utilities
that were used to calculate the expectation are reliable, and we anticipate
performing these actions frequently, then there is a good case to be made for
taking the action with the higher expected utility.  On the other hand, if the
underlying numbers are suspect, then the action with the guaranteed utility
might well be preferable.
\end{enumerate}

We see these difficulties not as ones that should prevent us from using
decision theory, but rather as directions for further research.
%joe11
It may be possible in many cases to learn a user's utility.  Moreover, we
expect that in many applications, except for a small region of doubt, the
choice of which decision to make will be quite robust, in that perturbations to
the probability and utility will not change the decision.  Even in cases where
perturbations do change the decision, both decisions will have roughly equal
expected utility.  Thus, as long as we can get somewhat reasonable estimates of
the probability and utility, decision theory may have something to offer.

%fcc25 Clearly another 
Another
important direction for research is to consider
\emph{qualitative decision theory}, where
%fcc23 both our measures of utility and probability and more qualitative,
both utility and likelihood are more qualitative,
and not necessarily real numbers.  This is, in fact, an active area of current
research,
%fcc11	The file no longer exists.  I cannot find it at stanford.edu.
%as the bibliography of 290 papers
%at http://walrus.stanford.edu/diglib\\/csbibliography/Ai/qualitative.decision.theory.html
%attests.
%fcc11	I may have found it else where, you may wish to check to see if
%this is right.
%joe14: looks good
%fcc23  spacing problems
%as the bibliography of over 290 papers at
%fcc25
as %the bibliography of over 290 papers at
%see 
{\tt http:/$\!$/www.medg.lcs.mit.edu/qdt/bib/unsorted.bib}
(a bibliography of over 290 papers) attests.
%for a bibliography of over 290 papers on this subject.
Note that once we use more qualitative notions, then we may not be able to
compute expected utilities at all (since utilities may not be numeric) let
alone take the action with maximum expected utility, so we will have to
consider other decision rules.

Finally, we might consider what would be an appropriate language to specify and
reason about utilities, both for the user and the system designer.

While it is clear that there is still a great deal of work to be done in order
to use decision-theoretic techniques in systems design and specification, we
hope that this discussion has convinced the reader of the
%joe22
%promise
utility
of the approach.

\section*{Acknowledgments}
We thank Sam Toueg for numerous discussions regarding heartbeats and Jim Gray
for giving us some insight on costs in database computations and for pointing
out the use of costs in deadlock detection.  We also thank the anonymous
referees for their helpful comments.

%joe22: you had cut this out, but I reinstated it.
\appendix
\section*{Appendix: Proofs}

We present the proofs of \mprop{p:comparison}
%joe22
%\mthm{thm:spc2},
and \mthm{thm:avg}.  We repeat the statements of the results for the
convenience of the reader.
%fcc24
%joe24
%We would like to remind the reader
Recall
that for \mprop{p:comparison}, we are
assuming that
%joe24
%$\frac{1}{\pf_p}$ and $\frac{1}{\pf_q}$ are both $O(\eps^{-1})$
$\pf_p$ and $\pf_q$ are both $\Theta(\eps)$,
%joe24: cut; say it when we use it
%(so that when one of them is multiplied by an $O(\eps^2)$ term, the
%result is $O(\eps)$ and can then be ignored)
and that for \mthm{thm:avg}, we are assuming
that $\sqrt{\pf_p}$ and $\sqrt{\pf_q}$ are both $O(\eps)$.

\opro{compSecNum}{compThmNum}
\pro
%fcc23 statement change
If $\pct_p = \pct_q = 0$,
then
\begin{center}
\begin{tabular}{lll}
%fcc24
%$\E^{\triv}(\tow) \aeq \frac{1}{\pf_p + \pf_q}$, &  &
$\E^{\triv}(\tow) = \frac{1-(\pf_p+\pf_q-\pf_p\pf_q)}{\pf_p + \pf_q-\pf_p\pf_q}$, &  &
$\E^{\triv}(\numprt) = 0$,\\
$\E^{\sdp}(\tow) \aeq \ttime$, &  &
$\E^{\sdp}(\numprt) \aeq \frac{(\ttime+1) \pf_q}{\msgt \pf_p} + 2 \cterm$,\\
$\E^{\rdp}(\tow) \aeq 2\ttime$, & &
$\E^{\rdp}(\numprt) \aeq \frac{(\ttime+1) \pf_p}{\msgt\pf_q} + 2 \cterm$.\\
\end{tabular}
\end{center}
\epro
\eopro

\prf
For $\triv$, note that $\numprt(r) = 0$ for all $r$,
%joe22: "whence" means "from where",
%whence
so
$\E^{\triv}(\numprt) = 0$.
We also have that $\tow(r)$ is the time of the first crash in $r$.
Since the probability of a crash during a time unit is $\pf = \pf_p + \pf_q -
\pf_p\pf_q$, we have that the expected time of the first crash,
%joe22
and hence $\E^{\triv}(\tow)$,
is
%fcc24
%$\frac{1}{\pf} \aeq \frac{1}{\pf_p+\pf_q}$.
$$\sum_{k=0}^\infty k (1-\pf)^k \pf = \frac{\pf(1-\pf)}{(1-(1-\pf))^2} =
\frac{1-\pf}{\pf} = \frac{1-(\pf_p+\pf_q-\pf_p\pf_q)}{\pf_p + \pf_q-\pf_p\pf_q}.$$
%joe24: added (I don't think it appears any more)
%fcc25 moved
%(Here and elsewhere in this proof we use the well-known fact that $sum_k
%k x^k = x/(1-x)^2$.)

For $\sdp$, we first show that $\esd(\tow) \aeq \ttime$.
%joe22: not quite
%Recall that
%fcc24  we don't really need this: this statement holds for any protocol
%       under our assumptions.
%Since it is easy to check that $\Pr(\tow(r) = \infty) = 0$ for the
%protocol $\sdp$, it follows that
Since $\pct_p=\pct_q=0$, $\Pr(\tow(r) = \infty) = 0$, thus
%fcc10
%joe11 I think this is unnecessary
%we have
$\esd(\tow) = \sum_{k=1}^{\infty} k \Pr(\tow = k)$.
%joe22
%by definition.
We
break the sum into three
%fcc23  (since we called these ``summands'')
%pieces---$\sum_{k=1}^{\ttime -1} k \Pr(\tow = k)$,
%$\ttime \Pr(\tow = \ttime)$, and $\sum_{k=\ttime+1}^{\infty} k \Pr(\tow
%= k)$---and
%fcc24 consistency
%pieces---
pieces,
%\bit
%\item $\ds \sum_{k=1}^{\ttime -1} k \Pr(\tow = k)$
%\item $\ds \ttime \Pr(\tow = \ttime)$
%\item $\ds \sum_{k=\ttime+1}^{\infty} k \Pr(\tow = k)$
%\eit
%fcc25 commas
\bit
\item $\ds \sum_{k=1}^{\ttime -1} k \Pr(\tow = k)$,
\item $\ds \Pr(\tow = \ttime)$, and
\item $\ds \sum_{k=\ttime+1}^{\infty} k \Pr(\tow = k)$,
\eit
and analyze each one separately.

%fcc25 summand -> part
%joe9
%First,
For the first part, note that the only way that $\tow = k$ for $1
\leq k < \ttime$ is for there to be a crash before $\ttime$.  Thus
$$\Pr(\tow = k) = ((1-\pf_p)(1-\pf_q))^k (\pf_p+\pf_q-\pf_p\pf_q)
%joe9: this does not involve approximations
< \pf_p + \pf_q.$$
%We have
%that the first summand is
%bounded above by
It follows that
%\begin{equation}\label{eq1}
$$\sum_{k=1}^{\ttime -1} k \Pr(\tow = k) <
(\pf_p+\pf_q)\sum_{k=1}^{\ttime-1} k =
(\pf_p+\pf_q)\frac{\ttime(\ttime-1)}
%fcc9
{2} \aeq 0.
$$
%\end{equation}
%joe9: unnecessary; it's an upper bound without any assumptions
%(This is an upper bound since we are treating $(1-\pf_p)(1-\pf_q)$ as 1.)
Thus we may drop the first part.
%, as it is $O(\eps)$.
%fcc9	t -> \ttime

%joe24: added paragraph break
For the second part, note that $\tow = \ttime$ if $p$ and $q$ are up
%fcc19 to
%joe20 till
until
$\ttime$ and
%fcc21 the first $\psend(\Hm)$ by $p$ is successful.
$q$ $\prcvd$ the first copy of $\Hm$ $p$ $\psent$.
(We may also have
$\tow = \ttime$ if one of $p$ or $q$ crashes at time $\ttime$.)  Thus,
%joe9: simplified
%$\Pr(\tow = \ttime)
%=((1-\pf_p)(1-\pf_q))^{\ttime}(\pf_p+\pf_q+(1-\lf)+O(\eps^2)) \aeq 1$.
%(The $O(\eps^2)$ terms are terms from the  inclusion-exclusion principle.)
%\begin{equation}\label{eq2}
$$\Pr(\tow = \ttime) \ge ((1-\pf_p)(1-\pf_q))^{\ttime}(1-\lf)
\aeq 1,$$
%\end{equation}\label{eq2}
so the second part is $\aeq \ttime$.
%Before we consider the third summand, note that $\sum_{i=m}{m+n} = m+\frac{n(2m+n+1)}{2}$
%joe9: simplified.  You have to be more ruthless with your
%approximations!

%joe24: added paragraph break
Finally, for the third part, if $k > \ttime$, then $k$ has the form
$\ttime + a\msgt+b$, where $a \ge 0$ and 
%fcc26 $0 \le b < \msgt$.  
$0 \le b < \msgt$ (and $a+b>0$).
If
%joe22
%$\tow(r) = k$, then
$\tow = k = \ttime + a \msgt + b$, then
$a+1$ messages are
%fcc23 not delivered,
lost by the link,
so $\Pr(\tow = k) \le \lf^{a+1}$.
%joe22: it's not *that* easy
%It easily follows that
A straightforward calculation shows that
%\begin{equation}\label{eq3}
%fcc9	\ttime^2
%fcc23  changed so that multi-line equations all have the same format
$$
%\begin{array}{ll}
\begin{array}{lll}
%&
\ds \sum_{k=\ttime +1}^\infty k \Pr(\tow = k)
%\\
&
%joe22: added details
%fcc26: corrections
%= &\ds \sum_{a=0}^\infty \sum_{b=0}^{\msgt-1} (\ttime + a \msgt + b)
%\Pr(\tow = \ttime + a \msgt + b)\\
= &\ds 
\sum_{b=1}^{\msgt-1} (\ttime+b) \Pr(\tow = \ttime+b) \\
& &+ \ds
\sum_{a=1}^\infty \sum_{b=0}^{\msgt-1} (\ttime + a \msgt + b)
\Pr(\tow = \ttime + a \msgt + b)\\
&
\le & \ds \sum_{a=0}^\infty \msgt (\ttime + (a+1) \msgt)
%fcc25 
%\Pr(\tow = \ttime + a \msgt)\\
\lf^{a+1}\\
%joe10; corrected
%\sum_{a=1}^\infty ((a+1)\ttime)^2 \lf^a \aeq \lf.$$
%\sum_{a=0}^\infty \left(\msgt (\ttime + (a+1) \msgt) + \frac{\msgt
%(\msgt-1)}{2}\right)\lf^{a+1} \aeq \lf.$$
%joe22: you seem to be getting an extra \msgt^2 term; please check
%\le &\sum_{a=0}^\infty ((a+1)\msgt^2 + \msgt(\msgt+\ttime))\lf^{a+1}
&
\le & \ds \sum_{a=0}^\infty ((a+1)\msgt^2 + \msgt\ttime)\lf^{a+1}\\
%joe22: \aeq \lf is the same as \aeq 0
%\aeq \lf.$$
%fcc25  (more details)
& = & \ds \msgt^2 \sum_{a=1}^{\infty} a \lf^a + \msgt \ttime
\sum_{a=1}^{\infty} \lf^{a} \\
&
\aeq & \ds 0.
\end{array}$$
%\end{equation}
%$\Pr(\tow(r) = k) = \lf^a ((1-\pf_p)(1-\pf_q))^k(\pf_p+\pf_q-\pf_p\pf_q)$,
%since the waiting must have ended by a crash; we
%also have that for $a \geq 2$,\footnote{We take $a \geq 2$ here since
%setting
%$a = 1$ yields the second summand.}
% $\Pr(\tow(r) = a\ttime) = \lf^{a-1}
%((1-\pf_p)(1-\pf_q))^{a\ttime}(\pf_p+\pf_q+(1-\lf) + O(\eps^2))$.  The
%third summand is bounded above by
%$\left((\pf_p+\pf_q-\pf_p\pf_q)\sum_{a=1} (C+(a-1)D)\lf^a  \right) +
%\left(\ttime(\pf_p+\pf_q+(1-\lf))\sum_{a=2}^{\infty}a\lf^{a-1}\right)$,
%where $C = \frac{(\ttime-1)(2\ttime+1)}{2}$ and $D = \ttime(\ttime-1)$.
%It is routine to check that this upper bound is in fact $O(\eps)$.
%We
%$\Pr(\tow(r) = \ttime) = (1-\pf_p)^\ttime(1-\pf_q)^\ttime (1-\lf) \aeq
%(1-\ttime(\pf_p + \pf_q)-\lf) \aeq 1$
%joe8: This does *not* follow.  The fact that $\Pr(\tow(r) = k) is
%$O(\eps)$ for all $k \ne\ttime$ does not imply that $\Pr(\tow(r) =
%\ttime) = 1 -\eps$.  You need to work out the expectation more
%carefully here.  Likewise $\E^{\sdp}(\numprt)
%joe22
%Thus, the third summand is also $O(\eps)$.
Thus, we can also ignore the third part.
This gives us $\E^{\sdp}(\tow)) \aeq \ttime$, as desired.
%joe10: this should be part of the next paragraph
%fcc9
%Now we turn to $\E^{\sdp}(\numprt(r))$.
%fcc8	UNCHANGED:  It seems that \Pr(\numprt(r) = k) is somewhat complicated
%in general.  For example, all k msgs could be send by the sender, in which
%case the receiver must crash before the first message that is not lost
%arrives and the sender must die before it is time to send the k+1st
%msg.  It is possible also that some are sent by the receiver, in which
%case we need the ``proper mix'' of msg losses and process crashes.
%The last time we did it I think we ignored cases like more than one
%message loss or process crashes.
%joe9: you need to be more ruthless.  An argument much in the spirit as
%the one I did for the third summand above shows that you can ignore all
%the cases where the protocol runs beyond time 2 \ttime.  Then you have
%to deal with all the cases up to then.  It's really not that hard.
%Think about it tonight and we can talk about it tomorrow.

%fcc23
\commentout{
%fcc9
%joe10: moved here
We now calculate $\E^{\sdp}(\numprt)$.  Consider the following partition of
the runs:
\begin{itemize}
%fcc11	changed \epsilon to \eps
%fcc19
%\item $A_1$, the set of runs where $q$ is up for at least $\ttime$ time units,
%$p$'s first message is not lost and $q$'s first $\ack$ is not lost;
%\item $A_2$, the set of runs where $q$ crashes before time $\ttime$;
%\item $A_3$,
%joe10: It wasn't a partition before
%the set of runs where $p$'s first message is lost and $q$ is up for at least
%$\ttime$ time units;
%\item $A_4$, the set of runs where $q$'s first $\ack$ is lost
%\item $A_4$, the remaining runs.
%and $q$ does not crash before time $\ttime$;
\item $A_1 = \{r : $ $q$ $\prcvs$ $\Hm$ at time $\ttime$ and successfully
$\psends$ $\ack(\Hm)$ at time $\ttime$ in $r\}$,\footnote{Note that in this
case $p$ must have successfully $\psent$ $\Hm$ at time 0.}
\item $A_2 = \{r : $ $q$ crashes before time $\ttime$ in $r\}$,
\item $A_3 = \{r : $ $q$ is up at time $\ttime$ but does not $\prcv$ $\Hm$
%joe22
at time $\ttime$
in $r\}$, and
\item $A_4 = \{r : $ $q$ $\prcvs$ $\Hm$ at time $\ttime$ but the
%joe22
corresponding
 $\ack(\Hm)$ is lost in $r\}$.
\end{itemize}
%joe10: rewrote and reordered.  You were missing the story
It is easy to see that
%joe22: itemized
\begin{itemize}
\item $\Pr(A_1) = (1-\pf_p)(1-\pf_q)^\ttime(1-\lf)^2
%joe10
%\aeq 1,
= 1 - \pf_p - \ttime \pf_q - 2 \lf + O(\eps^2)
%joe22: put back
\aeq 1$,
\item $\Pr(A_2) =
(1-(1-\pf_q)^\ttime)
%joe10: this is \aeq 0!  It's not what you really want to say
%\aeq \ttime \pf_q
= \ttime \pf_q + O(\eps^2)
%joe22
\aeq 0$,
%joe10
%$\Pr(A_3) = (1-\pf_p)\lf
%< \lf$, and $\Pr(A_4) = (1-\pf_p)(1-\pf_q)^\ttime(1-\lf)\lf < \lf$.
%$A_1$ is the most likely case.  $A_2$, $A_3$, and $A_4$ are all very
%unlikely compared to $A_1$.
%
%fcc19
%$\Pr(A_3) = \lf + \pf_p + O(\eps^2)$, and $\Pr(A_4) = \lf + O(\eps^2)$.
\item $\Pr(A_3) = ((1-\pf_p)\lf + \pf_p)(1-\pf_q)^{\ttime} = \lf + \pf_p
+ O(\eps^2)
%joe22
\aeq 0$, and
\item $\Pr(A_4) = (1-\pf_p)(1-\lf)\lf(1-\pf_q)^{\ttime} = \lf +
O(\eps^2)
%joe2
\aeq 0$.
\end{itemize}
Roughly speaking, what happens is that
%fcc21 in $A_1$, the expected value of $\numprt(r)$ is $2 \cterm$
$\E^{\sdp}(\numprt \gv A_1) \aeq 2 \cterm$;
that is why this term appears
%fcc21 deleted
%as a term
in
$\E^{\sdp}(\numprt)$.  In $A_2$, the expected value of $\numprt$ is very
large, so that despite the low probability of $A_2$, it contributes the term
$\frac{\ttime \pf_q}{\msgt \pf_p}$ to $\E^{\sdp}(\numprt)$.
%why this in $A_2$
%we expect $\numprt(r)$ to be  much higher, since $p$ will not stop
%until it crashes.
The expected value of $\numprt$ in $A_3$ or $A_4$ is not large enough to
overcome the low probability of these sets, and so does not affect
$\E^{\sdp}(\numprt)$.
%In $A_3$ and $A_4$, $\numprt(r)$ is expected to be bigger than in $A_1$
%but not as big as in $A_2$.
We now present this analysis this analysis in more detail.

For $A_1$, note that for $k > 2 \cterm$, $\Pr(\numprt=k \gv A_1) = 0$, since by
time $2\ttime$, either $p$ crashed or
%fcc19 it $\prcv$d an $\ack$.  For $k < 2
$p$ $\prcvd$ $\ack(\Hm)$. For $k < 2 \cterm$,
%fcc21
%$\Pr(\numprt(r)=k \gv A_1) \leq \ttime(\lf+\pf_p)$,
%joe10: here (and about 30 other times) I removed r from the scope of E.
%The expectation is independent of any particular run r
%crashed.  Thus $\E(\numprt \gv A_1) \aeq 2 \cterm$.
%fcc21
$\Pr(\numprt = k \gv A_1)
%joe22
%=
\le
1 - ((1-\pf_p)(1-\lf))^{\ttime}
%joe22
%= \ttime(\pf_p + \lf) + O(\eps^2)
\aeq 0$,
since either a message is lost or $p$ crashed.
Thus,
%joe22: you left out the superscript here and a few other places.  I
%hope that I added them everywhere; please check.
%$\E(\numprt \gv A_1) \aeq 2 \cterm$.
$\E^{\sdp}(\numprt \gv A_1) \aeq 2 \cterm$.

For $A_2$, we have $\Pr(\numprt = k  \gv A_2) = (1-\pf_p)^{(k-1)\msgt}
(1-(1-\pf_p)^{\msgt})$, since $p$ must live for at least $(k-1)\msgt$ time
units but less than $k\msgt$ time units.
%joe22: moved up from below
Using the well-known fact that $\sum_{k=1}^\infty k x^k =
\frac{x}{(1-x)^2}$, we get that
%joe10: this went by a little too quickly
%fcc10	reformat
$$\begin{array}{lll}
\E^{\sdp}(\numprt \gv A_2)
& = & \ds \sum_{k=1}^\infty
k (1-\pf_p)^{(k-1)\msgt}(1-(1-\pf_p)^{\msgt})\\
& = & \ds \frac{1-(1-\pf_p)^{\msgt}}{(1-\pf_p)^\msgt}
\sum_{k=1}^\infty
k ((1-\pf_p)^\msgt)^k\\
& = & \ds \frac{1}{1-(1-\pf_p)^\msgt}\\
& \aeq & \ds \frac{1}{\msgt\pf_p}.
\end{array}$$

%joe10: You need to explain where this is coming from in much more
%detail here.
%fcc10
%fcc19 this is not true
%We want to argue that $\E(\numprt \gv A_3) \aeq \E(\numprt)+1+
%\frac{\pf_q}{\pf_p}$.  To see this,
We now turn our attention to $A_3$.
Note that $\Pr(\numprt > 2k+1 \gv A_3) \le \lf^k$, since for every
two consecutive messages
%fcc24 sent
$\psent$
in a run $r \in A_3$ after the first one,
one must be a message of $p$ which is lost or one must be an $\ack$ by
$q$ which is lost.  Thus, $$\E^{\sdp}(\numprt \gv A_3) =
\sum_{k=1}^\infty k \Pr(\numprt = k \gv A_3) \le 6 +
\sum_{k=1}^\infty
(2k+2 + 2k+3) \lf^k \aeq 6.$$

An almost identical analysis shows that $\E^{\sdp}(\numprt \gv A_4) \le
10 + O(\eps)$.
Putting these calculations together, we get
$$\begin{array}{lll}
\E^{\sdp}(\numprt) & = & \Pr(A_1)\E^{\sdp}(\numprt \gv A_1) +
\Pr(A_2)\E^{\sdp}(\numprt \gv A_2) \\
& & + \Pr(A_3)\E^{\sdp}(\numprt \gv A_3) + \Pr(A_4)\E^{\sdp}(\numprt \gv
A_4) \\
&\aeq& 2 \cterm + \frac{\ttime \pf_q}{\msgt \pf_p}.
%fcc19
%2\lf(\E(\numprt)).
%\end{array}
%joe22: cut
%2\lf(\E(\numprt)).
\end{array}
$$
%joe22: cut this too
%regardless of whether $\msgt \geq \ttime$ or $\msgt < \ttime$, since
%$\Pr(A_3)$ and $\Pr(A_4)$ are both $O(\eps)$.
%Let $D = 2 \cterm + \frac{\ttime \pf_q}{\msgt \pf_p}$.
%%Then we see that $\E(\numprt) = C + 2\lf\E(\numprt)$, and so
%%$\E(\numprt) = C + 2\lf(\frac{C}{1-2\lf}) \aeq
%%C = 2 \cterm +\frac{\ttime \pf_q}{\msgt \pf_p}$.
%Since $\E(\numprt) \aeq D + 2\lf\E(\numprt)$, it follows that
%$\E(\numprt) \aeq D + 2\lf(\frac{D}{1-2\lf}) \aeq D$, as desired.
}%\end{commentout}
%fcc23 new proof
Now let us turn to $\E^{\sdp}(\numprt)$.  
%fcc25 added  (this is not the same as saying the message is received, since
%the recipient could crash.)
Let us say that a $\psend$ is \emph{successful} iff the link does not drop the
message (which could be an $\ack$).  
Consider the set of runs $A = \{r : $ $q$ successfully $\psends$ $\ack(\Hm)$
before crashing in $r\}$.  Roughly speaking, what happens is that
%fcc24
in runs of $A$, $p$ is $\prcv$s $\ack(\Hm)$ at time $2\ttime$ with probability
$\aeq 1$.  In the
%joe24
%mean time,
meantime,
$p$ has $\psent$ $\Hm$ exactly $\cterm$ times with
probability $\aeq 1$.  With
probability $\aeq 1$, all of these are $\prcvd$ by $q$; $q$ in turn
acknowledges all copies and thus $\E^{\sdp}(\numprt \gv A) \aeq 2 \cterm$; that
is why this term appears in $\E^{\sdp}(\numprt)$.  In $\cpl{A}$, the expected
value of $\numprt$ is very large, since $p$ will $\psend$ $\Hm$ until it
crashes, so despite the low probability of $\cpl{A}$, it contributes the term
$\frac{(\ttime+1) \pf_q}{\msgt \pf_p}$.  We now turn to the details.

We first compute $\Pr(A)$.  Note that $q$ can $\psend$ $\ack(\Hm)$ only at
times of the form $\ttime + k \msgt$.  Let $B_k = \{r : $ $q$ $\psends$ the
first successful $\ack(\Hm)$ at time $\ttime +  k \msgt\}$.  Note that $A =
\bigcup_{k=0}^\infty B_k$ and that $B_i \cap B_j = \emptyset$ if $i \neq j$.
Thus $\Pr(A) = \sum_{k=0}^\infty \Pr(B_k)$.  Since $q$ $\psends$ the first
successful $\ack(\Hm)$ at time $\ttime + k \msgt$ in runs of $B_k$, $p$ must
(successfully) $\psend$ $\Hm$ at time $k \msgt$ in runs of $B_k$.  Thus
\[
\Pr(B_k) = (1-\pf_p)^{k \msgt + 1} (1-\pf_q)^{\ttime + k\msgt + 1} (2\lf-\lf^2)^k (1-\lf)^2.
\]
The first factor reflects the fact that $p$ must have been up at time $k \msgt$
(to $\psend$ $\Hm$) while the second factor reflects the fact that $q$ must
have been up at time $\ttime + k \msgt$ (to $\prcv$ $\Hm$ and $\psend$
$\ack(\Hm)$).  The third factor reflects the fact that the previous $k$
attempts have failed: either $\Hm$ was lost or the corresponding $\ack(\Hm)$
was lost, which occurs with probability $(\lf + (1-\lf)\lf) = 2\lf - \lf^2$.
The final factor reflects the fact that the $(k+1)$st attempt succeeded: both
messages got through.  So
\[
%\begin{array}{ll}
\begin{array}{lll}
% &
\ds \Pr(A)
%\\
&
= & \ds \sum_{k=0}^\infty \Pr(B_k) \\
&
= & \ds \sum_{k=0}^\infty (1-\pf_p)^{k \msgt + 1} (1-\pf_q)^{\ttime + k \msgt +
1} (2\lf - \lf^2)^k (1-\lf)^2 \\
&
= & \ds (1-\pf_p) (1-\pf_q)^{\ttime+1} (1-\lf)^2
\sum_{k=0}^\infty (1-\pf_p)^{k \msgt} (1-\pf_q)^{k \msgt} (2\lf-\lf^2)^k\\
&
= & \ds (1-\pf_p) (1-\pf_q)^{\ttime+1} (1-\lf)^2 \frac{1}{1 - (1-\pf_p)^\msgt
(1-\pf_q)^\msgt (2\lf - \lf^2)} \\
&
= & \ds (1-\pf_p) (1-\pf_q)^{\ttime+1} (1-\lf)^2 ((1+2\lf) + O(\eps^2))\\
&
= & 1 - \pf_p - (\ttime+1) \pf_q + O(\eps^2) \\
&
\aeq & 1.\\
\end{array}
\]
We now want to compute $\E^{\sdp}(\numprt \gv A)$.
Again, we break $\E^{\sdp}(\numprt \gv A)$ into three
%fcc24---
pieces,
%joe24: I think it's easier to parse when it's itemized; reinstated
%fcc25: added commas
\bit
\item $\ds \sum_{k=0}^{2\cterm-1} k \Pr(\numprt = k \gv A)$,
\item $2\cterm \Pr(\numprt = 2\cterm \gv A)$, and 
\item $\ds \sum_{k=2\cterm+1}^\infty k \Pr(\numprt = k \gv A)$,
\eit
%\[
%\ds \sum_{k=0}^{2\cterm-1} k \Pr(\numprt = k \gv A) +
%2{\ts \cterm} \Pr(\numprt = 2{\ts \cterm} \gv A) +
%\ds \sum_{k=2\cterm+1}^\infty k \Pr(\numprt = k \gv A),
%\]
and compute each part separately.

Note that $\Pr(\numprt = k \gv A) \leq \pf_p + \pf_q + \lf$ for $k < 2\cterm$,
since either a process crashed or a message is lost.  Thus the first part is
no more than $2\cterm (\pf_p + \pf_q + \lf) \aeq 0$, so we may ignore it.
%Suppose $r$ is a run such that $p$ is up at $2\ttime$, $q$ is up at $\cterm
%\msgt + \ttime$, all copies of $m$ got through, and the first $\ack(\Hm)$ of
%$q$ got through. then $p$ will $\psend$ $\cterm$ messages.  If $p$ $\prcv$s
%$\ack(\Hm)$ at $2\ttime$, $p$ will stop.  If $q$ is up at $\cterm \msgt + \ttime$
For the second part, we have
\[
\ts \Pr(\numprt = 2 \cterm \gv A) \geq (1-\pf_p)^{2\ttime+1} (1-\pf_q)^{\cterm\msgt +
\ttime+1} (1-\lf)^{\cterm+1} \aeq 1,
\]
since if $p$ is up at time $2\ttime$, $q$ is up at time $\cterm \msgt +\ttime$,
all of $p$'s $\psend$s got through, and $q$'s first $\ack(\Hm)$ got through,
then $\numprt = 2 \cterm$;
thus the second part is $\aeq 2\cterm$.  We now turn our attention to the last
part.

%Let us say that a copy of $m$ is \emph{successful} if that copy of $m$ $\psent$
%by $p$ is delivered by the link and the corresponding $\ack(\Hm)$ was also
%delivered by the link.
Note that $p$ $\psends$ at least half the messages in every run $r$ (whether $r
\in A$ or $r \in \cpl{A}$).  Note also that,
after the first successful attempt
%joe24
%fcc25 Not quite: p may crash before receiving.  ``Successful'' just means the
%link didn't fail.  The reason I removed the first part of this paragraph is
%because the notion of a successful attempt was already discussed shortly
%before this.
%
(that is, after the first message $\psent$ by $p$ which is $\prcvd$ by
%$q$ whose corresponding acknowledgment is $\prcvd$ by $p$),
$q$ whose corresponding acknowledgment is not lost by the link),
$p$ will $\psend$ at most $\cterm$ messages, since $p$ would stop $\psending$
$2\ttime$ time units after the first successful attempt (either because $p$
$\prcvd$ $\ack(\Hm)$ or $p$ crashed).  Combining the above two observations, we
see that if $\numprt(r) = 2 \cterm + k$ for $k > 0$, then $p$ must have
$\psent$ at least $\cterm + \bceil{\frac{k}{2}}$ messages
%joe24
%which means that
and
there are at least $\bceil{\frac{k}{2}}$ unsuccessful attempts in $r$.
%  (Recall
%that the probability of an unsuccessful attempt is $\lf + (1-\lf)\lf = 2\lf -
%\lf^2$ for runs of $A$.)
% (since otherwise $p$ would
%$p$ must have $\psent$ a message at time
%$\cterm \msgt + \bceil{\frac{k}{2}} \msgt$ (since $p$ $\psends$ $\Hm$ every
%$\msgt$ time units).  This in turn implies that there are at least
%not $\psend$ any message at time $\cterm \msgt + \ktime \msgt$, either because
%it $\prcvd$ $\ack(\Hm)$ or because it crashed).
Thus, $\Pr(\numprt = 2 \cterm + k \gv A) \leq (2\lf-\lf^2)^{\bceil{\frac{k}{2}}}$.
So we have
\[
%\begin{array}{ll}
\begin{array}{lll}
%&
\ds \sum_{k=2\cterm+1}^\infty k \Pr(\numprt = k \gv A)
%\\
&
\leq & \ds \sum_{k=2\cterm+1}^\infty k (2\lf-\lf^2)^{\ktime} \\
&
= & \ds \sum_{k=\cterm}^\infty ((2k+1) + (2k+2)) (2\lf-\lf^2)^{k+1} \\
&
= & \ds \sum_{k=\cterm}^\infty (4k+3) (2\lf-\lf^2)^{k+1} \\
&
\aeq & 0.
\end{array}
\]
So we may ignore the last part as well.
Thus $\E^{\sdp}(\numprt \gv A) \aeq 2 \cterm$.
Since $\Pr(A) \aeq 1$, we have $\E^{\sdp}(\numprt \gv A) \Pr(A) \aeq 2 \cterm$.

We now focus on $\E^{\sdp}(\numprt \gv \cpl{A}) \Pr(\cpl{A})$.  Recall that for $r \in
\cpl{A}$, $q$ fails to successfully $\psend$ $\ack(\Hm)$ in $r$.
Consider the following three sets (which is a partition of the set of all runs):
\bit
%fcc24
%joe24: I preferred the original; reinstated
\item $C_1 = \{r : p$ crashes at time 0 in $r\}$,
\item $C_2 = \{r : p$ does not crash at time 0
%joe24
%but $q$ crashes at or before
and $q$ crashes at or before
time $\ttime$ in $r\}$, and
\item $C_3 = \{r : p$ does not crash at time 0 and $q$ does not crash at or before
time $\ttime$ in $r\}$.
%\item $C_1 = \{r : p$ crashes at time 0 in $r\}$.
%\item $C_2 = \{r : p$ does not crash at time 0 but $q$ crashes at or before
%time $\ttime$ in $r\}$.
%\item $C_3 = \{r : p$ does not crash at time 0 and $q$ does not crash
%at or before time $\ttime$ in $r\}$.
\eit
%joe24
%We will show
We now show
%fcc24 that
that these are their probabilities:
\bit
%\item $\Pr(C_1 \cap \cpl{A}) = \pf_p$,
%\item $\Pr(C_2 \cap \cpl{A}) = (1-\pf_p)(1-(1-\pf_q)^{\ttime+1}) =
%(\ttime+1)
%\pf_q + O(\eps^2)$, and
%\item $\Pr(C_3 \cap \cpl{A}) = O(\eps^2)$.
\item $\Pr(C_1 \cap \cpl{A}) = \pf_p$,
\item $\Pr(C_2 \cap \cpl{A}) = (1-\pf_p)(1-(1-\pf_q)^{\ttime+1}) =
(\ttime+1) \pf_q + O(\eps^2)$, and
\item $\Pr(C_3 \cap \cpl{A}) = O(\eps^2)$.
\eit

First note that $\Pr(C_1) = \pf_p$ and
$\Pr(C_2) = (1-\pf_p) (1-(1-\pf_q)^{\ttime+1}) = (\ttime+1)\pf_q + O(\eps^2)$.
Furthermore, $C_1 \cup C_2 \sbset \cpl{A}$, since if $r \in C_1 \cup C_2$, $q$
does not $\psend$ $\ack(\Hm)$ successfully before crashing.  Thus
$\Pr(C_1 \cap \cpl{A}) = \pf_p$ and
$\Pr(C_2 \cap \cpl{A}) = (\ttime+1) \pf_q + O(\eps^2)$.
%joe24: added next sentence
Since, as we showed earlier, $\Pr(A) = 1 - \pf_p - (\ttime+1) \pf_q +
O(\eps^2)$, it also follows that $\Pr(C_3 \inter \cpl{A}) = O(\eps^2)$.

Now that we have $\Pr(C_i \inter \cpl{A})$, let us turn to $\E^{\sdp}(\numprt
\gv C_i \inter \cpl{A})$.
Note that for $r \in \cpl{A}$, $p$ will $\psend$ messages until it crashes.
For $r \in C_1$, $p$ crashes immediately, so $\numprt(r) = 0$ for $r \in C_1$.
For $r \in C_2$, $q$ crashes before it can possibly $\psend$ any messages,
%joe24
%thus
so
all the messages are $\psent$ by $p$.  Thus
\[
%fcc24  off by one
%\Pr(\numprt = k \gv C_2) = (1-\pf_p)^{(k-1)\msgt} (1-(1-\pf_p)^{\msgt}),
\Pr(\numprt = k \gv C_2) = (1-\pf_p)^{(k-1)\msgt+1} (1-(1-\pf_p)^{\msgt}),
\]
since $p$
%joe24
must
be up at time $(k-1)\msgt$ and crash before time $k\msgt$
 to $\psend$ $\Hm$ exactly $k$ times.
%fcc24  This is not the first place we use this fact anymore.
%Using the well-known fact that
%$\sum_{k=1}^\infty k x^k = \frac{x}{(1-x)^2}$, we get that
So
\[
%\begin{array}{ll}
\begin{array}{lll}
%&
\E^{\sdp}(\numprt \gv C_2 \cap \cpl{A})
%\\
%fcc24 corrections for the above
&
 = & \ds \sum_{k=1}^\infty
k (1-\pf_p)^{(k-1)\msgt+1}(1-(1-\pf_p)^{\msgt})\\
&
 = & \ds \frac{(1-\pf_p)(1-(1-\pf_p)^{\msgt})}{(1-\pf_p)^{\msgt}}
\sum_{k=1}^\infty
k ((1-\pf_p)^\msgt)^k\\
&
 = & \ds \frac{(1-\pf_p)(1-(1-\pf_p)^{\msgt})}{(1-\pf_p)^{\msgt}}
\frac{(1-\pf_p)^\msgt}{(1-(1-\pf_p)^\msgt)^2}\\
&
 = & \ds \frac{1-\pf_p}{1-(1-\pf_p)^\msgt}\\
&
 = & \ds \frac{1}{\msgt\pf_p} + O(1).
\end{array}
\]
The $O(1)$ term is there because $\abs{\frac{1}{\msgt\pf_p} -
\frac{1-\pf_p}{1-(1-\pf_p)^\msgt}} = \abs{\frac{1}{\msgt\pf_p} - \frac{1-\pf_p}{\msgt \pf_p
+ O(\eps^2)}} = \abs{\frac{O(\eps^2)}{(\msgt\pf_p)^2 + O(\eps^3)}}$, which is
$O(1)$, since we assumed that
%joe24
%$\frac{1}{\pf_p}$ is $O(\eps^{-1})$ for this proposition.
$\pf_p$ is $\Theta(\eps)$ for this proposition.

%joe24: added paragraph break
For $r \in C_3 \inter \cpl{A}$, $q$ might $\psend$ messages
(none of which, however, will get through).  Let $E_k = \{r \in C_3 \inter
\cpl{A} : $ $p$ crashes at time $k\}$. We have $\Pr(E_k) \leq (1-\pf_p)^k
\pf_p$.  Furthermore, $\E^{\sdp}(\numprt \gv E_k) \leq
2\bceil{\frac{k}{\msgt}}$, since $p$ $\psends$ $\bceil{\frac{k}{\msgt}}$
messages in $E_k$ and $q$ $\psends$ at most that many messages.  So we have
\[
%\begin{array}{ll}
\begin{array}{lll}
%&
\ds \E^{\sdp}(\numprt \gv C_3 \inter \cpl{A})
%\\
&
= & \ds \sum_{k=1}^\infty \E^{\sdp}(\numprt \gv E_k) \Pr(E_k) \\
&
\leq & \ds \sum_{k=1}^\infty 2{\ts \bceil{\frac{k}{\msgt}}}
(1-\pf_p)^k \pf_p \\
%fcc25:  The reason I did this is because the following sum looks a lot like
%the sum for C_2.  That's why I didn't compute it explicitly: it's already done.
%joe24: I found it hard to follow this; there's a simpler argument anyway
%&
%= & \ds \sum_{k=1}^\infty 2k
%(1-\pf_p)^{(k-1)\msgt+1}(1-(1-\pf_p)^\msgt)\\
&
\leq & \ds \sum_{k=1}^\infty 2{\ts \left(\frac{k}{\msgt} + 1\right)}
(1-\pf_p)^k \pf_p \\
& = &\ds \frac{2 \pf_p}{\msgt}
\sum_{k=1}^\infty k (1-\pf_p)^k  + 2 \pf_p \sum_{k=1}^\infty
(1-\pf_p)^k\\
&= &\ds \frac{2 \pf_p}{\msgt} \frac{1-\pf_p}{\pf_p^2} + 2(1 - \pf_p)\\
&
%fcc24 \aeq
=
 & \ds \frac{2}{\msgt \pf_p} + O(1).\\
\end{array}
\]
%joe24: now unnecessary
%(The $O(1)$ term is there for the same reason as before.)
%joe24
%Since we assumed that $\frac{1}{\pf_p}$ is $O(\eps^{-1})$,
Since we assumed that $\pf_p$ is $\Theta(\eps)$,
$\E^{\sdp}(\numprt \gv C_3 \inter \cpl{A}) \Pr(C_3 \inter \cpl{A}) = O(\eps)$.
Recall that $\E^{\sdp}(\numprt \gv C_1 \inter \cpl{A}) = 0$, so
\[
\E^{\sdp}(\numprt \gv \cpl{A}) \Pr(\cpl{A})
\aeq  \ds \E^{\sdp}(\numprt \gv C_2 \inter \cpl{A}) \Pr(C_2 \inter \cpl{A})
%= & (\frac{1}{\msgt\pf_p} + O(1)) ((\ttime+1) \pf_q + O(\eps^2))\\
\aeq  \ds \frac{(\ttime+1)\pf_q}{\msgt \pf_p}.
\]
This gives us $\E^{\sdp}(\numprt) \aeq \frac{(\ttime+1)\pf_q}{\msgt\pf_p} + 2
\cterm$ as desired.

The reasoning for the $\rdp$ case is similar to the $\sdp$ case.  The only
major difference is that $q$ cannot possibly
%fcc19 $\rrcv$ $m$ before time $2\ttime$.
finish $\rrcving$ $\Hm$ before time $2\ttime$.
We leave details to the reader.~\eprf

%joe22
%\frac{1}{\hbt\spb}\cm$, where $0 < \lambda < 1$ is a constant that
\othm{avgSecNum}{avgThmNum}
\thm
Under the cost function $\csendav$, Protocol $\rb$ satisfies $\spc_3$.
Furthermore,
$\E(\csendav) \aeq ((1-\pct_p)(1-\pct_q)\lambda + \pct_p \pct_q)
\left(2\cterm\cm +
%fcc24
%\left(2\ttime + \frac{\hbt-1}{2}\right)\cw \right) +
\left(\ttime + \frac{\hbt-1}{2}\right)\cw \right) +
\frac{1}{\hbt\spb}\cm$, where $0 < \lambda < 1$.
%is a constant that depends only
%%fcc24
%%on $\pf_p$ and $\pf_q$.\footnote{Intuitively, $\lambda$ depends on the
%%probability that both $p$ and $q$ are up relative to the probability
%%that one is up and the other down.}
%on $\pf_p$, $\pf_q$, $\ttime$, and $\hbt$.
\ethm
\eothm

\prf Roughly speaking, the first summand corresponds to the expected
%joe22
%per invocation
per-invocation
cost of the protocol and the second corresponds to the expected
%joe22
per-invocation
cost of the heartbeats.  To do the analysis carefully, we divide the set of
runs into three subsets:
\begin{itemize}
%fcc17
%\item $C_1$, the set of runs where one process is correct and the other
%eventually crashes;
%\item $C_2$, the set of runs where both processes are correct;
%\item $C_3$, the set of runs where both processes crash.
%joe24: I preferred the original; reinstanted
\item $F_1 = \{r : $ one process is correct and the other eventually crashes in $r\}$,
\item $F_2 = \{r : $ both processes are correct in $r\}$, and
\item $F_3 = \{r : $ both processes eventually crash in $r\}$.
%\item $F_1 = \{r : $ one process is correct and the other eventually
%crashes in $r\}$.
%\item $F_2 = \{r : $ both processes are correct in $r\}$.
%\item $F_3 = \{r : $ both processes eventually crash in $r\}$.
\end{itemize}
%
%joe10
%fcc24 stylistic consistency
%Note that $\Pr(F_1) = \pct_p(1 - \pct_q) + \pct_q(1 - \pct_p)$, $\Pr(F_2) =
%\pct_p\pct_q$, and $\Pr(F_3) = (1 - \pct_p)(1 - \pct_q)$.
These are their probabilities:
\bit
%joe24: used commas:
\item $\Pr(F_1) = \pct_p(1 - \pct_q) + \pct_q(1 - \pct_p)$,
\item $\Pr(F_2) = \pct_p\pct_q$, and
\item $\Pr(F_3) = (1 - \pct_p)(1 - \pct_q)$.
\eit
For $r \in F_1$, we
expect the lone correct process to invoke $\rb$ infinitely often.  All but
finitely many of these invocations will take place after the other process
crashed.  Thus the average cost of an invocation in $r$ will be 0.  For $r \in
F_2$, on the other hand, both processes are expected to invoke $\rb$ infinitely
often and the average cost of the invocation
%joe13
in $r$
is expected to be close to the expected cost
%joe10
of a single invocation of $\rb$.
%joe10
%For $r \in C_3$ there are finitely many invocations
%altogether and we need to partition $C_3$ when doing our analysis.
The computation of the expected cost of an invocation in a run in $F_3$ is more
delicate.  We now examine the details.

%joe8: added paragraph break
%joe10: cut
%If one of the processes
%crashed before the start of an invocation (clearly the other one must
%be up),
%the cost of that particular invocation is 0, since no messages are sent and the
%time of waiting is 0.
%joe8
%Thus if one of the process is correct and the other
%joe9: cut; now said it above
%Let $C_1$ be the set of runs in which
%one of the process is correct and the other crashes,
%joe8: added next four lines
Let $G_1$ be the subset of $F_1$ consisting of runs $r$ in which the correct
process tries to invoke the protocol infinitely often.
%fcc8	balanced $
Clearly $\Pr(G_1 \gv F_1) = 1$, since the protocol is invoked with probability
$\spb$ at each time unit.  Moreover, for each run
%joe10
%$r \in B$, we have
$r \in G_1$, we have
%joe9
%$\lim_{t\rar\infty} \tcst(r,t)/\tcmp(r,t) = 0$,
%fcc19
%$\lim_{t\rar\infty} (\tcst(r,t)/(\tcmp(r,t)+1)) = 0$,
$$\lim_{t\rar\infty} \frac{\tcst(r,t)}{\tcmp(r,t)+1} = 0,$$
%joe8: This didn't make sense?  What wass this probability being taken
%over?
%with probability 1,
since there are only finitely many complete invocations with non-zero cost and
%fcc8 (with probability 1)
there are infinitely many complete invocations.
%joe10: added
Thus, $\E(\csendav \gv F_1) = 0$.

%joe8: added
%If both processes are correct, then
%joe9: rewrote; moved in material from first paragraph
%Let $C_2$ consist of all runs in which both processes are correct.
Let $G_2$ be the subset of $F_2$ where there are infinitely many
invocations of
$\rb$.  Clearly $\Pr(G_2 \gv F_2) = 1$.
%joe10
Let $\tempvar = 2\cterm\cm +
%fcc24 (2\ttime
\left(\ttime
+ \frac{\hbt-1}{2}\right)\cw$.  By the Law of
Large Numbers, for almost all runs $r$ of $G_2$, the analysis of
%joe22: we didn't do the proof of this theorem
%\mthm{t:impl2}
\mprop{p:comparison}
shows that
%fcc19
% $\lim_{t\rar\infty} (\tcst(r,t)/(\tcmp(r,t)+1))
%%fcc10 =
%\aeq
%%expected cost of each invocation of the protocol is
%\tempvar$.
$$\lim_{t\rar\infty} \frac{\tcst(r,t)}{\tcmp(r,t)+1} \aeq
\tempvar.$$
%fcc24  This is not correct:  I've corrected the statement of the thm.
%(The extra $\frac{\hbt-1}{2}$ term
%joe10
%in $\tempvar$ compared to the expression in \mthm{t:impl2} is due to the fact
%that the first $\hbmsg$ is not necessarily  sent at the start of the invocation.
%On average, $\frac{\hbt-1}{2}$ time will elapse before it is sent.)
%fcc25 rewrote
\commentout{
(We have $\ttime+\frac{\hbt-1}{2}$ instead of $2\ttime$ in $Z$, compared to the
expression in \mthm{t:impl2}, since $\hbmsg$s are $\psent$ every $\hbt$ time
%joe24
%units, so on average a $\hbmsg$ will arrive every $\hbt$ time units.
units so, on average, a $\hbmsg$ will arrive every $\hbt$ time units.
Thus, the
first $\psend$ is expected to happen $\frac{\hbt-1}{2}$ time units after the
invocation started.
%joe24
% After the first $\psend$ takes place, the
%receiver is
%expected to $\prcv$ $m$ $\ttime$ time units later, so the expected
%waiting time is $\ttime+ \frac{\hbt-1}{2}$ instead of $2\ttime$.)
By way of contrast, in \mthm{t:impl2}, the first $\hbmsg$ is sent at
time 0 and the first $\psend$ is expected to happen at roughly time
$\ttime$, when it arrives.  In both cases, the waiting time after the
first $\hbmsg$ is expected to be a further $\ttime$.)
}%\end{commentout}
(Note that we have $\ttime + \frac{\hbt-1}{2}$ instead of $2\ttime$ as in
\mthm{t:impl2}.  This is because in the current setting, the expected amount of
time elapsed between the start of an invocation and the arrival of the first
$\hbmsg$ is $\frac{\hbt-1}{2}$.  In the setting of \mthm{t:impl2}, however, the
first $\hbmsg$ cannot arrive until time $\ttime$, since the invocation starts
at time 0 and the first $\hbmsg$ is $\psent$ at time 0.  Note that in both
cases, the expected time of waiting is $\ttime$ plus the expected time elapsed
between the start of the invocation and the arrival of the next $\hbmsg$.)
%Thus, for almost all runs $r \in C_2$,
%$\lim_{T\rar\infty}
%\tcst(r,t)/(\tcmp(r,t)+1) = 2\cterm\cm + (2\ttime + \frac{\hbt-1}{2})\cw$
%with probability 1, by the (weak) Law of Large Numbers.
Thus $\Pr(
%joe10
%\lim_{t\rar\infty} (\tcst(r,t)/(\tcmp(r,t)+1))
%= 2\cterm\cm + \left(2\ttime + \frac{\hbt-1}{2}\right)\cw \gv F_2) = 1.
\csendav(r)
%fcc10 =
\aeq \tempvar \gv F_2) = 1$.

%joe8: again, you don't mean probability 1; you mean probability 1
%conditional on C_2.  Cut
%(Note that in this case,
%with probability 1, both the numerator and denominator are unbounded.)

%joe9: I think this was right before you changed it, but I'm cutting the
%sentence anyway.
%Finally, let $C_3$
%%%fcc8 consist
%%consists
%consist
%of those runs where both processes eventually crash.
%
%fcc15 Finally, for a run $r \in C_3$,
%joe24: simplified argument; cut old argument completely
\commentout{
 Finally, we consider runs in $F_3$.
%fcc24
%Let $r \in F_3$.
%joe24: unnecessary
%Let $Z$ be defined as above.
%joe24: this doesn't make sense.  Invocations have costs, not expected
%costs.
Intuitively, for most invocations in most runs of $F_3$, the expected cost is
$Z$.
However, since each run has only finitely many invocations, we cannot
resort to the Law of Large Numbers and the analysis is more complicated;
we now turn to the details.
Let $r \in F_3$ and $i$ be a particular invocation in $r$.
%joe24: yuck
%(We denote this relation, by a slight abuse of notation, via $i \in r$.)
%joe24: I added r to all your random variables; it seems you intended i
%to be unique to run r, but most people won't read it that way
%Let $\start(i)$
Let $\start(r,i)$
denote the time that $i$ starts and $\finish(r,i)$ denote the time that
$i$ finishes.  Let
$\sd(r,i)$ be the process that started $i$.
%joe24
%(since $\sdrc$ is a sender-driven protocol).
Let $\numprt(r,i,t)$ denote the number of messages $\psent$ by
invocation $i$ up to time $t$
%joe24
%(which is $0$ before $i$ starts)
($\numprt(i,t) = 0$ for $t < \start(i)$)
%joe24: this is very hard to parse
%and $\tow(i, t)$
%be $\tow$ for invocation $i$ up to time $t$ (which we take to be $0$
%before $i$ starts).
and let $\tow(r, i, t) = \max(\start(r,i),\min(t_p,t_q,t_f,t)) -
\start(r,i)$
%be the time of waiting for invocation $i$ up to time $t$ (which we take
%to be $0$ before $i$ starts).
(where $t_p$, $t_q$, and $t_f$ are as in the original definition of
$\tow(r)$.
Let $\tcst(r,i,t) = \numprt(r,i,t) \cm + \tow(r,i,t) \cw$.  Note
that
$\tcst(r,t) = \sum_{i \in r} \tcst(r,i, t)$.
%joe24: \tcst(i) was undefined
%and that $\tcst(i) = \tcst(i,\infty)$.

Let $t_{c(r)}$ denote the time of the first crash in $r$.  Let $x(r)$ be the
process that crashes first and let $y(r)$ denote the other process.  (If both
%joe24
%process crashes
processes crash
simultaneously at $t_{c(r)}$, let $x(r)$ be $p$ and
$y(r)$ be $q$.)
%Note that the expected value of $\tcst(i)$ for $i \in r \in F_3$ depends
%on the relation of $\start(i)$ and $t_{c(r)}$.  The intuition is that for
%invocations that start long before $t_{c(r)}$ the expected cost is $\aeq Z$ and
%for invocations that start shortly before or start after $t_{c(r)}$, the expected
%cost is less than $Z$ (and how much less depends on exactly when they start).
%
%the argument in the proof of
%\mprop{p:comparison} applies while for invocations that start shortly before or
%start after $t_{c(r)}$, the argument in the proof of \mprop{p:comparison}
\commentout{
For example, for $i$ such that
%$\start(i) > \max\left\{\bfloor{\frac{t_{c(r)}}{\hbt}}
%\hbt + \ttime, t_{c(r)} \right\}$,
$\start(i) > t_{c(r)}+\hbt$, we have $\tcst(i) = 0$, since $\start(i) > t_{c(r)}$
implies $\tow(i) = 0$ and $\start(i) > t_{c(r)}+\hbt$ implies that $\sd(i)$ will
never $\prcv$ $\hbmsg$s from the other process after $\start(i)$ and thus
$\numprt(i) = 0$.  For $i$ with $\sd(i) = x(r)$ and
%$\start(i) \leq t_{x(r)} =
%\bfloor{\frac{t_{c(r)}-3\ttime}{\hbt}} \hbt + \ttime$,
$\start(i) \leq t_{x(r)} = t_{c(r)} - 2\ttime - \hbt$, the expected value of
$\tcst(i)$ is $\aeq Z$ (which can be established by an argument similar to the
one in the proof of \mprop{p:comparison}).
%, since with probability $\aeq 1$,
%$\numprt(i) = 2\cterm$ and $\tow(i) = \ttime + \frac{\hbt-1}{2}$.
%To see this,
%note that with probability $\aeq 1$, $x$ will $\psend$ $m$ for the first time
%at or before $t_x$, since $x$ will get a $\hbmsg$ from $y$ at or before $t_x$
%with probability $\aeq 1$.  Note that $t_x + 2\ttime \leq t_c$, so $x$ will
%$\psend$ $m$ at least $\cterm$ times during $i$ (and, with probability $\aeq
%1$, $x$ will $\psend$ $m$ exactly $\cterm$ times during $i$).
%With probability $\aeq 1$, $2\ttime$ time units and
%$\cterm$ messages later (which is still no later than $t_c$), $x$ will $\prcv$ the
%first $\ack(\Hm)$ \emph{before} it crashes.
%With probability $\aeq 1$, $y$
%will $\psend$ $\ack$ for all the copies of $m$.  The point is that $x$ lives
%long enough to $\psend$ all $\cterm$ copies of $m$ before it crashes.  With a
%similar
Similarly, for $i$ with $\sd(i) = y(r)$ and
%$\start(i) \leq t_{y(r)} =
%\bfloor{\frac{t_{c(r)}-4\ttime}{\hbt}} \hbt + \ttime$,
$\start(i) \leq t_{y(r)} = t_{c(r)} - 3\ttime - \hbt$, the expected value of
$\tcst(i)$ is $\aeq Z$.  (If both processes crash simultaneously at $t_{c(r)}$,
let $t_{x(r)} = t_{y(r)} = t_{c(r)} - 3\ttime - \hbt$.)  }%\end{commentout}
Let
$t_{x(r)} = t_{c(r)} - 2\ttime - \hbt$ and $t_{y(r)} = t_{c(r)} - 3\ttime -
\hbt$.  (If both processes crash simultaneously at $t_{c(r)}$, let $t_{x(r)} =
t_{c(r)} - 3\ttime - \hbt$ as well.)  Let us divide the invocations in a given
run $r \in F_3$ into four categories:
\bit
\item An invocation $i$ is \emph{too late} if $\start(i) > t_{c(r)} + \ttime$.
\item An invocation $i$ is \emph{late} if $t_{\sd(i)} < \start(i) \leq t_{c(r)}+\ttime$.
\item An invocation $i$ is \emph{early} if $\ttime \leq \start(i) \leq t_{\sd(i)}$.
\item An invocation $i$ is \emph{too early} if $\start(i) < \min\{\ttime - \hbt, t_{\sd(i)}\}$.
\eit
In general, for a given run $r \in F_3$, some of these categories could be
empty.  Note that if $i$ is too late, then $\tcst(i) = 0$.  If $i$ is late,
then the expected cost of $i$ is lower than $Z$ (by more than $O(\eps)$) but
higher than $0$ (again, by more than $O(\eps)$).  If $i$ is early, then the
expected cost of $i$ is $\aeq Z$.  If $i$ is too early, then the expected cost
of $i$ is higher than $Z$ (by more than $O(\eps)$).  Let $Z_{\ell}$ be the
expected cost of late invocations and let $Z_{e}$ be the expected cost of
invocations that are too early.  Let $F_3(j_1, j_2, j_3, j_4) = \{ r \in F_3 :
r$ has exactly $j_1$ invocations that are too early, $j_2$ early invocations,
$j_3$ late invocations, and $j_4$ invocations that are too late$\}$.
We have that
\[
\E(\csendav \gv F_3(j_1, j_2, j_3, j_4)) \aeq
\frac{j_1 Z_e + j_2 Z + j_3 Z_{\ell}}{j_1+j_2+j_3+j_4}.
\]
Letting $z_e = \frac{Z_e}{Z}$ and $z_\ell = \frac{Z_\ell}{Z}$, we get
\[
\E(\csendav) \aeq \sum_{j_1,j_2,j_3,j_4} \frac{j_1 z_e + j_2 + j_3
z_{\ell}}{j_1+j_2+j_3+j_4} \Pr(F_3(j_1,j_2,j_3,j_4)) Z.
\]
Let
\[
\lambda = \sum_{j_1,j_2,j_3,j_4} \frac{j_1 z_e + j_2 + j_3
z_\ell}{j_1+j_2+j_3+j_4} \Pr(F_3(j_1,j_2,j_3,j_4)).
\]
\commentout{
%These statements can all be established with arguments similar to
%those found in the proof of \mprop{p:comparison}.

%Note that $t_{x(r)} - t_{c(r)}$ is
%between $2\ttime$ and $2\ttime+\hbt-1$.  This is because we need $x$ to be
alive long enough after $\start(i)$ so that it will $\psend$ $m$ exactly
$\cterm$ times with probability $\aeq 1$.
%Since $x$ does not start $\psending$
%$\Hm$ until it $\prcvs$ a $\hbmsg$ from $y$,
%it might be slightly more than $2\ttime$ since $x$ has to
%wait for a $\hbmsg$ from $y$ before $\psending$ $m$ on behalf of $i$.  We need
%$x$ to be up $2\ttime$ time units after $\start(i)$ so that $x$ will $\psend$
%$m$ exactly $\cterm$ times with probability $\aeq 1$.
Similarly, $t_{y(r)} - t_{c(r)}$ is roughly $3\ttime$.  The reason that it is
$\ttime$ longer is because now we need $x$ to $\psend$ $\ack(\Hm)$ exactly
$\cterm$ times with probability $\aeq 1$ and $x$ must $\prcv$ $m$ before it can
start $\psending$ $\ack(\Hm)$.

We say that an invocation
$i$ is \emph{early} iff $\start(i) \leq \bfloor{\frac{t_c-3\ttime}{\hbt}}
\hbt + \ttime$.  An invocation is \emph{late} iff it is not early.
An invocation $i$ is \emph{very late} iff
$\start(i) > \bfloor{\frac{t_{c(r)}}{\hbt}} \hbt$.
Note that if $i$ is very late then $\tcst(i) = 0$, since $\tow(i) = 0$ and
$\numprt(i) = 0$ (because no $\hbmsg$ can arrive after $i$ starts).
If $i$ is early, then $\E(\tcst(i)) \aeq Z$, by arguments similar to those in
the proof of \mprop{p:comparison}.
If $i$ is late but not very late, we have
$\numprt(i) \leq 2\cterm - 1$,

The
intuition is that early invocations start long before the first crash, so their
expected cost is $\aeq Z$.

Suppose the first crash in $r$ happens at time $t_c(r)$ and $x(r)$ is the process that
crashes and $y(r)$ is the other process.  Note that $x(r)$ $\psends$ the last
$\hbmsg$ at time $t_h(r) = \bfloor{\frac{t_c(r)}{\hbt}} \hbt$ and that $\hbmsg$
arrives at time $t_e(r) = t_h(r) + \ttime$ (if $y(r)$ is up).  So $\tcst(i) = 0$ for $i$
such that $\start(i) > t_e(r)$, since $\sd(i) = y(r)$ for these invocations and $y(r)$
will never $\prcv$ $\hbmsg$s from $x(r)$ after time $t_e(r)$.  Let $t_x = $
%be the earlist time such that
% $$\ts \Pr(\numprt(i) < 2\cterm \gv \sd(i) = x(r) \wedge
%\start(i) > t_x(r)) = 1$$
and let $t_y(r)$ be the earliest time such that $$\ts \Pr(\numprt(i) < 2\cterm \gv
\sd(i) = y(r) \wedge \start(i) > t_y(r)) = 1.$$
For $i$ with $\sd(i) = x(r)$, the expected cost is $\aeq Z$ if $\start(i) \leq t_x(r)$
and for $i$ with $\sd(i) = y(r)$, the expected cost is also $\aeq Z$ if $\start(i)
\leq t_y(r)$.  (This can be established via arguments similar to those of the
proof of \mprop{p:comparison}.)  The intuition is that the invocations started
``long enough'' before the first crash will have $Z$ as their expected cost,
where long enough means that the processes have a chance to $\psend$ $2\cterm$
messages.  (For invocations starting too close to the crash, the processes will
not have a chance to $\psend$ $2\cterm$, independent of the behavior of the
link; thus their expected cost is lower.)
%For invocations made by $x$ between
%$t_x$ and $t_c-1$ and invocations made by $y$ between $t_y$ and $t_e$, the
%expected cost is (strictly) between $Z$ and $0$.
Let $F^i_3 = \{ i \in r \in F_3 : \sd(i) = x(r) \wedge t_x(r) \leq \start(i)$
or $ \sd(i)=y(r) \wedge t_y(r) \leq \start(i) \leq t_e(r) \}.$
%(In runs in which both
%processes crash simultaneously, take $t_x(r) = t_y(r) = \max \left\{0,
%\bfloor{\frac{t_c-3\ttime}{\hbt}}\hbt + \ttime\right\}$.)
Let $\kappa = \frac{\E(\tcst(i) \gv F^i_3)}{Z}.$

The proof of
\mprop{p:comparison} shows that $\E(\tcst(i)) \aeq Z$, where $Z$ is defined as
above.  However, we see that $\E(\tcst(i) \gv \start(i) \geq t_c - 3\ttime) \not
\aeq Z$, since $\numprt(i) < 2\cterm$ given that $i$ starts after $t_c -
3\ttime$ (since $x$ would crash before $\psending$ all the messages

Suppose $x$ is the first process to crash in $r$ and $x$ crashes at time $t_c$.
Let $y$ be the other process (\eg, if $x$ is $p$ then $y$ is $q$).
Note that each process can only get $\hbmsg$s at times of the form $a \hbt +
\ttime$, so each process will only $\psend$ messages (that are not $\ack$s) at
times of the abovementioned form.  Thus if $y$ starts an invocation after time
$t_y = \bfloor{\frac{t_c-4\ttime}{\hbt}} \hbt + \ttime$ (which is essentially
the last time of the form $a \hbt + \ttime$ before $t_c - 3\ttime$), $2\cterm -
\numprt = O(1)$ for that invocation, since $x$ does not live long enough to
$\psend$ (all) the $\ack$s.  Let us call invocations $y$ makes before Similarly, if
$x$ starts an invocation after time $t_x = \bfloor{\frac{t_c-3\ttime}{\hbt}}
\hbt + \ttime$ (the last time of the form $a \hbt + \ttime$ before $t_c -
2\ttime$), $2 \cterm - \numprt = O(1)$ for that invocation, since $x$ would
crash (and stop $\psending$ $m$) before $\prcving$ $\ack(\Hm)$.
For the invocations made by $x$ ($y$) before time $t_x$ ($t_y$), our previous
calculations apply and we have that the expected cost of these invocations are
$\aeq Z$, where $Z$ is defined as above.
Let $X$ be the expected cost of the invocations given that the invocation is
made by $x$ \emph{after} time $t_x$.  Let $Y$ be the expected cost of the
invocations
made by $y$ after time $t_y$ and \emph{before} time $\bfloor{\frac{t_x}{\hbt}}
\hbt + \ttime$ (the time $y$ would $\prcv$ the last $\hbmsg$ from $x$ if $y$ is
still up).

Let $\lambda_x =
\frac{X}{Z}$ and $\lambda_y = \frac{Y}{Z}$.  Let $F_3(i_1, i_x, i_y, i_0)$ be
the subset of $F_3$ such that $i_1$ is the number of invocations $x$ made
before $t_x$ plus the number of invocations $y$ made before $t_y$,

Note that
the $\numprt < 2\cterm$ for the invocations $y$ makes that is too close to
$t_c$, since $x$ would crash before it is able to $\psend$ all the $\ack$s.
How close is too close?  Let us consider a failure-free invocation of $\sdrc$
in which $y$ is the sender. Once $y$ starts an invocation, it has to wait for a
$\hbmsg$ from $x$ before $\psending$ $m$.  Since $\hbmsg$s are $\psent$ every
$\hbt$ time units, $y$ will $\prcv$ $\hbmsg$s only at times of the form $a \hbt
+ \ttime$.  So an invocation started at time $t_0$ will $\psend$ its first
message at time $\bceil{\fract{t_0-\ttime}{\hbt}}\hbt + \ttime$.
Thus $x$ will $\prcv$ $m$ for the first time at time
$\bceil{\fract{t_0-\ttime}{\hbt}}\hbt + 2\ttime$, which means $y$
will $\prcv$ the first $\ack(m)$ at time $t_1 =
\bceil{\fract{t_0-\ttime}{\hbt}}\hbt + 3\ttime$.  (In the mean
time, $y$ has $\psent$ $\cterm$ copies of $m$.)  So $x$ will $\prcv$ the last
copy of $m$ at time $t_2 = \bfloor{\fract{t_1-\ttime}{\hbt}}\hbt +
\ttime$; that is when $x$ will $\psend$ $\ack(\Hm)$ for the last time.  Suppose
the invocation happens at time

Note that $x$ can only $\psend$ $\ack$s at times of
the form $a\msgt+2\ttime$, since these are the only times messages from $y$
could arrive.

after time
$\bfloor{\frac{t_c - 3\ttime}{\msgt}} \msgt$, since $x$ would crash before it is
able to $\psend$ all the $\ack$s.  Similarly, $\numprt < 2\cterm$ for the
invocations $x$ makes after time $\bfloor{\frac{t_c-2\ttime}{\msgt}} \msgt$,
since $x$ would stop $\psending$ $m$ because it crashed.  $\tow < \ttime$ for
the invocations made by either $x$ or $y$ after time ${t_c-\ttime}$,

There are four different kinds of invocation that interests us.  Let
Also, the expected cost of invocations starting shortly before the first crash
will be less than the expected cost of invocations that start long before the
first crash, since $\numprt$ and $\tow$ must be low given that the
Invocations that occur shortly before the first crash will have less expected
cost, since we could have $\tow - \ttime$ and $\numprt - 2\cterm$ both be $O(1)$.
%joe8 \lambda needs to be more carefully defined; what's the space?
%fcc8
%Given a run $r \in C_3$, we have that
%both processes eventually crash.
Given a particular invocation of $\rb$ in $r$, it started either when both
processes are up or when only one process is up.
%fcc9
%fcc24
%Let $\cjk$ be the subset of $F_3$ where $j$ invocations are started when both
%processes are up, and $k$ invocations are started when only one is up.
Let $\cijk$ be the subset of $F_3$ where $i$ invocations are started at least
$3\ttime$ time units before the first crash, $j$ invocations are started within
$3\ttime$ time units before the first crash, and $k$ invocations are started at
or after the first crash.  Note that the expected cost of the first group of
invocations is $\aeq Z$, where $Z$ is defined as above.  The cost of the last
group of invocations is $0$, given our definitions.  The expected cost of the
second group is more complicated: the expected cost is less than $Z$ but the
difference between $Z$ and the expected cost is $O(1)$.  To see this, note that
an invocation that is started $1$ time unit before the first crash has cost
$\leq 2 \cm + \cw$, independent of $\msgt$ and $\ttime$ (since $\tow = 1$ and
at most two messages are $\psent$ on behave of this invocation).
%fcc24
%We have that $\E(\csendav \gv \cjk) = \frac{j}{j+k}
We have that $\E(\csendav \gv \cjk) \aeq \frac{j}{j+k}
%joe10
%\E(\csend).
%fcc21 \tempvar
\tempvar$, where $\tempvar$ is as defined as above.
Thus $\E(\csendav) = \sum_{j,k}\frac{j}{j+k} \Pr(\cjk)
%\E(\csend);
\tempvar$.
Let $\lambda = \sum_{j,k}\frac{j}{j+k}\Pr(\cjk)$.
}%\end{commentout}
%fcc9
%Let $i_1(r)$ be the number of invocations started when both processes are up and
%$i_2(r)$ be the total number of invocations in $r$.
%joe9: Francis, this isn't right; $\Pr(r) = 0$ for every run r.
%Let $\lambda = \sum_{r\in
%C_3} \Pr(r) \frac{i_1(r)}{i_2(r)}$.
%Intuitively,  $\lambda$ is the probability that
%an invocations starts when both processes are still up.
%Thus the expected
%number of invocations with non-trivial cost is $\lambda \tcmp(r)$.
%Each of
%these invocations, as before, has expected cost $\left(2\cterm\cm + (2\ttime +
%\frac{\hbt-1}{2})\cw\right)$.
Then the
expected
%fcc10
per-invocation
cost of the protocol is
%fcc10 indeed
%fcc24 added \aeq
$\aeq ((1-\pct_p)(1-\pct_q)\lambda + \pct_p \pct_q) \tempvar$, as desired.
%fcc10 		moved for syntax coloring
%joe10:
%\left(2\cterm\cm + ( 2\ttime + \frac{\hbt-1}{2})\cw\right)$.
}%end{commentout}
%joe24: New argument
%fcc25: rewrote
\commentout{
There are three types of invocations $i$ of $\rb$ in runs of $F_3$.
We say that
\begin{itemize}
\item $i$ is {\em proper\/} if both processes are up when $i$ starts;
\item $i$ is {\em improper\/} if one process crashes less than
$\ttime$ units before $i$ starts;
\item $i$ is {\em one-sided\/} if one process crashes more than $\ttime$
units before $i$ starts.
\end{itemize}
It is easy to see that the cost of a one-sided invocation is 0 and the
expected cost of a proper invocation is $\aeq \tempvar$.  The cost of an
improper invocation is $k \cm$ for some $k \le
\bceil{\frac{\ttime}{\hbt}}$, since the receiver will $\psend$ a message
for each of the at most $\bceil{\frac{\ttime}{\hbt}}$ $\hbmsg$s it
$\prcv$s.  Thus,
$\E(\csendav \gv F_3) = \lambda \tempvar$ for some constant $\lambda$
with $0 < \lambda < 1$ (where $\lambda$ clearly depends on the relative
probability of the three types of invocations and the expected number of
$\hbmsg$s received by the sender in an improper invocation).
}%\end{commentout}
%fcc25
We now turn our attention to $F_3$.
Let $F_3(t_1, t_2, i_1, i_2, i_3)$ be a subset of $F_3$ with the following
properties:
\bit
\item the first crash in $r$ happens at time $t_1$,
\item the second crash in $r$ happens at time $t_2$,
\item the number of invocations starting before time $t_1 - 3\ttime - \hbt$ is
$i_1$,
\item the number of invocations starting between times $t_1 - 3\ttime - \hbt$ and $t_1 +
\ttime$ is $i_2$, and
\item the number of invocations starting after time $t_1 + \ttime$ is $i_3$.
\eit
%where
%\bit 
%\item $0 \leq i_1 \leq \ell_1 = \max \{0, t_1 - 3\ttime - \hbt\}$,
%\item  $0 \leq i_2 \leq \ell_2 = 4\ttime+\hbt$, and
%\item $0 \leq i_3 \leq \ell_3 = \max\{0, t_2 - t_1 - \ttime\}$.\footnote{Without these
%restrictions, some of the sets would be empty.}
%\eit
It is clear that each of these sets are measurable.  (Some of them are empty,
so they will have probability 0; we could introduce restrictions to rule out
the empty ones, but leaving them in is not a problem.)
%Let $\pf = \pf_p + \pf_q - \pf_p \pf_q$.  
%Note that $$
%\begin{array}{lll}
%\Pr(F_3(t_1, t_2, i_1, i_2, i_3)) 
%& = & (1-\pf)^{t_1} (\pf_p (1-\pf_q)^{t_2 - t_1} \pf_q + \pf_q
%(1-\pf_p)^{t_2-t_1} - \pf_p \pf_q) \\
%& & \cdot \cb{\ell_1}{i_1}\spb^{i_1} (1-\spb)^{\ell_1 - i_1} 
% \cb{\ell_2}{i_2}\spb^{i_2} (1-\spb)^{\ell_2 - i_2} 
% \cb{\ell_3}{i_3}\spb^{i_3} (1-\spb)^{\ell_3 - i_3}.\\
%\end{array}
%$$
%The first line is the probability of the first crash happening at time $t_1$
%and the second crash happening at time $t_2$.  The second line is the
%probability of having the number of invocations in each of the time periods.

Suppose  $F_3(t_1,t_2,i_1,i_2,i_3)$ is not empty. Then 
\[
\E(\csendav \gv F_3(t_1, t_2, i_1, i_2, i_3)) \aeq 
\frac{i_1 + \kappa(t_1,t_2,i_1,i_2,i_3) i_2}{i_1+i_2+i_3 + 1} Z,\]
where $0 < \kappa(t_1,t_2,i_1,i_2,i_3) < 1$.  
%joe25: technically, it doesn't make sense to talk about the expected
%cost of an invocation.
%To see this, note that the
%expected cost of an invocation in the first group is 
%in the interval
%\left[Z-\frac{\hbt-1}{2}\cw, Z+\frac{\hbt-1}{2}\cw\right]$, 
Roughly speaking, the expected cost of an invocation in the first group
is $Z$, since if no
messages are lost (which happens with probability $\aeq 1$), the number of
messages $\psent$ is exactly $2\cterm$ and the time of waiting is between
$\ttime$ and $\ttime + \hbt - 1$,  
%joe25
%The exact time of waiting depends on when
depending on when
the first $\hbmsg$ arrives after the invocation starts.  If no messages are
lost, a $\hbmsg$ is $\prcvd$ every $\hbt$ time units, so the wait for a
$\hbmsg$ is $\frac{\hbt-1}{2}$ on average.  Thus the first group of invocations
contribute $i_1 Z$ to $\tcst(r)$, on average.
As for the second group, they contribute something less than $i_2 Z$ to
$\tcst(r)$ on average;
%joe25
%, since they started too late (in the sense that the
%pending process crash would affect the cost of these invocations, 
%by reducing
in many of these invocation, the first process crash (which happens at
most $3 \ttime + \hbt$ after the beginning of an invocation in the
second group) may reduce
the time of waiting or the number of messages $\psent$.  That is why we have a
multiplicative constant $\kappa(t_1,t_2,i_1,i_2,i_3)$ in front of $i_2$.
The last group of invocations all have zero cost, since by the time they
started, the surviving process (which must be the invoker) will never $\prcv$
any new $\hbmsg$s from the crashed process; so the time of waiting and the
number of messages $\psent$ are both zero.

Thus we have
\[
\begin{array}{lll}
\E(\csendav \gv F_3) 
&=& \ds \sum_{t_1, t_2, i_1, i_2, i_3} \E(\csendav \gv F_3(t_1, t_2, i_1, i_2,
i_3)) \Pr(F_3(t_1, t_2, i_1, i_2, i_3)) \\
& \aeq & \ds Z 
\sum_{t_1, t_2, i_1, i_2, i_3} \frac{i_1 +
\kappa(t_1,t_2,i_1,i_2,i_3)i_2}{i_1+i_2+i_3+1} \Pr(F_3(t_1,t_2,i_1,i_2,i_3)).\\
\end{array}
\]
Let
\[
\lambda = \sum_{t_1, t_2, i_1, i_2, i_3} \frac{i_1 +
\kappa(t_1,t_2,i_1,i_2,i_3)i_2}{i_1+i_2+i_3+1}
\Pr(F_3(t_1,t_2,i_1,i_2,i_
%joe25
%3)),\\
3)).\\
\]
%joe25
Clearly $\lambda < 1$
and $\E(\csendav \gv F_3) \aeq \lambda Z$, as desired.

%joe9: This paragraph needs to be rewritten more carefully as above,
%using B_2 and C_2
Now we turn to the expected heartbeat costs per invocation.  Each process will
$\psend$ a $\hbmsg$ every $\hbt$ time units for as long as it is up.
%joe9: isn't this \floor rather than \ceil?
So if in $r$ a process is up at time $t$, then it
%fcc24 sent
$\psent$
$\bceil{\fract{t}{\hbt}}$ $\hbmsg$s in $r$
%joe8:
up to time $t$.
%joe9: this needs to be rewritten
%fcc9
%Consider a run in which both processes are correct.  Then
Suppose $r \in F_2$.
%joe10: this false;  the number of complete invocations is not
%necessarily $2T\sigma$.
Then,
%fcc24 added
$\nhb(r,t) = 2 \bceil{\fract{t}{\hbt}}$, and
by the Law of Large Numbers, for all %fcc10 $\delta > 0$, %fcc10   the
%fcc10	 \delta -> \eta (\hbt is \delta)
$\eta > 0$,
%fcc10: need to take lim
%$\Pr(\abs{\tcmp(r,T) - 2T \spb} \le \delta t \gv C_2) = 1$.  Thus,
%fcc19
%$\Pr(\lim_{t\rar\infty}\abs{\tcmp(r,t) - 2t \spb} \le \eta t \gv C_2) = 1$.  Thus,
$$\Pr\left(\lim_{t\rar\infty}\abs{\tcmp(r,t) - 2t \spb} \le \eta t \lgv
F_2\right) = 1.$$ Thus,
%$\nhb(r,t)/(\tcmp(r,t)+1) = \bceil{\frac{2t}{\hbt}}/(2t \spb)$, so
%fcc9 $\lim_{t\rar\infty} \nhb(r,t)/(\tcmp(r,t)+1) = \frac{1}{\hbt\spb}$.
%fcc19
%$\Pr(\lim_{t\rar\infty} (\nhb(r,t)/(\tcmp(r,t)+1)) = \frac{1}{\hbt\spb} \gv C_2) = 1$.
$$\Pr\left(\lim_{t\rar\infty} \frac{\nhb(r,t)}{\tcmp(r,t)+1} =
%joe22: I think a factor of 2 is missing; this caused the change in the
%statement of the theorem.  Please check.
%\frac{1}{\hbt\spb} \lgv C_2\right) = 1.$$
%fcc24:  there are two processes so it's fine
%\frac{1}{2\hbt\spb} \lgv F_2\right) = 1.$$
\frac{1}{\hbt\spb} \lgv F_2\right) = 1.$$
%joe9: cut the next line; unnecessary
%with probability 1, by the Law of Large Numbers (applied to the denominator).
%[[ACTUALLY HERE WE CAN CLAIM EQUALITY SINCE WE ARE TAKING LIMITS.]]
%fcc9

Next, suppose $r \in F_1$.
%If only one process (say $p$) eventually crashes, then $p$
%will send finitely many $\hbmsg$s and invoke $\rb$ finitely often.  Thus the
%limit is not affected by $p$'s actions.  After $p$ crashes, we have
Then one of the processes will
$\psend$ only finitely many $\hbmsg$s and invoke
$\rb$ finitely often.  Thus after the crash, we have
%cc19   c_1 -> H, c_2 -> I_1, c_3 -> I_2
%fcc10	add I_2
%fcc19 $\nhb(r,t)/(\tcmp(r,t)+1) = (\bceil{\frac{t}{\hbt}}+H)/(I_2 + I_1 +1)$,
\[
\frac{\nhb(r,t)}{\tcmp(r,t)+1} =
\frac{\bceil{\fract{t}{\hbt}}+H}{I_2 + I_1 + 1},
\]
where $H$ is the number of
%fcc19 $\hbmsg$s sent by
%joe20: again, I prefer the original; this doesn't read right.
%times the crashed process $\psend(\hbmsg)$
times the crashed process $\psends$ $\hbmsg$
%$p$
%the crashed process
%joe11
in $r$,
$I_1$ is the number of
times
%$p$
the crashed process invoked $\rb$
%joe11
in $r$,
and $I_2$ is the number of times the live
process invoked $\rb$
%joe11
in $r$.
For all $\eta > 0$, we have that
%fcc19 $\Pr(\lim_{t\rar\infty}\abs{I_2 - t\spb} \le \eta t \gv C_1) = 1$.
$$\Pr\left(\lim_{t\rar\infty}\abs{I_2 - t\spb} \le \eta t \lgv F_1\right) = 1.$$
Thus,
%joe11
%we see that
%fcc19 $\Pr(\lim_{t\rar\infty} (\nhb(r,t)/(\tcmp(r,t)+1)) = \frac{1}{\hbt\spb} \gv C_1) = 1$
$$\Pr\left(\lim_{t\rar\infty} \frac{\nhb(r,t)}{\tcmp(r,t)+1} =
\frac{1}{\hbt\spb}
\lgv F_1\right) = 1.$$
%fcc9
% with probability
%fcc9 	1.  (The case in which $q$ crashes is completely analogous.)
%1, conditional on $I$.

%joe9: rewrote;
%fcc9	C_3^{j,k} -> \cjkp
Finally, consider the set $F_3$, where both processes crash.
%fcc24
Again, the situation here is more complicated, since there are only finitely
many complete invocations and $\hbmsg$s in each run, so we cannot resort to the
Law of Large Numbers.
%fcc9 	moved
%joe10: cut; unnecessary
%We have that $\lim_{T\rar\infty} \nhb(r,t)$ and
%$\lim_{t\rar\infty} \tcmp(r,t)$ are both finite; let us call them
%$\nhb(r)$ and $\tcmp(r)$, respectively.
Let $\cjkp$ be the set of runs where $p$ crashes at time $j$ and $q$
crashes at
time $k$.  Clearly $\Pr(\cjkp \gv F_3) = (1-\pf_p)^j(1-\pf_q)^k\pf_p \pf_q$ and
the number of heartbeats
%fcc24 sent
$\psent$
in runs of
%joe10
%$C_3$
%fcc10	\delta -> \hbt
$\cjkp$
is $\bceil{\frac{j}{\hbt}} + \bceil{\frac{k}{\hbt}}$.
%fcc9	added
%joe22
%Let $n=j+k$.
%joe10
%As a preliminary result,
%joe22
Let $\nhbav(r) = \lim_{t\rar\infty} \frac{\nhb(r,t)}{\tcmp(r,t)+1}$.
Observe that
\[
%\begin{array}{ll}
\begin{array}{lll}
%fcc19 \E(\frac{1}{\tcmp+1}  \gv \cjkp)
%joe22: \tcmp(r) is undefined
%\ds \E\left(\frac{1}{\tcmp(r)+1}  \lgv \cjkp\right)
%&
\ds \E(\nhbav  \gv \cjkp)
%\\
&
 = & \ds
%joe22: replaced n by j+k throughout, since this is what we use
%\sum_{h=0}^n \frac{1}{h+1}\spb^h(1-\spb)^{n-h} \cb{n}{h} \\
\sum_{i=0}^{j+k}
%joe22: I think you were missing this term; added it throughout
%\frac{1}
\frac{\bceil{\frac{j}{\hbt}} + \bceil{\frac{k}{\hbt}}}
{i+1}\spb^i(1-\spb)^{j+k-i} \cb{j+k}{i} \\
&
 = & \ds
\frac{\bceil{\frac{j}{\hbt}} + \bceil{\frac{k}{\hbt}}}
%{\spb(n+1)} \sum_{h=0}^{n}
{\spb(j+k+1)} \sum_{i=0}^{j+k}
%fcc19 \spb^{n+1}
\spb^{i+1}
(1-\spb)^{j+k-i}
\cb{j+k+1}{i+1}
%fcc19 added
\\
%joe10: cut (just used h; this is standard)
%\end{array}
%\]
%Letting $h' = h+1$, we get
%\[
%\begin{array}{lll}
%\E(\frac{1}{\tcmp+1}  \gv \cjkp)
%& = & \ds \frac{1}{\spb(j+k+1)} \sum_{h=1}^{j+k+1}
%& = & \spb^{h'}(1-\spb)^{j+k+1-h'}\cb{j+k+1}{h'} \\
&
 = & \ds
\frac{\bceil{\frac{j}{\hbt}} + \bceil{\frac{k}{\hbt}}}
{\spb(j+k+1)} \sum_{i=1}^{j+k+1} \spb^{i}(1-\spb)^{j+k+1-i}
\cb{j+k+1}{i} \\
&
 = & \ds
%\frac{1}
\frac{\bceil{\frac{j}{\hbt}} + \bceil{\frac{k}{\hbt}}}
{\spb(j+k+1)}(1-(1-\spb)^{j+k+1}).
\end{array}
\]
%joe10
%Now we turn to $\E(\nhb/(\tcmp+1))$.
Thus,
%fcc10 	added \ds
\[
%\begin{array}{ll}
\begin{array}{lll}
%joe22
%\ds\E\left(\frac{\nhb(r)}{\tcmp(r)+1}
%&
\ds\E(\nhbav
%fcc19 added
 \gv F_3)
%\\
&
%\right)
%joe22: added
 = &\ds \sum_{j,k} \E(\nhbav \gv \cjkp) \Pr(\cjkp)\\
&
 = &\ds \sum_{j,k} \frac{\bceil{\frac{j}{\hbt}} +
\bceil{\frac{k}{\hbt}}}{\spb(j+k+1)}(1-(1-\spb)^{j+k+1})(1-\pf_p)^j(1-\pf_q)^k\pf_p\pf_q\\
%joe10
%\]
%We can break the above into two parts:
%\[
%\begin{array}{lll}
%\ds \E\left(\frac{\nhb}{\tcmp+1}\right)
&
 = &\ds \sum_{j,k}\frac{\bceil{\frac{j}{\hbt}} +
\bceil{\frac{k}{\hbt}}}{\spb(j+k+1)}(1-\pf_p)^j(1-\pf_q)^k\pf_p\pf_q\\
&
 & \ds
- \sum_{j,k} \frac{\bceil{\frac{j}{\hbt}}
+\bceil{\frac{k}{\hbt}}}{\spb(j+k+1)}(1-\spb)^{j+k+1}(1-\pf_p)^j(1-\pf_q)^k\pf_p\pf_q.
\end{array}
\]
Note that
%fcc19 $(\bceil{\frac{j}{\hbt}} + \bceil{\frac{k}{\hbt}})/((j+k+1)\spb) < D_1$
$$\frac{\bceil{\frac{j}{\hbt}} + \bceil{\frac{k}{\hbt}}}{\spb(j+k+1)} < L_1$$
for some constant $L_1$ (roughly $\frac{1}{\hbt\spb}$).  Thus the second
summand above is bounded above by
\[
%\begin{array}{ll}
\begin{array}{lll}
%&
\ds
L_1\pf_p\pf_q(1-\spb)\sum_{j,k}((1-\spb)(1-\pf_p))^j((1-\spb)(1-\pf_q))^k
%\\
&
%joe10: added
%fcc10 	added \ds
%joe22: I think two signs are wrong
%& = &\ds \frac{D_1\pf_p\pf_q(1-\spb)}{(\spb + \pf_p + \spb \pf_p)
%(\spb + \pf_q + \spb \pf_q)}\\
 = &\ds \frac{L_1\pf_p\pf_q(1-\spb)}{(\spb + \pf_p - \spb \pf_p)
(\spb + \pf_q - \spb \pf_q)}\\
&
 \aeq &\ds \frac{L_1\pf_p\pf_q(1-\spb)}{\spb^2},
\end{array}
\]
%joe10
%which is $O(\eps)$; thus
which is $O(\eps^2)$.  Thus
%joe22:
we can ignore the second summand.  Taking $L(j,k) =
\bceil{\frac{j}{\hbt}} + \bceil{\frac{k}{\hbt}} -
\frac{j+k+1}{\hbt}$, we get that
\[
%\begin{array}{ll}
\begin{array}{lll}
%joe22
%\ds\E\left(\frac{\nhb(r)}{\tcmp(r)+1}
% \lgv C_3 \right)
%&
\E(\nhbav \gv F_3)
%\\
&
\aeq & \ds \sum_{j,k}
\frac{\bceil{\frac{j}{\hbt}} + \bceil{\frac{k}{\hbt}}}{\spb(j+k+1)}
(1-\pf_p)^j(1-\pf_q)^k\pf_p\pf_q \\
%fcc21 more detail
&
= & \ds
\sum_{j,k}\frac{1}{\spb\hbt}
(1-\pf_p)^j(1-\pf_q)^k\pf_p\pf_q \\
& & +
\ds
\sum_{j,k}\frac{L(j,k)}{\spb(j+k+1)}(1-\pf_p)^j(1-\pf_q)^k\pf_p\pf_q\\
&
= & \ds \frac{1}{\spb\hbt} +
\frac{1}{\spb}\sum_{j,k}\frac{L(j,k)}{j+k+1}(1-\pf_p)^j(1-\pf_q)^k\pf_p\pf_q.
\end{array}
\]
%joe22
%where $D_2$ is between $\frac{-1}{\hbt}$ and
%%fcc21 more details
%%$2$.
%$2$, since $D_2$ is essentially
%$\bceil{\frac{j}{\hbt}} + \bceil{\frac{k}{\hbt}}
%- (\frac{j}{\hbt} + \frac{k}{\hbt} + \frac{1}{\hbt})$.
%joe22
%If we can show that the second
%summand above is $O(\eps)$, we would be done.
It clearly suffices to show that the second summand above is $O(\eps)$.
%joe10: moved back
%To this end, we assume that $\sqrt{\pf_p}$ and
%$\sqrt{\pf_q}$ are also $O(\eps)$.
%Then
%fcc10
Note that $\frac{1}{j+k+1} < \sqrt{\pf_p}$ if $j > \frac{1}{{\sqrt{\pf_p}}}$;
similarly, $\frac{1}{j+k+1} < \sqrt{\pf_q}$ if $k > \frac{1}{\sqrt{\pf_q}}$.  Finally,
it is clear that $\frac{1}{j+k+1} \leq 1$ for all $j, k \geq 0$.  Call
the second summand above $S$.
%joe22
Since $L(j,k) < 2$, we have that
%Then
%joe11
%we see that
%fcc10
%joe10: need a few more details here; it goes by a bit too fast.
\[
%\begin{array}{ll}
\begin{array}{lll}
%fcc10
%\ds \sum_{j,k}\frac{D_2}{j+k+1}(1-\pf_p)^j(1-\pf_q)^k\pf_p\pf_q
%fcc21 added $\spb$
%S
%&
\spb S
%\\
&
%\begin{array}{lll}
%joe22: replacing D_2 by 2
%& \leq & \ds D_2\sqrt{\pf_p}
 \leq & \ds 2\sqrt{\pf_p}
%joe10
%\sum_{j>\sqrt{\pf_p},k}
%fcc24
%\sum_{j>\sqrt{\pf_p}}\sum_k
\sum_{j>\frac{1}{\sqrt{\pf_p}}}\sum_k
(1-\pf_p)^j(1-\pf_q)^k\pf_p\pf_q
\\
&
& \ds + 2\sqrt{\pf_q}
%fcc24
%\sum_j \sum_{k>\sqrt{\pf_q}}
\sum_j \sum_{k>\frac{1}{\sqrt{\pf_q}}}
(1-\pf_p)^j(1-\pf_q)^k\pf_p\pf_q \\
%fcc24
%& \ds +\sum_{j\leq\sqrt{\pf_p},k\leq\sqrt{\pf_q}} 2\pf_p\pf_q \\
&
&
\ds +\sum_{j\leq\frac{1}{\sqrt{\pf_p}}} \sum_{k\leq\frac{1}{\sqrt{\pf_q}}} 2\pf_p\pf_q \\
&
 \leq &  2(\sqrt{\pf_p}+\sqrt{\pf_q})+ 2\sqrt{\pf_p\pf_q}.
\end{array}
\]
Since we assumed that $\sqrt{\pf_p}$ and $\sqrt{\pf_q}$ are both $O(\eps)$ for
this theorem, the second summand above is $O(\eps)$.
%fcc24
Thus,
%\[
%joe22
%\E\left(\frac{\nhb(r)}{\tcmp(r)+1}  \lgv C_3 \right)
%fcc24
%\E(\nhbav) \aeq \frac{1 + \pct_p \pct_q}{\spb\hbt}
$\E(\nhbav \gv F_3) \aeq \frac{1}{\spb\hbt}$.
%joe24
%so
It follows that
%\frac{\pct_p \pct_q}{\spb\hbt} +
%\frac{\pct_p+\pct_q-2\pct_p\pct_q}{\spb\hbt} +
%\frac{1-\pct_p-\pct_q+\pct_p\pct_q}{\spb\hbt} = \frac{1}{\spb\hbt}
%\]
$\E(\nhbav) \aeq \frac{1}{\spb\hbt}$,
as desired.~\eprf
%joe22: this repeats the previous line.  What did you intend?
%fcc24: this is unconditional whereas the line before was conditioned
%on C_3
%and we have that
%\[
%\E\left(\frac{\nhb(r)}{\tcmp(r)+1}\right) \aeq
%\frac{1}{\spb\hbt}\mbox{. \eprf}
%\]
%%fcc11

%joe25
%\bibliographystyle{plain}
\bibliographystyle{alpha}
\bibliography{z,joe}
\end{document}